\documentclass[longauth]{aa}  

\usepackage{graphicx}
\usepackage{txfonts}
\usepackage[dvipsnames]{xcolor}
\usepackage{float}
\usepackage{latexsym}
\usepackage{graphicx}
\usepackage{amssymb}
\usepackage{amsmath}
\usepackage{amsfonts}
\usepackage{color}
\usepackage{cool}
\usepackage{subfig}
\usepackage{dcolumn}
\usepackage{soul}
\usepackage[hidelinks,colorlinks=true,linkcolor=blue,citecolor=blue]{hyperref}
\usepackage[utf8]{inputenc}
\usepackage[english]{babel}
\usepackage{import}
\usepackage[dvipsnames]{xcolor}
\usepackage{booktabs}
\usepackage{natbib}
\bibpunct{(}{)}{;}{a}{}{,}

\graphicspath{ {./images/} }

\newcommand{\hi}{H{\sc i}}

\begin{document}

\title{The BINGO project I:}
\subtitle{Baryon acoustic oscillations from integrated neutral gas observations }

\author{Elcio Abdalla\inst{1}\thanks{eabdalla@usp.br},
Elisa G. M. Ferreira\inst{1,2},
Ricardo G. Landim\inst{3},
Andre A. Costa\inst{4},
Karin S. F. Fornazier\inst{1},
Filipe B. Abdalla\inst{1,5,6,7},
Luciano Barosi\inst{8},
Francisco A. Brito\inst{8},
Amilcar R. Queiroz\inst{8},
Thyrso Villela\inst{5,9,10},
Bin Wang\inst{4,11},
Carlos A. Wuensche\inst{5},
Alessandro Marins\inst{1},
Camila P. Novaes\inst{5},
Vincenzo Liccardo\inst{5},
Chenxi Shan\inst{12},
Jiajun Zhang\inst{13},
Zhongli Zhang\inst{14},
Zhenghao Zhu\inst{12},
Ian Browne\inst{15},
Jacques Delabrouille\inst{16,17,18},
Larissa Santos\inst{4,11},
Marcelo V. dos Santos\inst{8},
Haiguang Xu\inst{12},
Sonia Anton\inst{19},
Richard Battye\inst{15},
Tianyue Chen\inst{15,20},
Clive Dickinson\inst{15},
Yin-Zhe Ma\inst{21,22},
Bruno Maffei \inst{23},
Eduardo J. de Mericia\inst{5},
Pablo Motta\inst{1}, 
Carlos H. N. Otobone\inst{1},
Michael W. Peel\inst{24,25},
Sambit Roychowdhury \inst{15},
Mathieu Remazeilles\inst{15},
Rafael M. Ribeiro \inst{1},
Yu Sang\inst{4},
Joao R. L. Santos\inst{8},
Juliana F. R. dos Santos \inst{1},
Gustavo B. Silva \inst{1},
Frederico Vieira\inst{5},
Jordany Vieira \inst{1},
Linfeng Xiao\inst{11},
Xue Zhang\inst{4},
\and
Yongkai Zhu\inst{12}}

\institute{Instituto de F\'isica, Universidade de S\~ao Paulo - C.P. 66318, CEP: 05315-970, S\~ao Paulo, Brazil\\
\email{bingotelescope@usp.br}
\and 
 Max-Planck-Institut f{\"u}r Astrophysik, Karl-Schwarzschild Str. 1, 85741 Garching, Germany
\and 
 Technische Universit\"at M\"unchen, Physik-Department T70, James-Franck-Stra\text{$\beta$}e 1, 85748 Garching, Germany
\and 
 Center for Gravitation and Cosmology, College of Physical Science and Technology, Yangzhou University, 225009, China
\and 
Instituto Nacional de Pesquisas Espaciais, Divis\~ao de Astrof\'isica,  Av. dos Astronautas, 1758, 12227-010 - S\~ao Jos\'e dos Campos, SP, Brazil
\and 
Department of Physics and Astronomy, University College London, Gower Street, London,WC1E 6BT, UK 
\and 
Department of Physics and Electronics, Rhodes University, PO Box 94, Grahamstown, 6140, South Africa
\and 
Unidade Acad\^emica de F\'{i}sica, Univ. Federal de Campina Grande, R. Apr\'{i}gio Veloso, 58429-900 - Campina Grande, Brazil
\and 
Instituto de F\'{i}sica, Universidade de Bras\'{i}lia, Bras\'{i}lia, DF, Brazil 
 \and 
 Centro de Gest\~ao e Estudos Estrat\'egicos - CGEE,
SCS Quadra 9, Lote C, Torre C S/N Salas 401 - 405, 70308-200 - Bras\'ilia, DF, Brazil
 \and 
School of Aeronautics and Astronautics, Shanghai Jiao Tong University, Shanghai 200240, China 
\and 
School of Physics and Astronomy, Shanghai Jiao Tong University, 800 Dongchuan Road, Minhang, Shanghai 200240, China
\and 
Center for Theoretical Physics of the Universe, Institute for Basic Science (IBS), Daejeon 34126, Korea
\and 
Shanghai Astronomical Observatory, Chinese Academy of Sciences, 80 Nandan Road, Shanghai 200030, China  
\and 
Jodrell Bank Centre for Astrophysics, The University of Manchester, Oxford Road, Manchester, M13 9PL, U.K. 
\and 
Laboratoire Astroparticule et Cosmologie (APC), CNRS/IN2P3, Universit\'e Paris Diderot, 75205 Paris Cedex 13, France 
\and 
IRFU, CEA, Universit\'e Paris Saclay, 91191 Gif-sur-Yvette, France 
\and 
Department of Astronomy, School of Physical Sciences, University of Science and Technology of China, Hefei, Anhui 230026
\and 
Center for Research and Development in Mathematics and Applications – CIDMA, Campus de Santiago, 3810-183 Aveiro, Portugal
\and 
MIT Kavli Institute for Astrophysics and Space Research, Massachusetts Institute of Technology, 77 Massachusetts Ave, Cambridge, MA 02139, USA
\and 
School of Chemistry and Physics, University of KwaZulu-Natal, Westville Campus, Private Bag X54001, Durban 4000, South Africa
\and 
NAOC-UKZN Computational Astrophysics Centre (NUCAC), University of KwaZulu-Natal, Durban, 4000, South Africa
\and 
Institut d’Astrophysique Spatiale, Orsay (CNRS-INSU), France
\and 
Instituto de Astrof\'{i}sica de Canarias, 38200, La Laguna, Tenerife, Canary Islands, Spain 
\and 
Departamento de Astrof\'{i}sica, Universidad de La Laguna (ULL), 38206, La Laguna, Tenerife, Spain
}
              
\authorrunning{E. Abdalla et al.}
\titlerunning{The BINGO Project I: Baryon Acoustic Oscillations from Integrated Neutral Gas Observations}

\abstract
{Observations of the redshifted 21-cm line of neutral hydrogen (\hi)  are a new and powerful window of observation that offers us the possibility to map the spatial distribution of cosmic \hi\, and learn about cosmology.  \textbf{B}aryon Acoustic Oscillations from \textbf{I}ntegrated \textbf{N}eutral \textbf{G}as \textbf{O}bservations (BINGO) is a new 
unique radio telescope designed to be one of the first to probe baryon acoustic oscillations (BAO) at radio frequencies.}
{BINGO has two science goals: cosmology and astrophysics. Cosmology is the main science goal and the driver for BINGO's design and strategy. The key of BINGO is to detect the low redshift BAO to put strong constraints on the dark sector models and test the $\Lambda$CDM (cold dark matter) model. Given the versatility of the BINGO telescope, a secondary goal is astrophysics, where BINGO can help discover and study fast radio bursts (FRB) and other transients,  as well as study Galactic and extragalactic science. In this  paper, we introduce the latest progress of the BINGO project, its science goals, describing the scientific potential of the project for each goal and the new developments obtained by the collaboration.}
{BINGO is a single dish transit telescope that will measure the BAO at low-z by making a 3D map of the \hi\, distribution through the technique of intensity mapping over a large area of the sky. In order to achieve the project's goals, a science strategy and a specific pipeline for cleaning and analyzing the produced maps and mock maps was developed by the BINGO team, which we generally summarize here.}
{We introduce the BINGO project and its science goals and give a general summary of recent developments in construction, science potential, and pipeline development obtained by the BINGO collaboration in the past few years. We show that BINGO will be able to obtain competitive constraints for the dark sector. It also has the potential to discover several FRBs in the southern hemisphere. The capacity of BINGO in obtaining information from 21-cm is also tested in the pipeline introduced here. Following these developments, the construction and observational strategies of BINGO have been defined.}
{There is still no measurement of the BAO in radio, and studying cosmology in this new window of observations is one of the most promising advances in the field. The BINGO project is a radio telescope that has the goal to be one of the first to perform this measurement and it is currently being built in the northeast of Brazil. This paper is the first of a series of papers that describe in detail each part of the development of the BINGO project.}


\maketitle



\section{Introduction}
This is the first of a series of papers presenting the Baryon Acoustic Oscillations [BAO] from Integrated Neutral Gas Observations (BINGO) project. Here the project is presented from a theoretical point of view, showing the idea behind it as well as the plan of the whole construction of the BINGO radio telescope, which evolved from ideas previously discussed elsewhere \citep{Battye:2012,Wuensche:2018,Wuensche:2019,Peel:2019,Battye:2016qhf}. We also outline the structure of the project as a whole. The companion papers II through VII show the instrument \citep{BINGO.Instrument:2020}, the optics design \citep{2020_optical_design}, the mission simulation \citep{2020_sky_simulation}, the component separation analysis \citep{2020_component_separation}, the mock \citep{2020_mock_simulations}, and forecasts on several models \citep{2020_forecast}, respectively. 
The BINGO concept was presented in \cite{Battye:2013} with later updates presented in \cite{Dickinson:2014}, \cite{Bigot-Sazy:2015jaa}, \cite{Battye:2016qhf}, and \cite{Wuensche:2018}. 

The BINGO project  is a single dish (strictly, it revolves around a single dish telescope, see paper II (\citealt{BINGO.Instrument:2020}) and paper III (\citealt{2020_optical_design}) for further details) radio telescope aiming at the observation of the 21-cm line, corresponding to a hyperfine interaction of the atomic hydrogen.  It  will survey a sky area of 6000 square degrees in a redshift range from 0.127 to 0.449 (corresponding to a frequency span of 980 to 1260 MHz) with an angular resolution of $\approx 40$ arcmin.   The phase 1 of the project will have 28 feed horns and receivers with dual polarization. It is designed to have a system temperature of $\approx 70$ K (see, e.g., \citealt {Battye:2013,Bigot-Sazy:2015jaa}). These numbers will be finalized depending on financial and technical achievements. We expect to increase the number of feed horns, up to 56, in phase 2, also updating the receivers aiming at a smaller system temperature.

Hydrogen is the most common element of the Universe. We assume that, barring bias, a hydrogen map is representative of the full matter content in the Universe. BINGO will be able to deliver detailed maps of the matter distribution in its redshift range. We can thus extract BAO measurements comparable to those in the optical observations \citep{Eisenstein:2005su}. Moreover, in view of the telescope characteristics, it will be able to describe phenomena at very short time scales, being thus an interesting 
device to study pulsars and fast radio bursts  \citep[FRBs;][]{Lorimer:2007qn,Lorimer:2012mn}.


The standard approach to probe the large scale structure (LSS) is to perform a large redshift survey. This yields the positions and redshifts of a large number of galaxies, that can be used to infer their density contrasts and two-point correlation functions. Essentially, galaxies are being used as tracers of the underlying total matter distribution. This approach has been successfully carried out in the optical band (e.g., SDSS, \citealt{Blanton:2017qot} or 2dF, \citealt{Lewis:2002yk}). In the radio band the natural tracer is the 21-cm line of neutral hydrogen (\hi), but the volume emissivity associated with this line is low, meaning that detecting individual galaxies at $z \sim 1$ requires a very substantial collecting area. This was the original motivation for the Square Kilometre Array \citep[SKA;][]{SKA:Blake2004,SKA:Wilkinson2004,Abdalla:2004ah}. 

However, 
measuring each galaxy individually is an expensive and inefficient way of mapping cosmological volume, and just a few percent of the Universe has been mapped with such a type of technique.
It has been proposed \citep{Battye:2004re,Chang:2007xk,Loeb:2008hg,Sethi:2005gv,Visbal:2008rg} that mapping the Universe by measuring the collective 21-cm emission of the underlying matter can be significantly more efficient than the previous technique using individual galaxies. Such a technique is called intensity mapping  \citep[IM;][]{Madau:1996cs,Bharadwaj:2001vs}. Therefore, exciting progress can be made with much smaller scale instruments. The novel idea is to exploit the broad beam at low-frequencies to carry out IM \citep{Battye:2004re,Loeb:2008hg,Masui:2013,Switzer:2013,Peterson:2006bx} and hence measure the overall integrated \hi\, brightness temperature of a large number of galaxies, taken together as a tracer of the LSS.

A number of approaches has been proposed to conduct IM surveys using an interferometer array rather than in total intensity with single or double dish telescopes \citep[see, for instance,][]{vanBemmel:2012cj,Pober:2012zz,Baker:2011jy}. This approach can have a number of advantages, but it also requires complicated, and hence expensive, electronics to make the correlations. Using a single telescope with a stable receiver system is the lowest-cost approach to IM measurements of BAO at low redshifts \citep{Battye:2013}, in particular when compared with IM experiments that use interferometer arrays. Interferometer arrays are likely to be the best approach to probe higher redshifts at $z \sim 1$, where an angular resolution of $\sim 0.1^\circ$ is required. BINGO will be the first telescope in the world operating in its frequency range whose goal is to study BAO with 21-cm IM.

In recent years new observational results have provided insight into various aspects of general relativity (GR) and cosmology, especially concerning the astonishing contribution provided by dark matter (DM) and dark energy (DE) to the total energy budget of the Universe, contributions that form an overwhelming majority of the mass of the Universe \citep[see, e.g.,][]{Abdalla:2020ypg}. Therefore, it is expected that a new physical picture will emerge once more is found about the structure of the dark sector  \citep{Abdalla:2009wr,Wang:2006qw,Wang:2016lxa}.  There are many observational programs underway aimed at elucidating the dark sector of the Universe, involving many different approaches. Among these, the study of BAO is recognized as one of the most powerful probes of the properties of DE  \citep[see e.g.,][]{Albrecht:2006um} -- others include weak  gravitational lensing  \citep{Bartelmann:1999yn,Refregier:2003ct}, cluster counts \citep{He:2010ta} and supernova's luminosity distance inference  \citep{Feng:2008fx,He:2010im}. To date, BAO have only been detected by performing large galaxy redshift surveys in the optical waveband. However, given the profound implications of these measurements, it is important that they are confirmed in other wavebands, where systematic effects are different, and measured over a wide range of redshifts. The radio band provides a unique and complementary observational window for the understanding of DE via the redshifted 21-cm \hi\,  emission line from distant galaxies \citep{Abdalla:2009wr}.

Following the first detection of BAO using the 2dF and SDSS surveys \citep[e.g.,][]{Eisenstein:2005su,Percival:2009xn}, BAO have been measured with several new generation galaxy and quasar redshift surveys. In particular, the 10 recent measurements with the BOSS survey achieved a precision in the BAO scale of a few percent in the redshift range $z = [0, 2]$ \citep{Anderson:2013zyy,Alam:2016hwk}\footnote{For a compilation of all the results from BOSS, see \url{https://www.sdss.org/science/final-bao-and-rsd-measurements/}.}. In the future, several large experiments designed to measure BAO's in the optical band, such as DESI \citep{Aghamousa:2016zmz,Aghamousa:2016sne}, Euclid \citep{EUCLID:Refregier2010}, J-PAS \citep{Benitez:2008ts,J-PAS:2014hgg,Bonoli:2020ciz} and WFIRST \citep{Levi:2011dq,Green:2011zi}  \citep[as well as LSST for imaging;][]{2002TysonAPS}, aim to achieve a precision below the percent level at higher redshifts. These experiments are large undertakings and will have their first light during the following decade or are already under commission, like the J-PAS telescope.

Radio \hi\, IM experiments such as BINGO are very promising complementary routes to study BAO. Indeed, optical experiments are required to gather data with very high angular resolution (about an arcsecond) to detect individual galaxies, while the BAO scale is typically of the order of a degree, depending on the redshift in question. Radio BAO experiments on the other hand have angular resolutions that are well matched to the BAO scale performing the above mentioned experiments in a very  cost-effective way. Also, an experiment such as BINGO will pave the way to the extension of the use of \hi\, IM to high redshifts and open a new window on cosmological epochs, which cannot be probed in the optical or near-infrared bands. Moreover, since radio experiments use \hi\, as opposed to stellar light as a tracer of the LSS, they are subject to different astrophysical effects.   IM experiments are  more direct tracers of DM, being less sensitive to ``peak statistics'' bias and 
thus complementary to optical surveys, with which they can be combined to reduce systematic errors in both.

The main goal in the present project is to detect the BAO present in the maps as well as to map the 3D distribution of \hi\,, with observations spanning a few years. Extensive simulations of the expected BAO signal and of how it can be extracted from the observations are presented in the companion papers IV and V \citep{2020_sky_simulation,2020_component_separation}. At frequencies below a few GHz the future of radio astronomy will be dominated by the ambitious SKA project that will constrain cosmological models \citep{Maartens:2015mra} with unprecedented accuracy. Its design has been driven by the aim to probe the \hi\, 21-cm line over a broad range of radio frequencies. The high importance of the IM technique is highlighted by the decision to change the baseline design of the first phase of the SKA instrument to enable it to do IM \citep{SKAWG:2020}. The SKA will use very similar techniques to ours and having developed bespoke analysis techniques, the knowledge involved in BINGO will be in a perfect startup to take advantage of the increase in the signal-to-noise ratio that the SKA will yield \citep{Bull:2015nra,Abdalla:2015kra}.

In particular, one of the big challenges for those measurements is that large Galactic as well as extragalactic foregrounds have to be removed. The \hi\, signal amplitude is typically $\sim 100 \,\mu$K whereas the foreground continuum emission from the Galaxy is $\sim 1 $ K in the BINGO frequency band. Fortunately, the integrated 21-cm emission exhibits characteristic variations as a function of frequency whereas the continuum emission has a very smooth spectrum; this allows for the two signals to be separated \citep{Chang:2007xk,Liu:2011hh,Chang:2010jp,Olivari:2015}. The BINGO project is an excellent ground for testing such ideas using IM as its method of approach (see paper IV and V of this series: \citealt{2020_sky_simulation,2020_component_separation}).

%

The telescope will be located in the hills of ``{\it Serra da Catarina}'' in the municipality of Aguiar, Para\'{\i}ba, Brazil. The measured radio frequency interference (RFI) is at the lowest amplitude we could observe among all measurements performed in South America, thus excellent for the project (see  Section~\ref{sec:instrument} and the respective BINGO paper on the instrument \citep{BINGO.Instrument:2020}. The telescope consists of two dishes concentrating the radiation on to the set of horns; this radiation then follows through a waveguide to the receiver, which is also described in \citet{BINGO.Instrument:2020} . 

The main science goal of BINGO is cosmology –– see Section~\ref{sec:science}, where we describe how BINGO will measure BAO and redshift space distortion (RSD) to constrain cosmological parameters in different cosmological models. Nonetheless, the BINGO telescope, as designed, also allows studying different astrophysical phenomena, from transient objects, such as FRBs and pulsars to Galactic and extragalactic science. Thus, such astrophysical issues are further scientific goals of the BINGO telescope and are discussed in Section~\ref{sec:astro}. In Section~\ref{sec:pipeline} we consider the pipeline, which analyses the signals gathered by the backends, after which we have the material for the science cases, cosmology and astrophysics.  In Section \ref{sec:comp} we summarize other IM experiments already in operation or planned for the near future. Section \ref{sec:conclusions} is reserved for conclusions.


\section{Instrument overview} \label{sec:instrument}

Detecting signals of $\sim 100  \mu$K with a noncryogenic receiver of standard performance requires that every visited sky pixel contains an accumulated integration time of more than 1 day over the course of the observing campaign. Gain variations should be very small on time scales of about $\sim$ 20 minutes, to allow for longer integration times. The design of the optical and electronics systems for BINGO, as well as its observation strategy, try to address these concerns from start, and the overall project has been designed to keep the subsystems as simple as possible, using, when available, ``off-the-shelf'' components to minimize the final cost and allow for a quicker construction phase. This section contains a summary of the instrument, with a more detailed description presented in the instrument companion paper \citep{BINGO.Instrument:2020}.

BINGO is a fixed transit telescope, employing the rotation of the Earth to observe the sky as it drifts across the instrument's field of view (FoV), in the frequency band $980 - 1260$ MHz. In  doing so it meets the criterion of maximizing the observation time by revisiting the same patch of the sky every day. The BINGO optical design follows a crossed-Dragone configuration \citep{Dragone:1978}, with an instantaneous FoV of $\sim 14.75^{\circ} \times 6.0^{\circ}$ (DEC, RA), covered by 28 horns in a double rectangular, non-redundant array. The telescope structure is aligned north-south and the primary reflector points to $\delta=-15^{\circ}$.  The final studies produced very clean optics, with low sidelobes and good cross-polarization rejection. The telescope focal length is 63\,m, with a full beam of $0.67^{\circ}$, achieved with slight under-illumination of the secondary mirror to avoid stray radiation. Table \ref{tab:bingo_param} contains a summary of the BINGO main parameters, being a subset of Table 1 of \cite{BINGO.Instrument:2020}.

The engineering project for the focal plane allows the vertical, elevation and azimuth, and longitudinal displacement of all feed horns aiming toward a complete and more uniform coverage of the observed sky area. This focal arrangement and clean beam are essential for the full success of the BINGO experiment, as they allow the instrument to resolve structures corresponding to a linear scale of around 150 Mpc in the BINGO redshift range. The chosen declination strip also allows for overlapping with important optical surveys, such as DES, Pan-Starss, LSST and 6dF. The details of the optical design are reported in~\citet{2020_optical_design}.

BINGO receivers will operate at room temperature in a correlation receiver (CR) mode to avoid gain variations and $1/f$ noise levels that might hamper its capability to detect the \hi\, signal. The CR radiometer chains were initially measured in Manchester and later at INPE, and we expect to operate at nominal $ T_{\rm{sys}} \lesssim 70 $ K. The prototypes for the front end, including feed horns, polarizers and magic tees were fully manufactured and successfully tested in Brazil. The design, construction and testing of the prototype corrugated horn and polarizer are described in \cite{BINGO.Instrument:2020}. 

\begin{figure}[ht]
\centering
\includegraphics[width=0.48\textwidth]{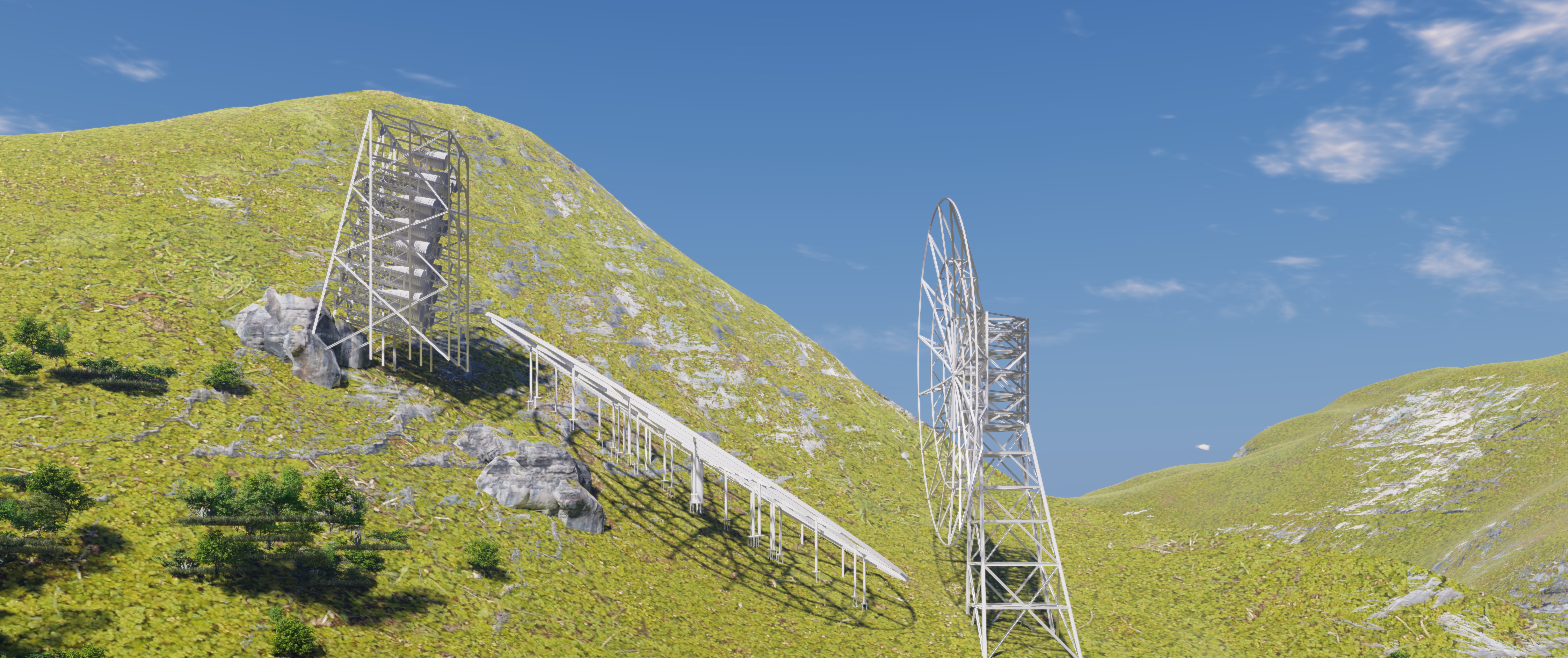}

\caption{BINGO telescope site - artist's view. In this figure we show the hill and the use of the natural inclination of the terrain to place mirrors and horn structure. In this conception the control cabin is located under the service deck. } 
\label{fig:bingo_artistic}
\end{figure}

The telescope is being built in a very isolated area in Aguiar, Para\'\i ba, northeastern Brazil (Lat: 7$^{\circ}$\, 2'\, 29''\, S Long: 38$^{\circ}$\, 16'\,  5''\, W), in an area surrounded by hills, with very small nearby population and hardly any mobile signal in the surroundings. A request for a ``radio-quiet'' zone has been placed to the local authorities and to ANATEL, the Brazilian regulator agency for telecommunications. The site selection process and RFI measurements are described in \citet{Peel:2019}. Fig. \ref{fig:bingo_artistic} presents an artist's view of the telescope. The picture displays a west-east view (left-right) of the site, with the sheltering hill in the background and the access road ending in a service deck. The current design contemplates a control building located under the deck.

\begin{table}[ht]
 \scriptsize
\caption{Summary of the BINGO telescope parameters - Phase 1}
\label{tab:instrument}
\centering
\begin{tabular}{lc}
\hline\hline
Parameter & Value \\
\hline
T$_{\rm sys}$ (K)			            		    & 70		 \\
Lower frequency cutoff  (MHz) - $f_{\rm{min}}$           & 980	 \\
Higher frequency cutoff (MHz) - $f_{\rm{max}}$	        & 1260	 \\
Frequency band - B (MHz)		                    & 280 	 	\\
Number of redshift bands\tablefootmark{a}    			        & 30		  \\
Estimated sensitivity\tablefootmark{b} ($\mu$K)  & 206  \\ 
Number of circular polarizations 			        & 2	\\
Focal length (m)			            		    & 63.2 \\
Number of feed horns (Phase 1)			            & 28		 \\
Primary dish main diameter (m)                 		& $\sim$ 40.0  \\
Secondary dish main diameter (m) 			        & $\sim$ 35.6 \\
Focal plane area ($\textrm{m}^2$)                			    & $ 18.6 \times 7.8 $	 \\
Instantaneous focal plane area ($\textrm{deg}^{\circ}$)   & $14.75 \times 6.0$    \\
Optics FWHM (deg, at 1.1 GHz) 				        & 0.67	 \\
Pixel solid angle (sqr deg) - $\Omega_{\rm{pix}}$		& 0.35	 \\
Telescope  area (m$^2$) - $A_{\rm tot}$	            & 1602 \\
Telescope effective area (m$^2$) - $A_{\rm eff}$	& $\sim 1120 $ \\
Telescope azimuth orientation (deg)	& 0 (North)\\
Primary mirror declination pointing (deg) & -15 \\
Declination strip - instantaneous (deg)                   		    &  14.75$^{\circ}$ \\
Full survey area - 5 yr(square deg)	        		    & $\sim 5324 $	 \\
Mission duration (Phase 1)	                        & 5 years	\\
Duty cycle (estimated, Phase 1)		                & 60 - 90\% of the time	\\
\hline

\end{tabular}
\tablefoot{\scriptsize \\
\tablefoottext{a}{Assumed for science simulations, operational value will be downsampled from the digital backend output} \\
\tablefoottext{b}{Computed for 1 pixel, full 280 MHz band, 1 year of observations and a 60\% duty cycle} \\
}
\label{tab:bingo_param}
\end{table}

BINGO is funded by several Brazilian and Chinese agencies. From the Brazilian side, the major supporter is FAPESP, with 60\% of the funds. China and the UK have also contributed in terms of parts and components.

As far as the project timeline is concerned, soil moving services, as land cleaning and topography, have already been carried out. The project received a green light to go ahead after a major project review in July 2019. We had a positive report from five panelists, including two foreign specialists. The engineering project for the telescope structure, as well as optical project, have been completed, along the main receiver prototype. Some of the construction aspects were impacted in 2020 due to the Covid-19 pandemic, but we had a major advance in early 2021 regarding terrain prospection, roadwork, and the area cleaning.

We have a major part of funding (around 70\%) already approved and available for use from the FAPESP, the S\~ao Paulo State Funding Agency. We also received contributions from our Chinese partners regarding electronic components (circulators, isolators, connectors, cables).


\section{Scientific goals I: cosmology} \label{sec:science}
 
In this section we describe the \hi\  IM signal for cosmology and how it can be used as a tool to measure BAO and RSDs. As a new observation, plenty of work has to be done to understand the foregrounds and systematics involved in the experiment, leading to the development of simulation and analysis pipelines. This work is important to guarantee that BINGO measurements can indeed place competitive constraints on cosmological parameters. With these tools, we show how the BAO and RSD measured from BINGO can be used to constrain cosmological models and test extensions of the $\Lambda$CDM model \citep{ 2020_forecast}.  

The aim of this section is to present the 21-cm signal and the corresponding \hi\, angular  power spectrum. Then using the Fisher matrix formalism we obtain forecasts for  several extensions of the $\Lambda$CDM model and alternative cosmologies, such as dynamical DE, interacting DE and modified gravity. Bounds on the sum of neutrino masses and on the 21-cm parameters are also presented. For further  details interested readers can find more information in the companion paper VII \citep{2020_forecast}.

\subsection{The \hi\, intensity mapping signal and power spectrum} \label{sec:HI_signal}
    
The 21-cm line emission from \hi\, is a technique already used for many years in astronomy to measure the rotation curves of galaxies \citep{Huchtmeier:1975,Bosma:1981,Corbelli:1989,Puche:1991,Olling:1996}. Since hydrogen is the most abundant component in our Universe, measuring its 21-cm emission line can be used as a tracer of the LSS. This is the idea of the 21-cm  \hi\, as a cosmological probe. However, after reionization 
there is little hydrogen left in neutral form. \hi\, is mostly hosted inside dense clouds in damped Ly-$\alpha$ (DLA) systems, that provides shielding to preserve the \hi\, neutral, with optically-thin Ly-$\alpha$ absorbers in low-density regions being responsible for a small part of the signal \citep{SDSS_Lya,Curran:2006md}. Inside those hosting clouds, the excited energy level is more populated given that they have higher temperatures than the energy difference between the hyperfine levels, although below the DLA transition energy. 
    
The 21-cm  emission is a highly forbidden process, with a transition rate of $2.9 \times 10^{-15} \mathrm{s}^{-1}$, however, the amount of hydrogen in galaxies is very high, allowing for the observation of the emission of  the 21-cm  line. In view of this faint signal, resolving individual galaxies requires much longer integration times.  Instead, we can perform a low angular resolution survey that contains a large number of sources and map the integrated 21-cm emission of those unresolved targets. This is the 21-cm IM technique, allowing to probe large volumes very efficiently.
    
Since the wavelength is redshifted by the expansion of the Universe, we have a precise determination of the redshift of the emitting source. We can thus map the distribution of the \hi\, in the Universe at different redshifts, performing a tomographic analysis  that allows studying the evolution of the matter distribution.  This precise and free measurement of the redshift allows for a precise determination of the RSD signal. These features make 21-cm IM surveys very efficient for determining RSDs.
     
The IM technique measures the full intensity field, the brightness temperature, from the 21-cm emission. 
In the Rayleigh-Jeans limit and at low redshifts, the brightness temperature can be calculated as the sum of all the emitted photons, received by an observer along the line of sight. In a Friedmann-Lema\^{\i}tre-Robertson-Walker (FLRW) universe in the absence of perturbations, the brightness temperature is given by \citep{Hall:2012wd} 
\begin{align}         \bar{T}_b(z)=\frac{3}{32 \pi} \frac{(2\pi \hbar  c)^3 n_{\rm{HI}} A_{21}}{k_B \,E_{21}^2 (1+z) H(z)}\,, \end{align}
where  $E_{21} = 5.88 \, \mu\mathrm{eV}$ is the  energy of a 21-cm photon, $A_{21} \simeq 2.869 \times 10^{-15} \, \mathrm{s}^{-1}$ is the spontaneous emission coefficient, $n_{\rm{HI}}=\rho_{\rm{HI}}/m_{\rm{HI}}$  is the number density of \hi\, atoms of mass $m_{\rm{HI}}$,  $\rho_{\rm{HI}}$ the \hi\, average energy density. For a homogeneous and isotropic  background energy density, we have $\bar{\rho}_{\rm{HI}}(z)=\Omega_{\rm{HI}} (z) \rho_{\rm{crit}} (1+z)^3$, where $\Omega_{\rm{HI}}=\rho_{\rm{HI}}/\rho_{\rm{crit}}$, $\rho_{\rm{crit}}=3M^2_{\rm{pl}}H_0^2$ is the critical energy density and $M_{\rm{pl}}$ is the reduced Planck mass. The brightness temperature is thus given by
\begin{equation}
\bar{T}_{b} (z)= 44 \, \mu \mathrm{K} \, \left( \frac{\Omega_{\rm{HI}}(z) h}{2.45 \times 10^{-4}} \right) \frac{(1+z)^2}{E(z)}\,, \label{eq:mean_Tb}
    \end{equation}
where  $H_0=100 h \, \mathrm{km \, s}^{-1}\mathrm{Mpc}^{-1}$ and $E(z)=H(z)/H_0$. This expression describes the evolution of the brightness temperature with redshift, given a specified cosmology. 

The perturbed regime can be obtained considering the metric expanded in the conformal Newtonian gauge,
\begin{equation} ds^2 = a^2(\eta)\Big[(1+2\Psi(\eta,\Vec{x}))d\eta^2-(1-2\Phi(\eta,\Vec{x}))d\Vec{x}^2\Big]\,,
\label{eq:line_element}    \end{equation}
where the functions $\Phi$ and $\Psi$ determine the space-time gravitational potentials.
In the presence of perturbations, the brightness temperature receives corrections coming from $n_{\rm{HI}}(z,\hat{\mathbf{n}})=\bar{n}_{\rm{HI}}(1+\delta_n)$, from the change in the redshift measured by the observer in this perturbed metric and the perturbed conformal time $\delta \eta (z,\hat{\mathbf{n}})= \bar{\eta}_z + \delta \eta$ around the unperturbed value $\bar{\eta}_z$. To quantify these changes to first order, we expand the observed brightness temperature,
\begin{equation} T_b(z,\hat{\mathbf{n}})=\bar{T}_b \left( 1+\Delta_{T_b} (z,\hat{\mathbf{n}}) \right)\,,    \end{equation}
where the mean brightness temperature $\bar{T}_b$ is given in Eq. (\ref{eq:mean_Tb}) and $\Delta_{T_b} (z,\hat{\mathbf{n}})$ is the first-order fractional perturbation to the brightness temperature. Taking into consideration all the corrections, the first-order fractional perturbation of $T_b$ is given by
\begin{align}
 \Delta_{T_b} (z,\hat{\mathbf{n}})&=\delta_n - \frac{1}{\mathcal{H}} \, \hat{\mathbf{n}} \cdot \left( \hat{\mathbf{n}} \cdot \nabla \mathbf{v} \right) + \left( \frac{d \ln (a^3 \bar{n}_{\rm{HI}})}{d \eta} -\frac{\dot{\mathcal{H}}}{\mathcal{H}} -2 \mathcal{H} \right) \delta \eta \nonumber\\&+ \frac{1}{\mathcal{H}} \dot{\Phi}+\Psi \,.    
 \end{align}
This expression is fully relativistic and each term in this expression corresponds to a different physical effect~\citep{Hall:2012wd}. The first term is the perturbation of the \hi\, density from discrete sources that host the \hi\, in Newtonian gauge. The second term is the RSD term that comes from peculiar velocities of the sources; the third term encodes the effects of the zero-order brightness temperature calculated at the perturbed time of the observed redshift; and the last two terms are related to the integrated Sachs-Wolf (ISW) effect and conversion between increments in redshift to radial distances in the gas frame, respectively.

\vspace{0.5cm}
        
Now, we are ready to define the \hi \, power spectrum. In many papers in the literature of 21-cm \hi\, IM, the 3D $k$-space power spectrum is used  \citep[see e.g.,][]{Bull:2015nra}. However, there are many advantages of using a tomographic 2D angular power spectrum that are specially important for the BINGO science goals, survey strategy and the way the experiment is designed. Among those advantages is the fact that this approach is independent of the cosmological model, since it is not necessary to assume a cosmology to convert into distance. It also allows the inclusion of lensing effect and wide angle correlations, and it can be easily used to make cross correlations with other LSS tracers  \cite[for a discussion of the advantages see][]{Shaw:2008aa,DiDio:2013sea,Tansella:2017rpi,Camera:2018jys}. We  evaluate the \hi\, angular power spectra from the first order brightness temperature fluctuations. First, we  decompose the brightness temperature fluctuation in spherical harmonics,
\begin{equation}
  \Delta_{T_b} (z,\hat{\mathbf{n}})= \sum_{\ell m} \Delta_{T_b, \ell m} (z) Y_{\ell m}(\hat{\mathbf{n}})\,.
\end{equation}
We then Fourier transform the coefficient $\Delta_{T_b, \ell m} (z)=4 \pi i^\ell \int \frac{d^3 \mathbf{k}}{(2\pi)^{3/2}} \Delta_{T_b, \ell} (\mathbf{k}, z) Y^{*}_{\ell m}(\hat{\mathbf{k}})$ and re-write the brightness temperature perturbation as \citep{Hall:2012wd}
\begin{align}
\Delta_{T_b, \ell} (\mathbf{k}, z) &= \delta_n j_\ell(k \chi)+\frac{kv}{\mathcal{H}} j_\ell^{''}(k \chi) + \left( \frac{1}{\mathcal{H}} \dot{\Phi} + \Psi \right) j_\ell(k \chi) \nonumber\\
&- \left( \frac{1}{\mathcal{H}} \frac{d \ln (a^3 \bar{n}_{\rm{HI}})}{d \eta} - \frac{\dot{\mathcal{H}}}{\mathcal{H}^2} - 2 \right) \nonumber\\&\times\left[ \Psi j_\ell(k \chi) + v j_\ell^{'} (k \chi) + \int_0^{\chi} (\dot{\Psi}+\dot{\Phi}) j_\ell(k \chi^{'}) d\chi^{'} \right]\,.
\end{align}
Here, $\chi$ is the conformal distance to redshift $z$, `` $'$ '' denotes the derivative with respect to $k \chi$ and $j_\ell$ is the spherical Bessel function, ignoring the monopole and dipole terms.
       
We thus define the tomographic angular power spectrum,
\begin{equation}
C_\ell (z_i, z_j) = 4 \pi \int d \ln k \mathcal{P}_{\mathcal{R}}(k) \Delta^W_{T_b, \ell} (k, z) \Delta^{W^{'}}_{T_b, \ell} (k, z)\,,
\label{eq:HI_power_spectrum}
\end{equation}
where in $\Delta^W_{T_b, \ell} (\mathbf{k}, z)=\int_0^{\infty} dz \bar{T}_b(z) W(z) \Delta_{T_b, \ell} (\mathbf{k}, z) $ we integrate the temperature fluctuation over the redshift bin with the normalized window function $W(z)$; $\Delta^W_{T_b, \ell} (k, z)=\Delta^W_{T_b, \ell} (\mathbf{k}, z)/\mathcal{R}(\mathbf{k})$ is the transfer function divided by the primordial curvature perturbation $\mathcal{R}(\mathbf{k})$, with dimensionless power spectrum given by $\mathcal{P}_{\mathcal{R}}(k)$.
       

\subsection{Baryon acoustic oscillations and redshift space distortion}
    
\begin{figure}[h!]
\centering 
\includegraphics[scale=0.45]{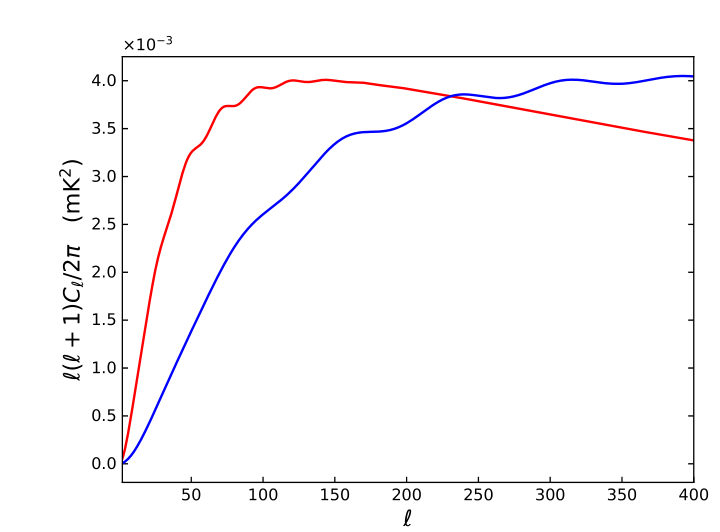}
\caption{\hi\,angular power spectra with a bandwidth of 9.33 MHz at redshifts $z = 0.127$ (red) and $z = 0.449$ (blue) showing the BAO feature.}
\label{Fig:plot_cl_bao}.  
\end{figure}

The BAO are signatures in the matter distribution from the recombination epoch and have become   a prime cosmological probe currently in use. They are based on small oscillations in the photon baryon cosmic fluid that are imprinted on matter perturbations after matter and radiation decouple at $z \approx 1090$ and show up in the distribution of galaxies at all redshifts. The scale of these oscillations ($\approx 150 \, \textrm{Mpc}$) 
is fixed in comoving coordinates and can be used as a standard ruler to measure the geometry of the Universe, thus probing the effects of DE  \citep{Percival:2009xn,Beutler:2011hx,Anderson:2013zyy,Delubac:2014aqe}. In Fourier space the BAO feature appears as wiggles in the power spectrum. Fig. \ref{Fig:plot_cl_bao} shows the BAO wiggles in the \hi\, angular power spectra for the two  redshift boundaries of BINGO.

The RSD is one of the most powerful probes of the rate of growth of structure \citep{Costa:2016tpb}. We have to consider $f \sigma_8$, where $f=-d\ln \delta (z)/d \ln (1+z)$ is the growth rate and $\sigma_8$  measures the amplitude of the (linear) power spectrum on the scale of $8 h^{-1} \, \mathrm{Mpc}$. RSD are effects that appear when we probe observables in 3D space, where the radial (comoving) distance is determined by the observed redshift. It receives contributions from peculiar velocities and  makes the redshift-space clustering anisotropic. 21-cm IM experiments are specially good to determine the RSD signal given its precise method for determining the redshift. Including RSD information leads to the breaking of degeneracies between parameters and the improvement in the constraint of many cosmological parameters, as we show below.


\subsection{Constraining cosmology}
BINGO is a telescope that will measure BAO and RSD using \hi\, as a tracer of the matter density in our Universe. This will provide precise measurements of the expansion history and growth of structures, which can be used to study the evolution of our Universe, properties of DE, extensions to the standard $\Lambda$CDM model and \hi\, physics.


Alternative cosmological models are usually based on either introducing an exotic cosmological component or by modifying GR \citep{Copeland:2006wr,Wang:2016lxa}.
Each of those classes encompasses many different models with different properties and parameters. To distinguish between those models, new and precise measurements of the expansion history and structure growth are necessary, which is the aim of BINGO.

BINGO will be particularly  powerful to improve the constraints on current cosmological parameters in combination with other experiments, such as optical surveys and CMB measurements. This will allow to put tighter constraints on extensions of the $\Lambda$CDM model, which are competitive with current and future LSS experiments. We have forecast BINGO capacity in several different scenarios in the companion paper VII \citep{2020_forecast}. The forecast was done using the Fisher matrix formalism considering the instrumental setup as described in Table~\ref{tab:instrument} but for an observational time of 1 year (and sky coverage with Galactic mask of 2900 deg)~\citep{2020_forecast}, and fiducial cosmological parameters listed in Table \ref{tab:baseline}. In this section, we highlight the main results.
\begin{table}[t]
\caption{Baseline values assumed for the cosmological parameters used in the Fisher forecast analysis, from \cite{Switzer:2013,Aghanim:2018eyx}. The last two parameters describe the \hi\, physics.}
\label{tab:baseline}
\centering
\begin{tabular}{lc}
\toprule\toprule
Parameter & Baseline\\
$\Omega_b h^2$ & 0.022383 \\
$\Omega_c h^2$ & 0.12011 \\
$h$ & 0.6732\\
$n_s$ & 0.96605 \\
$A_s$ & $2.1\times 10^{-9}$ \\
$w_0$ & -1 \\
$w_a$ &  0 \\ 
\midrule
$b_{\rm{HI}}$ & 1 \\ 
$\Omega_{\rm{HI}}$ & $6.2\times 10^{-4}$ \\
\toprule
\end{tabular}
\end{table}

\subsubsection{$\Lambda$CDM model}
The $\Lambda$CDM model is the basis for the current description of the evolution and composition of our Universe. This phenomenological model is our concordance model in cosmology, exhibiting exquisite agreement with current precise large scale observations. We assume here that the Universe is spatially flat  and seeded by adiabatic nearly scale invariant perturbations.\footnote{Spatial flatness can be relaxed.} This model  is composed of baryons, radiation, and two unknown components that are responsible for the majority of the energy budget of the Universe today, DM and DE. In this model, DM is the component responsible for the clustering of perturbations and the formation of structures, and it is described in the hydrodynamical limit by a cold, presureless fluid that is at most weakly interacting with baryons, the CDM. DE is the component responsible for the acceleration of the Universe and it is compatible, in the $\Lambda$CDM model, with a cosmological constant.
        
Such a simple and successful model is parametrized by only six parameters,\footnotemark
         \begin{equation}
              \left\{ \Omega_b h^2,\, \Omega_c h^2,\, \tau,\, A_s,\, n_s, \, h \right\}\,,
              \label{eq.:LCDM}
         \end{equation}
\footnotetext{Sometimes, instead of $h$, the acoustic angular scale $\theta_*$ is used.}
where $\Omega_i=\rho_i/\rho_{\rm{crit}}$ are the baryonic ($i=b$) and CDM ($i=c$) density parameters today, $\tau$ is the reionization optical depth, $A_s$ is the amplitude of the primordial curvature perturbation, $n_s$ is the spectral index of scalar perturbations and $h$ is the Hubble parameter, defined through the Hubble constant $H_0$ by $H_0 = 100 h $ km s$^{-1} $ Mpc$^{-1}$. These parameters can be constrained at the level of percent to subpercent by current cosmological observations \citep{Aghanim:2018eyx}. For 21-cm probes, we have to consider two additional parameters in the analysis describing the \hi\, signal: the \hi\, density parameter and the bias, $\Omega_{\rm{HI}}$ and $b_{\rm{HI}}$, respectively.
    
Although BINGO alone does not provide competitive constraints on the $\Lambda$CDM parameters as compared to \textit{Planck} \citep{Aghanim:2018eyx}, its combination with \textit{Planck} measurements can improve the cosmological fits, especially in the Hubble constant and matter density (by $\sim 25\%$). This breaking of the degeneracy present in the \textit{Planck} data is not new and has been discussed extensively in the data including in \citet{Aghanim:2018eyx} and the main reason for it is the presence of a distance indicator such as the BAOs in the nearby Universe. Besides, BINGO can help solving the tension in $H_0$ between \textit{Planck} and low redshift measurements \citep{Riess:2019cxk}.

\subsubsection{Dynamical dark energy}
In the $\Lambda$CDM scenario, the current acceleration of the Universe is explained by the introduction of a cosmological constant, also interpreted as a component with equation of state equal to $w=p/\rho =-1$. Although simple and well motivated observationally, this description is challenging from a theoretical point of view. 


Moreover, recent data from low redshift measurements present tensions when compared with \textit{Planck} results under the $\Lambda$CDM scenario. This motivates the search for alternative models to explain the current acceleration \citep{Copeland:2006wr,Wang:2016lxa}. Those models consider that DE can be a dynamical component. The simplest phenomenological modification from the cosmological constant is a dynamical model where the equation of state is a constant, $w=w_0$, which is now a free parameter, called the $w$CDM model. Our forecast shows that BINGO alone could already constrain this parameter with a better precision than \textit{Planck} ($\sigma w_0^{\rm{BINGO}} = 17\%$ against $\sigma w_0^{\rm{Planck}} = 26\%$). The combination of BINGO and \textit{Planck} has shown to be very powerful to constrain $w_0$, with $\sigma w_0^{\rm{comb.}} = 3.3\%$, an improvement of more than $87\%$ in the constraint of the DE equation of state.


\begin{figure}[h!]
    \centering
    \includegraphics[scale=0.6]{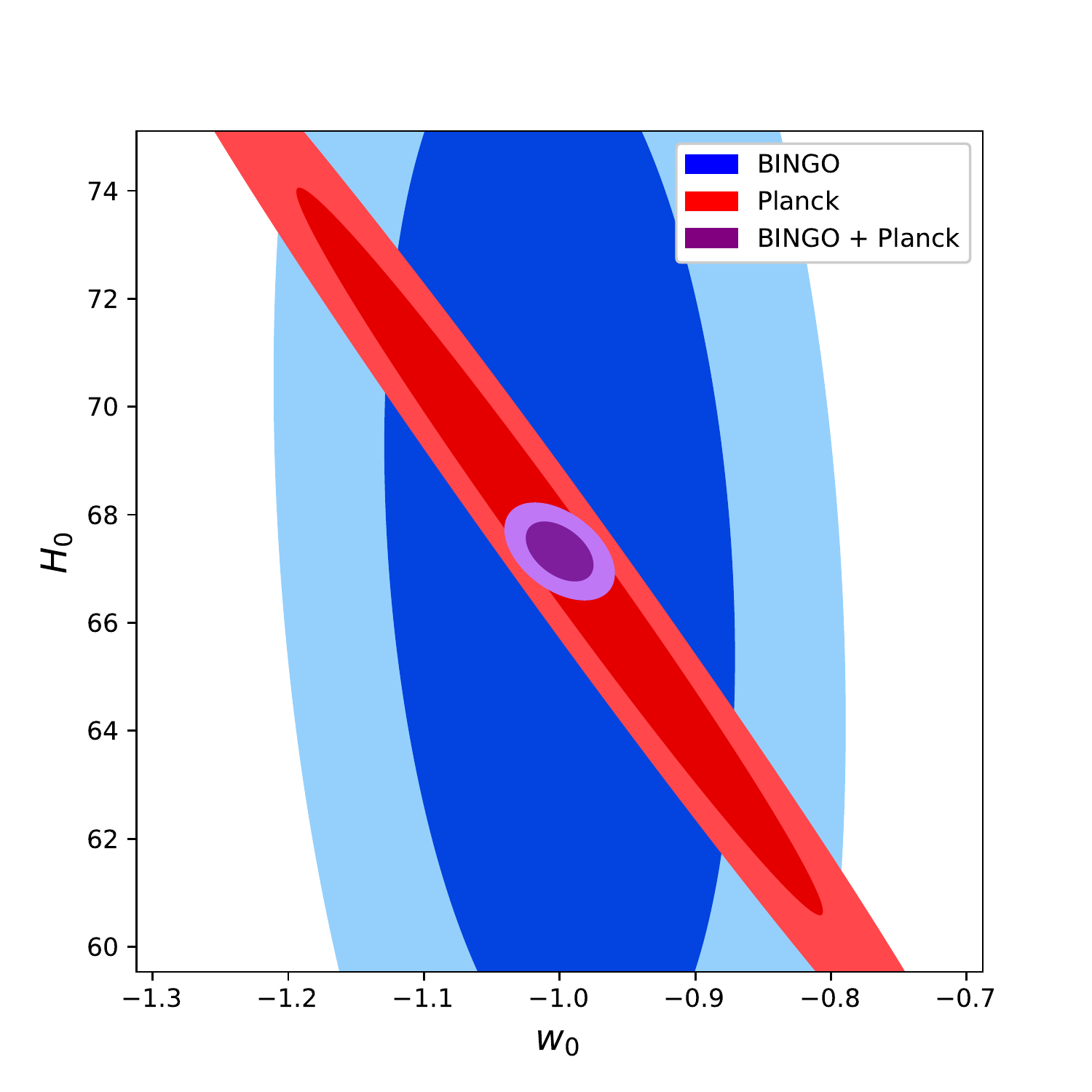}
        \caption{Contour plot showing the $68\%$ and $95\%$ confidence level constraints on the Hubble constant and DE equation of state parameters using BINGO, \textit{Planck} and BINGO + \textit{Planck}, for the dynamical DE model ($w$CDM).}
        \label{Fig:plot_h_w}
\end{figure}

In general, we can consider that the equation of state varies with time (or with the scale factor). Although there are more sophisticated models leading to that behavior, we can probe the evolution of $w(z)$ in a phenomenological way by parametrizing its evolution, as \citep{Chevallier:2000qy,Linder:2002et} 
\begin{equation}
 w(z)=w_0+w_a \frac{z}{1+z}\,,
\end{equation}
known as the CPL parametrization. This model is also called the $w_0w_a$CDM model and has two extra free parameters, $w_0$ and $w_a$, changing the normalized Hubble parameter \citep[see companion paper VII, ][]{2020_forecast}. 
Even tough we introduce two new parameters, the model is well behaved for small redshifts and bounded for high redshifts, which allows a manageable exploration of its parameter space. One of its advantages is that it can be used in many scalar field models of DE and, in this sense, it is a quite general description of a dynamical DE component. Answering the question whether DE is evolving is one of the biggest questions in cosmology, thus testing these models with BINGO is an important goal.

\begin{figure}[h!]
\centering \includegraphics[scale=0.60]{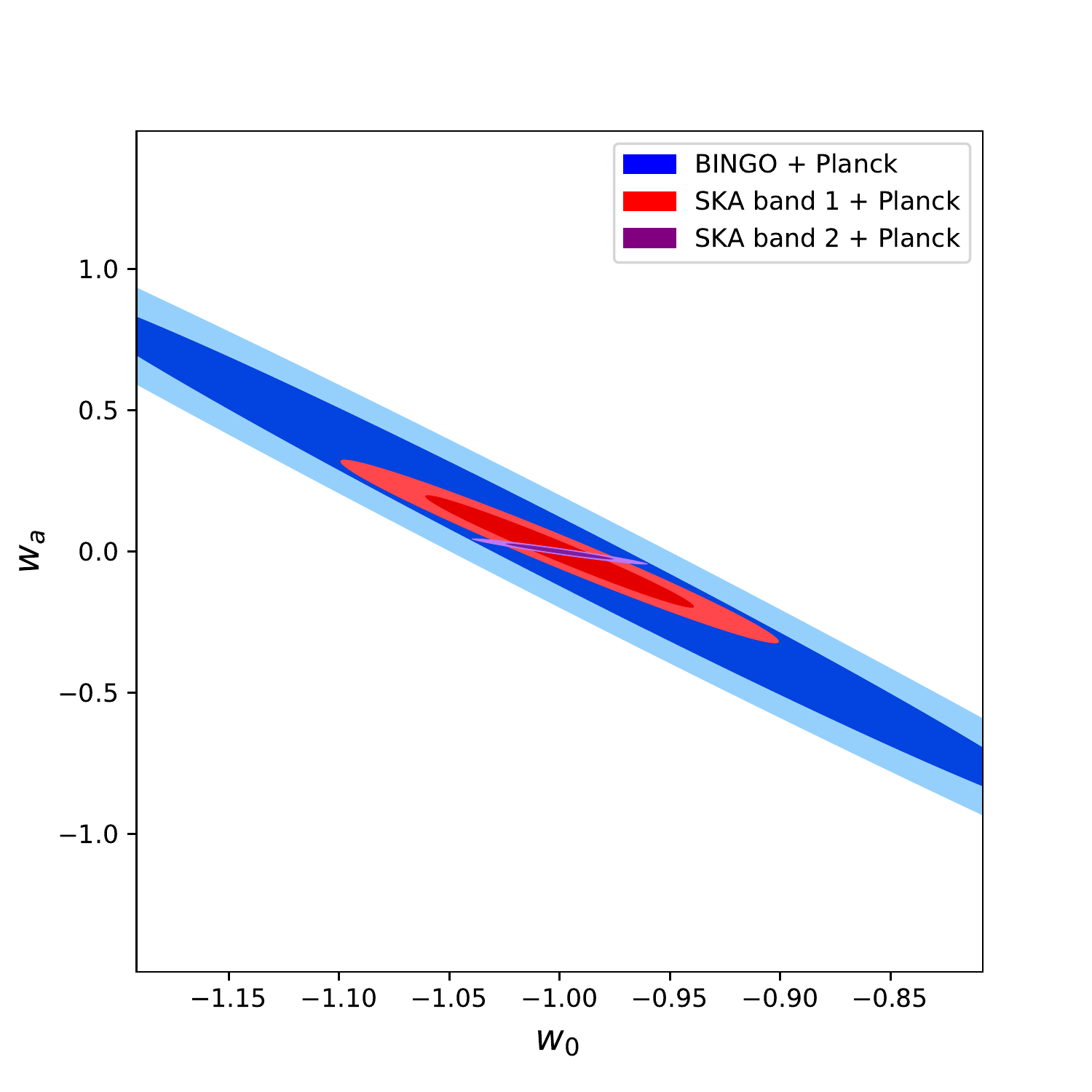}
\caption{Comparison between BINGO, SKA Band 1 and SKA Band 2 for the $68\%$ and $95\%$ confidence level constraints on the DE equation of state parameters for the CPL parametrization in combination with \textit{Planck}. }
\label{Fig:plot_w_wa}
\end{figure}

In Figures \ref{Fig:plot_h_w} and \ref{Fig:plot_w_wa} we show the 68\% and 95\% confidence level marginalized constraints for $H_0$ and $w_0$ (for the wCDM model) and for the equation of state parameters $w_0$ and $w_a$ (for the CPL parametrization), respectively. The combination BINGO + \textit{Planck} improves considerably the constraints on those parameters. In the case of the constraints on the DE equation of state, we compare BINGO's performance with the one from SKA band 1 and band 2, which correspond to different frequency (redshift) bands of SKA  \citep{Bacon:2018dui}. The constraints on the DE equation of state are the ones that present the most significant improvement for SKA in comparison to BINGO. The band 2 of SKA, which corresponds to lower redshifts than band 1 ($z =0-0.5$), presents a much stronger constraint than BINGO. The other cosmological parameters show only a small improvement when SKA is considered, as it can be seen in the in the companion paper VII \citep{2020_forecast}, together with further details. 
These results show the scientific potential of the BINGO project.


\subsubsection{Modified gravity}

The two new components, DM and DE, have motivated extensions of the standard theory, such as Brans-Dicke, Hornsdeski and $f(R)$ gravity. They generally predict a different perturbation dynamics. Scalar perturbations on the FLRW background are described in the Newtonian gauge by Eq. (\ref{eq:line_element}) with the Poisson equation
\begin{equation}
k^2\Psi=-4\pi G a^2\mu(a,k)\rho\Delta\,.
    \end{equation}
The parameters $\mu(a,k)$ and $\gamma(a,k)=\frac{\Psi}{\Phi}$ parametrize deviations from GR, and are $\gamma=\mu=1$ in GR.
    
To forecast constraints on modified gravity models, in the companion paper VII \citep{2020_forecast} a specific form for the parametrization above is considered, which is related to $f(R)$ theories of gravity. The $B_0$-parametrization of $f(R)$ gravity provides a good approximation on quasi-static scales \citep{Hu:2007nk,Giannantonio:2009gi,Hojjati:2012rf} and is expressed as \begin{align}
\mu(a,k) & = (1 - 1.4\times 10^{-8}|\lambda/{\rm Mpc}|^2a^3)^{-1}\frac{1 + 4\lambda^2k^2a^4/3}{1 + \lambda^2k^2a^4}\,, \\
\gamma(a,k) & = \frac{1 + 2\lambda^2k^2a^4/3}{1 + 4\lambda^2k^2a^4/3}\,, \end{align}
where $B_0 \equiv 2H_0^2 \lambda^2$. Considering the $\Lambda$CDM parameters plus $B_0$, we obtained the constraints
    \begin{align}
        \sigma_{B_0} & = 3.1\times 10^{-5}\quad \textrm{(BINGO)} \\
        \sigma_{B_0} & = 5.3\times 10^{-2}\quad \textrm{(\textit{Planck})} \\
        \sigma_{B_0} & = 1.1\times 10^{-5}\quad \textrm{(BINGO + \textit{Planck})}\,.
    \end{align}
BINGO alone can already put a better constraint on $B_0$ as compared to \textit{Planck} and the combination of the two surveys can improve it by  $\sim 64\%$.

\subsubsection{Interacting dark energy}
It has also been proposed that DM and DE interact \citep{Amendola:1999er,Zimdahl:2001ar,Chimento:2003iea}. Such interacting models are challenging because of the unknown nature of both components: it becomes hard to describe the origin of such an interaction from first principles. Despite that, many different models have been studied in the literature \citep[for a review see][]{Wang:2016lxa}.
        
These models are generally described by the divergence of the energy-momentum tensor of the dark sector components, $ \nabla _{\mu} T_{(i)}^{\mu\nu} = Q_{(i)}^{\nu}$. The  quadri-vector $Q_{(i)}^{\nu}$ provides the energy-momentum transfer between DM and DE, such that the total conservation law is still obeyed, $\sum _{i} Q_{(i)}^{\nu}=0$.
    
In order to forecast the power of BINGO in constraining those kinds of models, we consider a phenomenological interaction described by the DM-DE energy transfer as
\begin{equation} Q=3H (\xi_{dm} \rho_{dm}+ \xi_{de}\rho_{de})\,, \end{equation} 
where $\xi_{dm}$ and $\xi_{de}$ are the coupling constants. Here we present the general results for these models, while we refer readers to the more detailed study of the BINGO forecasts in our companion paper VII \citep{2020_forecast}.
    
We can consider two distinct situations: an interaction proportional to the DM density or an interaction proportional to the DE density. Because of regions of instabilities, the last one should be analyzed in two different regions, $w > -1$ and $w < -1$ \citep{He:2008si,He:2010im}. BINGO will provide complementary constraints on the interacting DE parameters and the equation of state. This will help break the degeneracy between them and provide additional information to the CMB data. Considering the BINGO fiducial setup in combination with \textit{Planck} data, we obtain
\begin{itemize}
\item $Q = 3H\xi_{dm}\rho_{dm}$:
\begin{align}
\sigma_w  & = 0.078 \,, \\
\sigma_{\xi_{dm}}  & = 0.0007 \,.
\end{align}
\item $Q = 3H\xi_{de}\rho_{de}\quad \hbox{and}\qquad  w > -1$:
\begin{align}
\sigma_w  & = 0.047 \,, \\
\sigma_{\xi_{de}} & = 0.018 \,.
\end{align} \item $Q = 3H\xi_{de}\rho_{de} \quad \hbox{and} \qquad w < -1$: \begin{align}
\sigma_w & = 0.056 \,, \\
\sigma_{\xi_{de}} & = 0.015 \,. \end{align}
\end{itemize}

The combination with BINGO improves the constraints from \textit{Planck} by at most $\sim 81\%$ for the DE EoS ($Q \propto \rho_{dm}$) and $\sim 77\%$ for the interacting parameter ($Q \propto \rho_{de}$, $w > -1$). Recently, \cite{Bachega:2019fki} have analyzed these models with the latest \textit{Planck} data and in combination with current BAO and SNIa data. Our results from BINGO + \textit{Planck} put competitive constraints to those obtained combining \textit{Planck} with five optical BAO data from 6dFGS ($z = 0.106$), SDSS-MGS ($z = 0.15$), and BOSS data release 12 ($z = 0.38, 0.51, 0.61$), or combining \textit{Planck} with the Pantheon SNIa data.

\subsubsection{Massive neutrinos}
We know today that neutrinos leave detectable imprints on the LSS of our Universe. This knowledge comes from results on neutrino oscillations experiments that show that neutrinos are massive, as a consequence of their oscillation between flavors \citep{Fukuda:1998mi,Ahmad:2002jz}.

Given their small mass and large thermal velocities, neutrinos can free stream away from dense regions before recombination and wash away structures on small scales, which can be measured as a suppression in the matter power spectrum. Although this effect is small, it can already be measured by current experiments and nowadays, with the precision that cosmological experiments have,  it is necessary to take into account the effect of massive neutrinos. The  upper limit on the total neutrino mass from combining \textit{Planck} temperature and polarization with BAO data is $\sum m_{\nu} < 0.13 \,\mathrm{eV}\, ( 95\% \,\text{CL})$ \citep{Aghanim:2018eyx}.  Future experiments have the goal to improve those bounds. 

The presence of massive neutrinos affects cosmology, which is sensitive to its mass and number of massive species. This change in the components of our cosmology modifies the angular diameter distance to last scattering. This can be probed using the angular scale of the CMB first peak. However, this measurement is degenerate with the density of DE and $H_0$ \citep{Olivari:2017}. Using LSS data, as the BAO peaks of the \hi\, power spectrum, can help reducing this degeneracy and improve the limits on the total mass of neutrinos. 

Including massive neutrino seems to be a necessary extension to $\Lambda$CDM. In this way we include the $\sum m_{\nu}$ as an extra free parameter and forecast the improvement BINGO will bring to the bounds of such a parameter. We estimate that BINGO in combination with \textit{Planck} data will be able to further constraint the total neutrino mass as
\begin{equation}
    \sigma_{\sum m_{\nu}} < 0.14 \, \textrm{eV} \quad (95\% \, \textrm{CL}) \,.
\end{equation}
This upper bound  is  slightly  higher  than  the one obtained by the \textit{Planck} Collaboration  \citep{Aghanim:2018eyx}. Therefore, BINGO will provide data through a  different tracer and can impose competitive constraints on the sum of the neutrino masses.

\subsubsection{Redshift space distortions}

\begin{figure}[h!]
\centering
\includegraphics[scale=0.5]{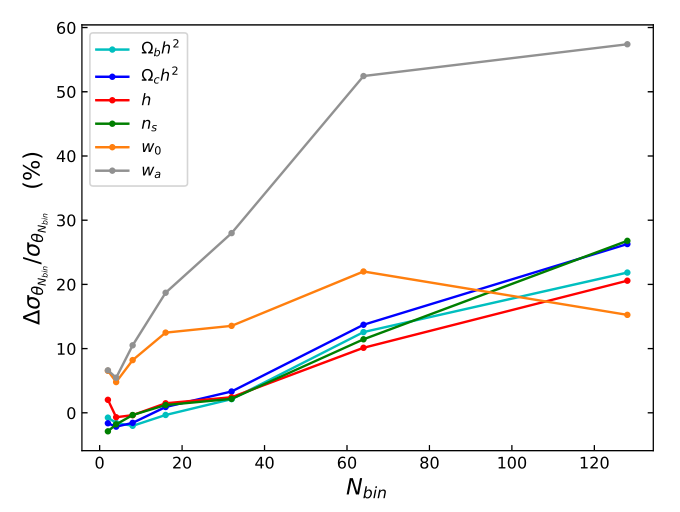}
\caption{\label{plot_rsd} Percentage difference between using or not information from RSD ($\Delta \sigma_{\theta}/\sigma_{\theta} =\sigma_{\theta\,\text{without} }/\sigma_{\theta\, \text{with} } - 1 $) in the cosmological parameter constraints from BINGO. $A_s$ and $b_{\textrm{\hi\,}}$ were not included because they go from complete ignorance to constraints of about percent level.}
\end{figure}

In this work, we have not considered disentangling the contribution of RSD from the \hi\, power spectrum and extracted the growth rate yet, nor its subsequent performance to constrain the cosmological parameters. However, in the companion paper VII~\citep{2020_forecast}, we have analyzed the impact RSD has on the total angular power spectra and how this affects our final constraints. First, including RSD information leads to the breaking of degeneracies between the primordial scalar amplitude $A_s$ and the \hi\, bias. Second, Fig.~\ref{plot_rsd} shows how the constraints on several cosmological parameters improve adding information from RSD as a function of the number of bins.

\subsubsection{HI history}
\begin{figure}[h!]
\centering
\includegraphics[scale=0.6]{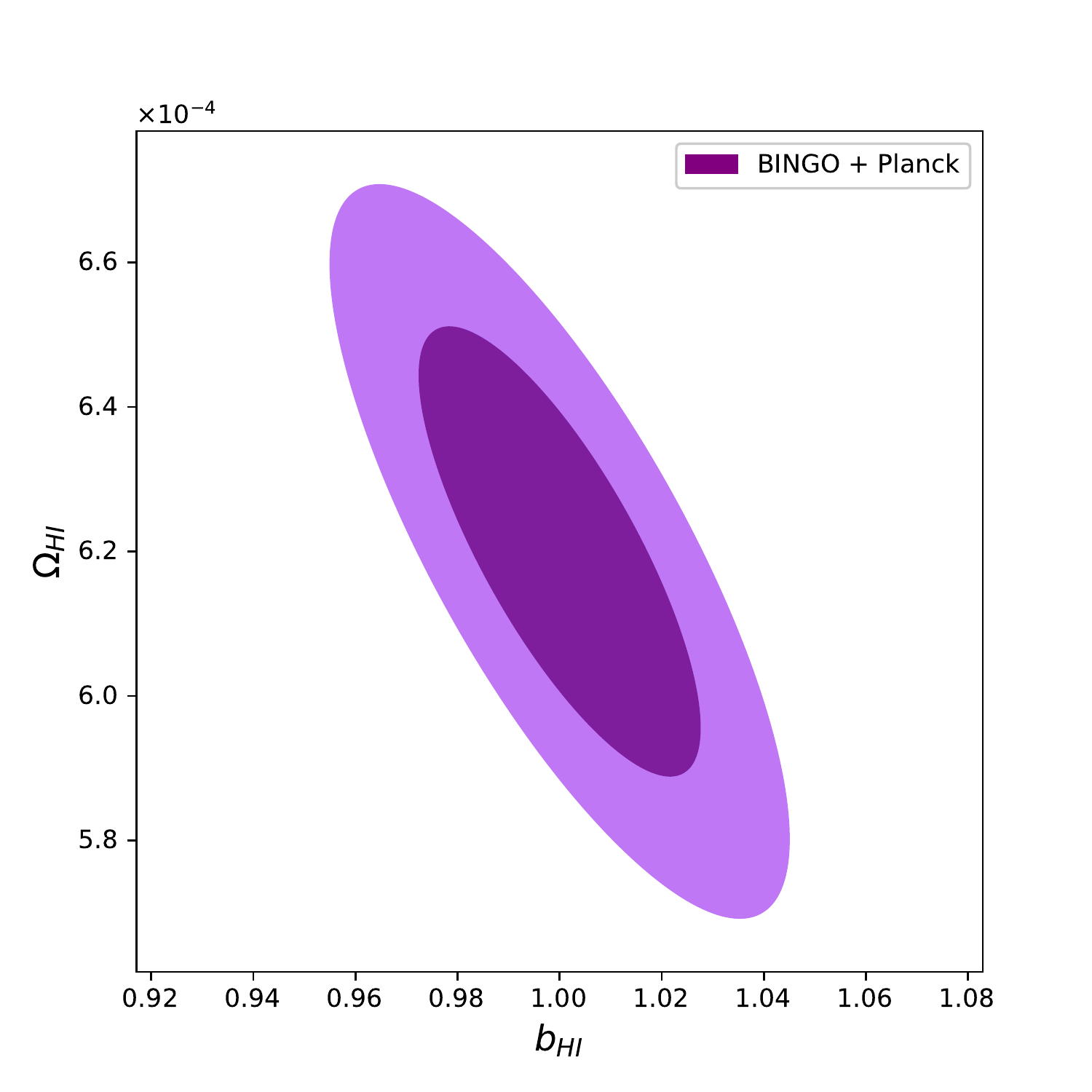}
\caption{Contour plot showing the $68\%$ and $95\%$ confidence level constraints on the \hi\, density parameter and bias using BINGO + \textit{Planck}.}
\label{Fig.:plot_bhi_omhi}
\end{figure}

The BINGO telescope will not only constrain the cosmological parameters as described in the previous sections, but it will also help shed light on the \hi\, evolution and distribution. Using the 21-cm line of neutral hydrogen as a tracer of the underlying matter distribution requires the knowledge of the \hi\, mean density and its bias. The simplest model assumes that they are constant, parametrized by $\Omega_{\rm{HI}}$ and $b_{\rm{HI}}$. Therefore, together with the cosmological parameters, BINGO will constrain these two additional \hi\, parameters. However, $\Omega_{\rm{HI}}$ and $b_{\rm{HI}}$ are strongly correlated. As we discuss above, we also have a complete degeneracy between $A_s$, and $\Omega_{\rm{HI}}$ and $b_{\rm{HI}}$. For this reason, in order to constrain these quantities independently we need external information. We show the constraints in $\Omega_{\rm{HI}}$ and $b_{\rm{HI}}$ in Fig.~\ref{Fig.:plot_bhi_omhi}, where we used BINGO in combination with \textit{Planck}. RSDs information can also help break this degeneracy, not only between $A_s$ and $\Omega_{\rm{HI}}$, but also between $\Omega_{\rm{HI}}$ and $b_{\rm{HI}}$.
In general, those parameters evolve with time and their measurement describes the \hi\, history.

In the companion paper VII, the ability of BINGO to constrain the \hi\, density and bias is considered under several circumstances. In Fig. \ref{Fig.:plot_bhi_omhi} the marginalized constraints on these two parameters is presented assuming they are constant over the BINGO redshift range.

\subsubsection{Systematic effects}

One of the main challenges of 21-cm experiments and of the BINGO project are astrophysical contamination and systematic effects. Systematic effects, present in the observed \hi\, signal might affect the constraints on cosmological parameters that the BINGO telescope can obtain. The main systematics affecting BINGO and any single-dish 21-cm experiment are $1/f$ noise, standing waves, atmospheric effects and RFI. Residuals from foreground removal will also be present and can strongly impact the measured \hi\,signal. These systematic and foreground effects are explained in Section \ref{sec:pipeline} in detail, together with the work made in the BINGO pipeline to mitigate them and their impact in recovering the \hi\, signal \citep[see also][]{Olivari:2017}.

\section{Scientific goals II: astrophysics}
\label{sec:astro}

Although the BINGO telescope is mainly designed to characterize BAO, it is an experiment that is versatile enough to allow for further scientific goals. The exceptionally long integration times and spectral stability will enable it to make important contributions in other fields, such as transient objects science, with the newly discovered and exciting FRBs and  pulsars, as well as Galactic science.

\subsection{Transient objects science}

The dynamic radio sky includes astrophysical phenomena that emit pulses in the radio frequency with a large variety of temporal and spectral features. The period range is from microseconds to a few minutes. Sometimes they can repeat, sometimes they are single events. The spectral index may show wide variations. Also diverse are the source objects, from neutron stars to magnetars (basically highly magnetized and rotating neutron stars) and flare stars. A better understanding of these phenomena will be profitable to many astrophysical and cosmological research fields, since they are related to non conventional emission mechanisms. They could be used as probes of the intergalactic medium (IGM), indicating the presence of matter and magnetic fields along the line of sight. The new member of this family, the most recently discovered, are the so called FRBs. We present here the capabilities of the BINGO telescope in studying such transients.

\subsubsection{Fast radio bursts}
Discovered in 2007 \citep{Lorimer:2007qn}, FRBs are  bright bursts ($0.05-100$ Jy) of radio photons  of unknown  origin (see \citealt{Pen:2018ilo,Platts:2018hiy, Popov:2018hkz,Petroff:2019tty} for  recent reviews), lasting from some microseconds to a few seconds ($80 \, \mu \text{s}-5 \text{ s}$). The radio waves are dispersed  by free electrons along the line-of-sight between the source and the observer, thus the  time delay of photons with different  frequencies can be measured. This time delay is proportional to the dispersion measure $\int n_e \, dl $ (where $n_e$ is the number of electrons per volume and $l$ is the distance along the line-of-sight), and it depends on the observed frequencies $\nu_1$ and $\nu_2$ as $\Delta t= e^2/(2 \pi m_e c)(\nu_2^{-2}-\nu_1^{-2})\int n_e dl $, where $m_e$ is the electron mass. The dispersion measure may have contributions from the vicinity of the FRBs source, the host galaxy and its halo, the IGM, and the Milky Way and its halo.

There are so far over 100 detected FRBs in a frequency range between 300 MHz and 8 GHz \citep{Gajjar:2018bth, Chawla:2020rds} with dispersion measure approximately ranging from $100$ pc cm$^{-3}$  to $2600$ pc cm$^{-3}$, where a few of them showed repeated bursts \citep{Spitler:2016dmz, Andersen:2019yex,Kumar:2019htf} and two appear to be periodic  \citep{Amiri:2020gno,rajwade2020possible}. Although their origin is still a mystery, most of them are extragalactic, with only one having been produced within the Milky Way by a magnetar \citep{Andersen:2020hvz}. Due to their still uncertain origin, several explanations have been proposed \citep{Platts:2018hiy}, including, but not limited to, mergers and interactions between compact objects  (such as neutron stars, black holes and white dwarfs), active galactic nuclei and supernovae remnants.  Due to the distinct repeating feature of some FRBs, it is possible that  non repeating and  repeating FRBs do not share the same  common progenitor.  

Additionally, FRBs can be very useful to various applications in cosmology and astrophysics, such as constraining cosmological parameters  \citep{Zhou:2014yta,Yang:2016zbm,Walters:2017afr,Li:2017mek,Yu:2017beg,Wei:2018cgd,Zitrin:2018let}, testing the Weak Equivalence Principle \citep{Wei:2015hwd,Tingay:2015wdv,Shao2017,Yu:2017xbb,Bertolami:2017opd,Yu:2018slt,Xing:2019geq}, estimating the photon mass \citep{Xing:2019geq,Wu2016,Wei:2016jgc,Bonetti:2016cpo,Bonetti:2017pym, Wei:2018pyh}, studying the properties of the IGM \citep{Deng:2013aga,Zhang:2013lta,Akahori:2016ami,Fujita:2016yve,Shull:2017eow,Ravi:2018ose}, and compact DM \citep{Munoz:2016tmg,Wang:2018ydd},  among others \citep{Linder:2020aru,Landim:2020ked}.

Several telescopes have been detecting FRBs, such as the  Parkes Radio Telescope   \citep{Lorimer:2007qn},  the Australian Square Kilometre Array Pathfinder   \citep[ASKAP;][]{Bannister:2017sie}, the Green Bank Telescope  \citep[GBT;][]{Masui:2015kmb}, the UTMOST telescope \citep{Caleb:2017vbk}, the Arecibo Observatory \citep{Spitler:2014fla}, the Canadian Hydrogen Intensity Mapping Experiment  \citep[CHIME;][]{Amiri:2018qsq}, the Apertif on the Westerbork Synthesis Radio  Telescope \citep{Connor:2020oay}, among others.\footnote{See \url{http://www.frbcat.org/} for a complete list of the detected FRBs.} 


BINGO has the potential to detect FRBs. Here we show an estimate of the detection rate for the BINGO telescope. The flux density depends  on the frequency as $S(\nu) = S_0 \nu^\alpha$, where $\alpha$ is the spectral index and can be either positive or negative. The minimum detectable flux density for a flat spectrum ($\alpha=0$) is
 \begin{equation}\label{eq:Smin}
        S_{\rm{rms}}=\text{S/N} \frac{ K \, T_{\rm{sys}}}{G\sqrt{n_p\Delta\nu\, \tau}}\,,
    \end{equation}
where $T_{\rm sys}$ is the system temperature, $G$ is the forward gain (in units of  K/Jy), $K=\sqrt{2}$ for a correlation receiver, $n_p$ is the number of polarizations, $\Delta \nu$ is the bandwidth, $\tau$ is the sampling time and S/N is the signal-to-noise ratio.
    
For a general spectral index the  detectable flux density (\ref{eq:Smin}) is the peak flux density $S_{\rm{peak}}=\text{S/N} \,T_{\rm{sys}}/(G\sqrt{n_p\Delta\nu\, \tau})$, which is written in terms of the bolometric luminosity $L_{\rm bol}$ as \citep{Lorimer:2013roa}
\begin{equation}\label{eq:Lbolspectral}
    S_{\rm{peak}}=\frac{1}{4\pi d_L(z)^2}\frac{L_{\rm{bol}}}{\nu_{\rm{high}}^{1+\alpha}-\nu_{\rm{low}}^{1+\alpha}}\left(\frac{\nu_2^{1+\alpha}-\nu_1^{1+\alpha}}{\nu_2-\nu_1}\right)\,,
\end{equation}
where $d_L(z)$ is the luminosity distance, $\nu_1$ and $\nu_2$ are the observed frequencies, and $\nu_{\rm{high}}$ and $\nu_{\rm{low}}$ are the highest and lowest frequencies, respectively, in which the source emits, as seen by the observer. 
   
\begin{figure}
        \centering
      \includegraphics[scale=0.4]{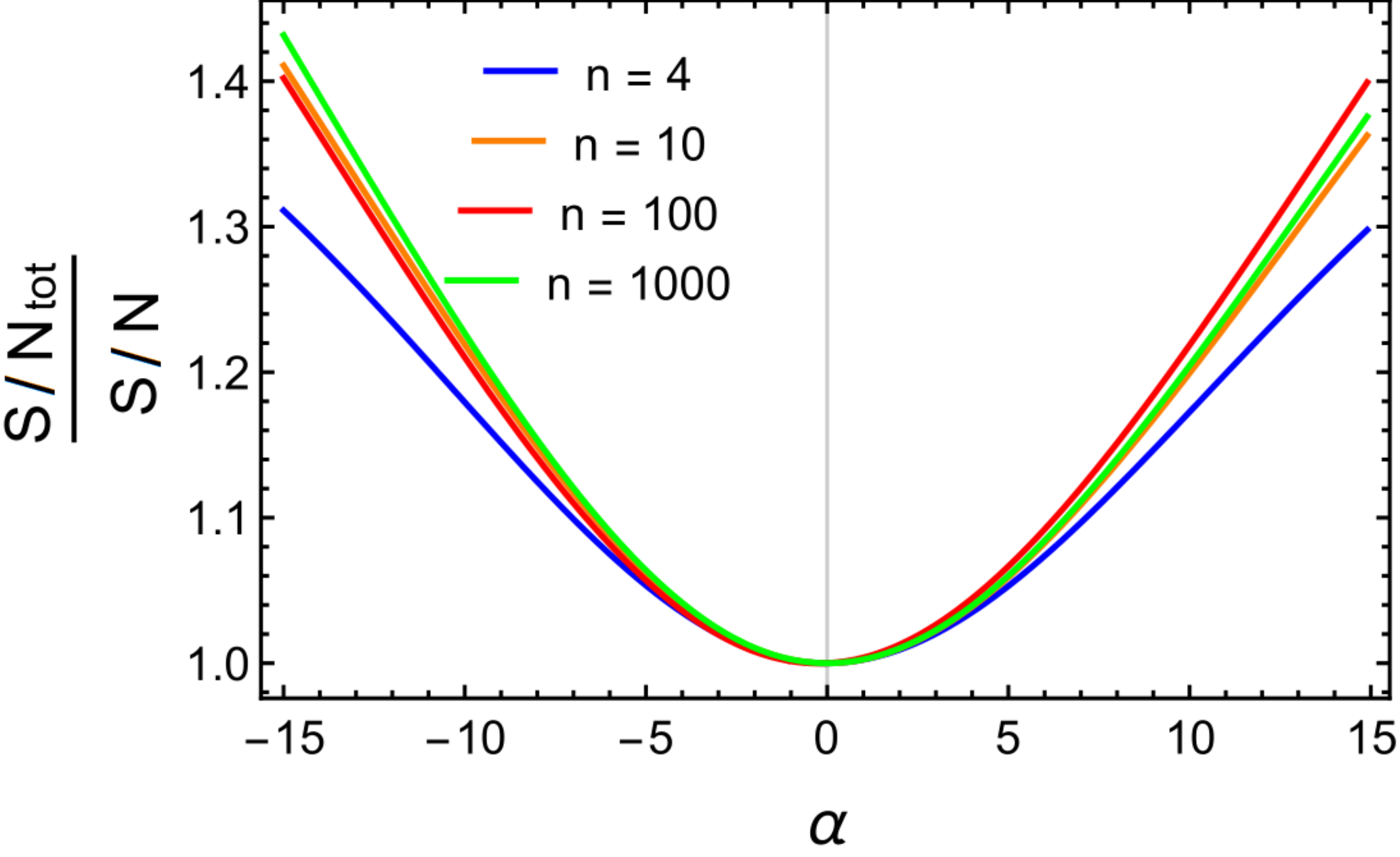}
      \caption{Deviation of Eq. (\ref{SNRtot}) from S/N (for a flat spectrum) as a function of the spectral index $\alpha$, for different number of channels $n$. It can be seen that the total S/N is only   $\sim$ 1.5 times larger for very large spectral index $|\alpha|$. }
      \label{fig:FRB2}
    \end{figure}

For a general spectral index the received power delivered by an antenna is
   \begin{align}\label{eq:Precalpha}
P_{\rm{rec}}&=\frac{1}{2}A_e\int_{\nu_1}^{\nu_2} S_0 \nu^\alpha\,d\nu\,,\nonumber\\
       &=\frac{1}{2}A_e S_0 \frac{\nu_2^{1+\alpha}-\nu_1^{1+\alpha}}{1+\alpha}\,,
        \end{align}
while the Johnson–Nyquist noise power remains the same as in the case of flat spectrum
    \begin{equation}\label{eq:nyqalpha}
      P_{\rm{noise}}=k_B T_A (\nu_2-\nu_1)\,,
    \end{equation}
where $T_A$ is the antenna temperature. Using the radiometer equation
\begin{equation}\label{eq:radiometer}
     \text{S/N}=\frac{T_A}{T_{sys}}\sqrt{\Delta\nu\, \tau}\,,
    \end{equation}
the total flux density for $n$ channels is
    \begin{equation}\label{eq:SminSNRtot}
      S_{\rm{tot}}=\text{S/N}_{\rm tot} \frac{ K\,  T_{\rm{sys}}}{G\sqrt{n_p\Delta\nu\, \tau}}\,,
    \end{equation}
where the total S/N is given by
      \begin{equation}\label{SNRtot}
        \text{S/N}_{\rm{tot}}= \text{S/N} \frac{\nu_2-\nu_1}{\nu_2^{1+\alpha}-\nu_1^{1+\alpha}}\frac{(1+\alpha)}{\sqrt{n}}\left(\sum_{p=1}^{n}\nu_p^{2\alpha}\right)^{1/2}\,.
    \end{equation}
The factor $\sqrt{n}$ comes from the division of bandwidth by the number of channels $n$  and
    \begin{equation}
       \nu_p\equiv\nu_1+\left(p-\frac{1}{2}\right)\frac{\Delta\nu}{n}\,.
    \end{equation}
The behavior of $\text{S/N}_{\rm{tot}}/ \text{S/N}$, from Eq.~(\ref{SNRtot}),  is shown in Fig. \ref{fig:FRB2} for the BINGO telescope. Since the peak flux density is also given by Eq. (\ref{eq:Smin}) the relation between $S_{\rm{peak}}$ and $S_{\rm{tot}}$ is 
     \begin{equation}
     S_{\rm{tot}}=\frac{\text{S/N}_{\rm{tot}}}{\text{S/N}}S_{\rm{peak}}\,.     \end{equation}
    
Having determined the  flux density for a given survey through Eq. (\ref{eq:Smin}), the detection rate of FRBs can be calculated using the same approach as \cite{Luo:2018tiy,Luo:2020wfx}, where the Schechter function \citep{Schechter:1976iz} was assumed as the distribution of the FRBs luminosity and the  intrinsic FRB width distribution is given by a log-normal function \citep{Connor:2019wnx} 
   \begin{align}\label{eq:frb_detec_rate}
       \lambda(\Omega,S_{\rm{peak}})=&\int_0^\Omega d\Omega\int_0^{z_{max}}dz \frac{c}{1+z}\frac{r(z)^2}{E(z)}\int_{w_i'}^\infty \frac{e^{-\frac{(\log w_i - \mu_w)^2}{2\sigma_w^2}}}{\sqrt{2\pi \sigma_w^2}}\,d\log w_i\nonumber\\
       &\int_{ L_{\rm min}}^{\infty}\phi^*\left(\frac{L}{L^*}\right)^\gamma e^{-\frac{L}{L^*}}\frac{dL}{L^*}\,,
   \end{align}
where $r(z)$ is the comoving distance, $w_i=w_0/(1+z)$, $w_0$ is the observed pulse width, and the 
 upper cut-off luminosity
$L_*$ and the normalization $\phi_*$  are parameters constrained in \cite{Luo:2020wfx}. The integral of the Schechter function gives the incomplete gamma function $\phi^*\Gamma(1+\gamma,L_{\rm{min}}/L^*)$, where a (Gaussian) beam response has already been integrated out \citep{Luo:2020wfx}. The minimum luminosity is given by the $L_{\rm{min}}=\max(L_0, L_{\rm{bol}})$, where the bolometric luminosity is given by Eq. (\ref{eq:Lbolspectral}) and we take the cut-off luminosity to be $L_0\leq 9.1\times 10^{41} \text{erg s}^{-1}$ \citep{Luo:2020wfx}.

\begin{figure}
    \centering
    \includegraphics[scale=0.29]{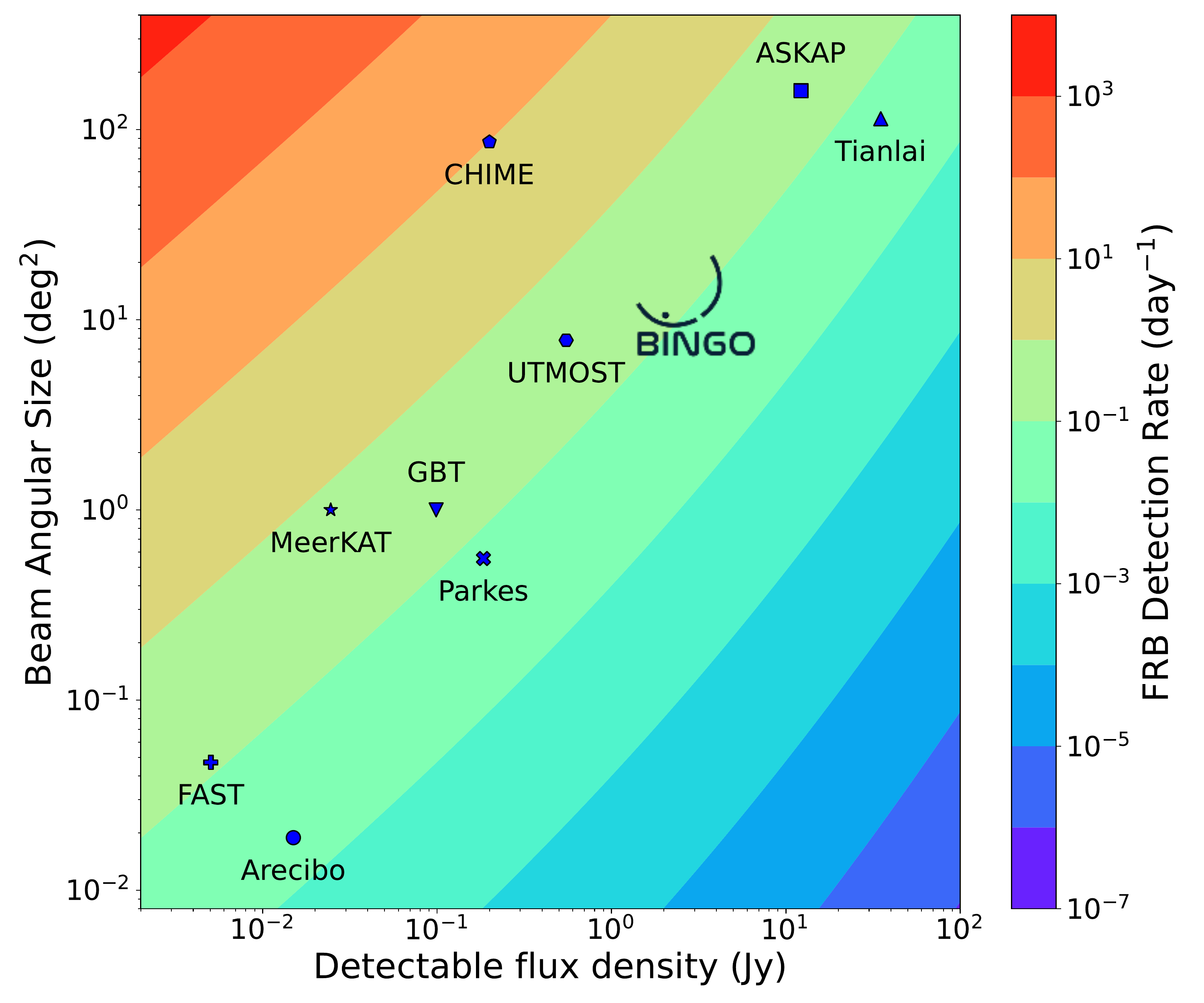}
    \caption{FRB detection rate per day (\ref{eq:frb_detec_rate}) as a function of the FoV and the minimum detectable flux density. Following \cite{Luo:2020wfx} we show the BINGO detection rate along with other surveys: Arecibo \citep{Spitler:2014fla}, ASKAP \citep{shannon2018dispersion}, CHIME \citep{Amiri:2018qsq}, FAST \citep{jiang2020fundamental}, GBT \citep{Masui:2015kmb}, MeerKAT \citep{bailes2020meerkat}, Parkes \citep{Keith:2010kk}, UTMOST \citep{caleb2016fast} and Tianlai \citep{li2020tianlai}. The BINGO detection rate is shown by the dot within its logo. The minimum flux density (corresponding to S/N$=1$) is 0.41 Jy and the FoV is 10.5 $\text{deg}^2$. 
    }
    \label{fig:frb-rates}
\end{figure}

Using the constrained parameters  found in \cite{Luo:2020wfx} ($L^*=2.9\times 10^{44} \text{erg s}^{-1}$, $\phi^*=339 \text{ Gpc}^{-3}\text{yr}^{-1}$, $\gamma=-1.79$, $\mu_w=0.13$ and $\sigma_w=0.33$), $z_{\rm{max}} =6$, $w_0=1$ ms and S/N $=5$ we show the detection rate for a flat spectrum for the BINGO telescope in Fig. \ref{fig:frb-rates}, along with the corresponding rates for other surveys. Those rates are also given in Table \ref{tab:frb} with different S/Ns. 

\begin{figure}
    \centering
    \includegraphics[scale=0.55]{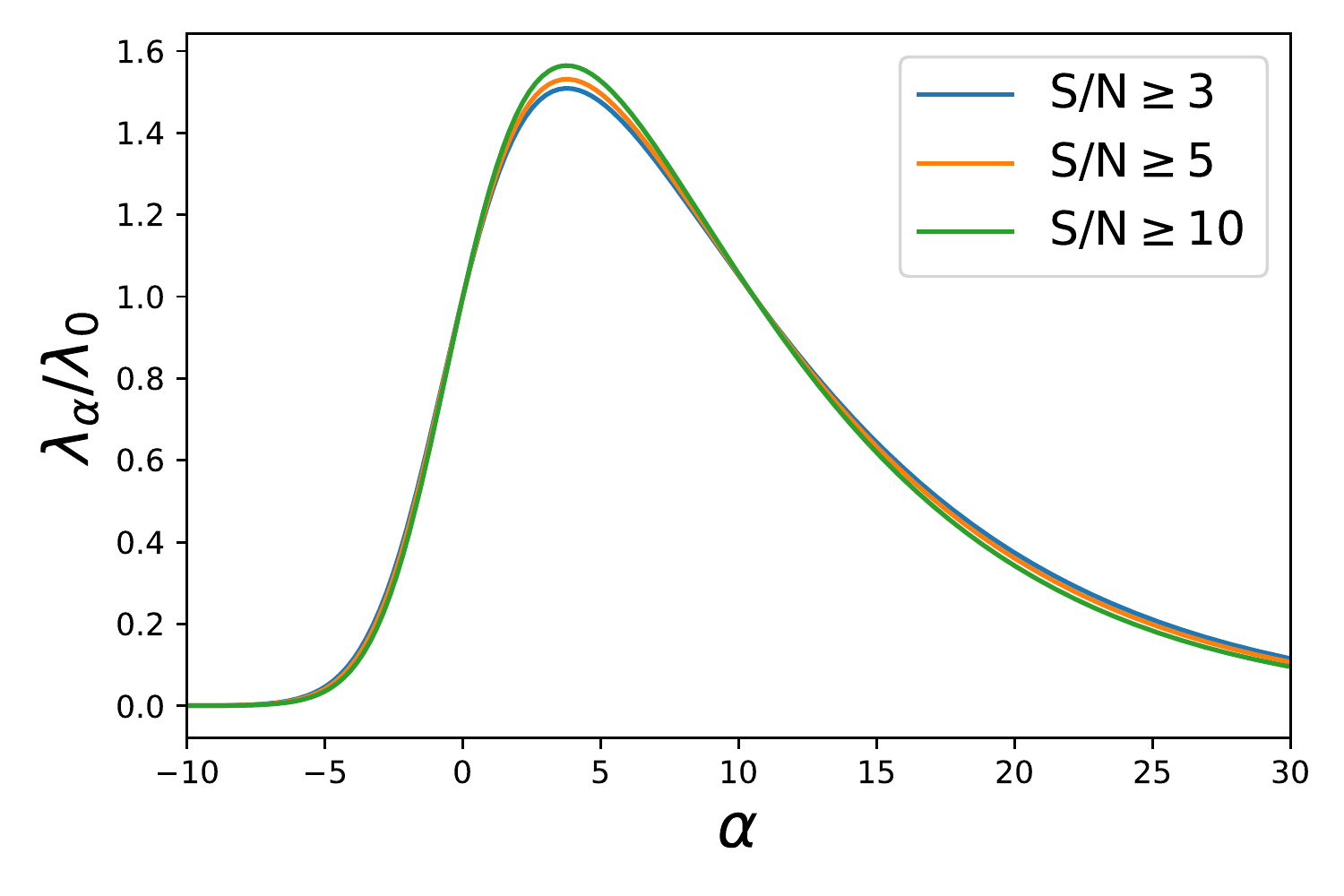}
    \caption{Detection rate for the BINGO telescope for different spectral indices and S/Ns, normalized to the corresponding values for $\alpha=0$. }
    \label{fig:dect_rate_alpha}
\end{figure}

In Fig. \ref{fig:dect_rate_alpha} we estimate the detection rate for BINGO, for a range of spectral indices between $-5\leq \alpha\leq +15$ for three values of S/N. The highest and lowest frequencies ($\nu_{\rm high}$ and $\nu_{\rm low}$) are chosen as $\nu_{\rm high}=1400$ MHz and $\nu_{\rm low}=400$ MHz to give illustrative results of the impact of the spectral index on the detection rate. The case $\alpha=0$ recovers the results presented in Fig. \ref{fig:frb-rates}  \citep{Luo:2020wfx}. According to the Schechter luminosity function, populations with larger luminosities are more suppressed than the ones with smaller luminosities, thus the detection rate (as seen in Fig. \ref{fig:dect_rate_alpha}) is smaller for more negative spectral indices, since they lead to larger luminosities.  In order to investigate the impact of $\nu_{\rm{high}}$ and $\nu_{\rm{low}}$ on the detection rate, we calculated the rate $\lambda_{\alpha}/\lambda_0$ for two other frequency ranges, such that their differences are still 1 GHz: $\nu_{\rm{high}}=1260 $ MHz and $\nu_{\rm{low}}= 260$ MHz, and $\nu_{\rm{high}}=1980 $ MHz and $\nu_{\rm{low}}= 980$ MHz. For the lower frequency band ($\nu_{\rm{high}}=1260 $ MHz and $\nu_{\rm{low}}= 260$ MHz) the peak flux density (\ref{eq:Lbolspectral}) is higher for positive spectrum indices than its corresponding value for the flat spectrum ($S_{\rm{peak},\alpha>0} >S_{\rm{peak}, \alpha=0}$), while for the higher frequency band ($\nu_{\rm{high}}=1980 $ MHz and $\nu_{\rm{\rm{low}}}= 980$ MHz) the peak flux density is higher for negative spectrum indices ($S_{\rm{peak},\alpha<0}>S_{\rm{peak},\alpha=0}$). Increasing the peak flux density increases the detection rate as well because given a S/N threshold a signal that would be smaller than this threshold now can be larger. For FRBs with spectral indices between $-5$ and $5$, for example, the difference in the detection rate from the lower band choice ($\nu_{\rm{high}}=1260 $ MHz and $\nu_{\rm{low}}= 260$ MHz) to the higher band choice ($\nu_{\rm{low}}= 260$ MHz, and $\nu_{\rm{high}}=1980 $ MHz) is only $\sim 2.5$.

\begin{table}
\caption{Mean time elapsed between two detections (in hours), for the surveys presented in Fig. \ref{fig:frb-rates}. }
\label{tab:frb}
\begin{tabular}{cccc}
\hline
Survey &	S/N $\geq$ 3& S/N $\geq$ 5&	S/N $\geq$ 10\\
\hline
CHIME	&1.5	&2.4	&4.6\\
MeerKAT	&22.6	&34.4	&61.6\\
UTMOST	&42.5	&68.7	&135.3\\
ASKAP	&50	&90.7	&210.2\\
GBT	&72.2	&112.2	&207.5\\
BINGO	&112.4	&189.8	&398.5\\
FAST	&133.5	&201.2	&352.5\\
Parkes	&223.3	&351.4	&662.5\\
Tianlai	&245.1	&463.9	&1139\\
Arecibo	&799.1	&1212.3	&2150.6\\
\hline
\end{tabular}
\end{table}

 In order to achieve the capability of resolving FRB spectral signatures, BINGO will need a complementary setup in the digital backend output, with a sampling time of $\thicksim 0.05 - 0.1$ ms (much shorter than what is needed for intensity mapping) and 1024 -- 2048 output channels. This fine resolution is needed for RFI flagging and data reduction. After the data cleaning and FRB searching process, data channels can be binned in the nominal value of 30, enough for BAO data analysis. The localization of FRBs on the sky  will be possible using outrigger telescopes and it will be explored in future work.


\subsubsection{Pulsars}

%
Pulsars are rapidly spinning highly magnetized neutron stars that emit beams of electromagnetic waves. They take their name from their periodic emission caused by the rotation frequency of the radiation beam in our line of sight, like a lighthouse. In most cases, the pulse profile is extremely stable, both in form and in the arrival phase at the rotational frequency. This stability allows for the precision timing of pulsars, with remarkable applications \citep{2015SSRv..191..207B}.

A fixed position in the sky will take $\sim$ 28 minutes to pass over the focal plane of BINGO. However, due to the sparse configuration of the horns (see Figures 4 and 5 of \citealt{BINGO.Instrument:2020}), considering the areas with responses higher than $50\%$ of the peak gain, a fixed position will pass over for $\sim$ 40 arc-minutes accumulated, which corresponds to $\sim$ 3 minutes and is taken as the integration time for one-day observation in the following estimations. 

Assuming that BINGO will have buffers large enough to store the pulsar data within 3 minutes, the sensitivity of pulsar detection ($S_{\rm{pulsar}}$) can be estimated via
\begin{equation}\label{eq:pulsar-sensi}
S_{\rm pulsar}=\rm{S/N}\frac{\sqrt{2}\,T_{sys}}{G\sqrt{n_p\Delta\nu\tau}}\,,
\end{equation}
where $T_{\rm sys}=70$ K is the system temperature, $G=0.32$ K/Jy is the gain, $n_p=2$ is the polarization number, $\Delta\nu=280$ MHz 
is the bandwidth, and $\tau=3$ minutes is the integration time. This gives a sensitivity of 
2.92 and 
4.87 mJy with S/N equals 3 and 5, respectively. In the BINGO FoV (strip between DEC$=-10\deg$ to DEC$=-25\deg$), there are 372 known pulsars in the Australia Telescope National Facility (ATNF) pulsar catalog\footnote{\url{https://www.atnf.csiro.au/research/pulsar/psrcat/}} \citep{2005AJ....129.1993M}. We estimated their fluxes at 1.1 GHz (the mid-frequency of BINGO) using a power-law fitting to their existing flux data in other frequencies, or a spectral index of $-1.6$ is assumed. The 
38 brightest known pulsars can be detected over S/N$=3$ with one-day observation, and 
4 of them are millisecond pulsars.

Taking the advantage of BINGO's daily monitoring of the same sky stripe for five years, we could at least focus on the following sciences with these $\sim$ 
40 brightest pulsars in the FoV. 

\textit{Refractive scintillation (RISS):} RISS of pulsars when their emissions pass through the interstellar medium (ISM) with large-scale inhomogeneities \citep{2021MNRAS.501.4490K}. The pulsar RISS phenomenon normally has a timescale of one or few weeks, which perfectly fits into long-term daily observations. The understanding of the RISS is still relatively poor although the underlying physics of RISS is clear, which is mainly because of the uncertainty about the ISM inhomogeneities in our Galaxy. The observational result from BINGO will help to improve the understanding the ISM distribution in our Galaxy. 

\textit{Monitor the change in pulsar periods:}  The pulsed emission of pulsar may appear to be completely ceased over the time scale of the pulsar period to several years, called pulse nulling (e.g., \citealt{Young15}). The physics behind pulse nulling has not been fully understood yet. Besides, some pulsars have quasi-periodic switching, which might be caused by magneto-spherical switching. Pulsar periods can also change due to the orbital motion on the timescale of several years (e.g., \citealt{Keane13}).

\textit{Pulsar glitch:}  The glitch of a pulsar refers to a sudden increase in its rotating frequency, and the exact reason that causes a glitch is still unknown \citep{2015IJMPD..2430008H}. Up to now, only few glitches are observed in Crab and Vela pulsars. BINGO has the penitential to discover more example of pulsar glitches (also for fainter pulsars in the FoV) and to study the underlying physical nature.

\textit{Rotating radio transients (RRATs):}  RRATs are a type of pulsar discovered through  single pulse searches, instead of the usual Fourier domain searches. Their detected radio pulses are more sporadically than the normal class. The reason for this anomalous feature is not yet known but could be due to fallback of material from a supernova debris disk, presence of an asteroid belt around the pulsar or just a normal pulsar showing extreme nulling \citep{McLaughlin:2005eq,Burke2012,Keane:2011,Shapiro2018}.


The pulsar science could be extended if a hydrogen maser is to be installed in the telescope. We will be able to combine observations of more than one day to achieve much higher sensitivities. For 10- and 100-day integrations, the $S_{\rm pulsar}$ goes down to 
0.92 mJy and 
0.29 mJy with S/N$=3$, which corresponds to the detection of 
103 and 
207 known pulsars in the FoV, respectively. We could at least study their pulse shape and polarization, although flux estimation would be difficult. With the hydrogen maser, precise pulsar timing could be done with the brightest pulsars as well. The millisecond pulsars among them would contribute to the pulsar timing array for the detection of the gravitational waves at nHz, which might be generated from super-massive black-hole binaries at a cosmological distance \citep{2017PhRvL.118o1104W}, and measure the gravitational constant as well as examine the gravitational theory  (e.g., \citealt{Hobbs2017}). 



\subsection{Galactic science}
Four satellite missions were designed to map CMB fluctuations: Relikt (1983) \citep{Strukov1984,Strukov1993}, COBE (1989 – 1993) \citep{COBE:Bennett1992}, WMAP (2001 – 2010) \citep{WMAP:Gold2011} and Planck (2009 – 2013) \citep{Akrami2018}. Three of them (COBE, WMAP, Planck) produced, as part of their ancillary science data, sets of maps with Galactic synchrotron, bremsstrahlung, and dust emission (unpolarized and some polarized). The lower frequencies where these satellites operated ($22$, $30$, $37$, $40$, $44$, $53$, $70$ GHz) cover the frequency band of minimum Galactic contribution to CMB measurements. 

However, around $1 \mathrm{GHz}$, where BINGO will operate, there will be a clear predominance of synchrotron emission dominating the Galactic foreground contribution \citep{1972A&AS....5..263B,Reich:1986,2001A&A...376..861R}, seconded by free-free and then, depending on the chosen model, by Anomalous Microwave emission (AME) \citep{Dickinson:2018}. Although BINGO will not map a vast area of the sky, we expect to collect data that can contribute to cross-check observations of ongoing sky mapping experiments. 

For instance, at 1,465 MHz, a frequency close to the BINGO band, the total sky brightness has been mapped in two adjacent 60-degree declination bands with the Galactic emission mapping (GEM) experiment, which employs a portable 5.5-m parabolic reflector to survey the radio continuum of the sky in decimeter and centimeter wavelengths \citep{Tello:2000}. The GEM experiment also mapped the sky at 2.3 GHz \citep{Tello:2013}. BINGO will add well-calibrated data to the Galactic microwave spectrum, contributing to a better determination of the spectral index of the synchrotron emission at the low-frequency microwave region.

\subsection{Extragalactic science -- associated \hi\ absorption}

Associated \hi\ absorption is detected against bright radio continuum emission sources with a compact core, like active galactic nuclei (AGNs) and nuclear starbursts, due to absorption by \hi\ clouds within the continuum sources themselves.
They are some of the best objects to study the interplay of AGN and stellar feedback on the interstellar medium of galaxies \citep{2018A&ARv..26....4M}.
There are less than 100 known detections of associated \hi\ absorption even with targeted searches, especially at $z>0.25$ \citep[e.g.,][]{1970ApJ...161L...9R,1986ApJ...300..190D,1989AJ.....97..708V,1998ApJ...494..175C,1999ApJ...524..684G,1999ApJ...510L..87M,2003A&A...404..871P,2003A&A...404..861V,2005A&A...444L...9M,2006MNRAS.373..972G,2011MNRAS.418.1787C,2011ApJ...742...60D,2012MNRAS.423.2601A,2013MNRAS.429.2380C,2014MNRAS.440..696A,2015A&A...575A..44G,2017A&A...604A..43M,2018MNRAS.473...59A,2018MNRAS.481.1578A,2019MNRAS.482.5597A}.

In a \hi\ IM survey such as BINGO, the bright radio continuum emission sources will typically by flagged out at the data reduction stage.
But given the high sensitivity of such surveys, many of the radio continuum sources observed by them will have detectable associated \hi\ absorption in them.
\citet{2019A&A...631A.115R} specifically calculated the expected number of associated \hi\ absorptions that will be detected with BINGO (see their Appendix B for details).
For the full survey, they expect the total number of \hi\ absorption detections will be almost twice the number known today.
Given the low spatial resolution of BINGO, each such detection should ideally be followed-up by interferometric observations.
Thus BINGO can provide targets to follow-up for the existing SKA precursors like ASKAP and MeerKAT, which are also observing the southern sky.

\section{Simulations and data analysis pipeline}
\label{sec:pipeline}

\begin{figure*}
\centering
\includegraphics[width=0.8\textwidth]{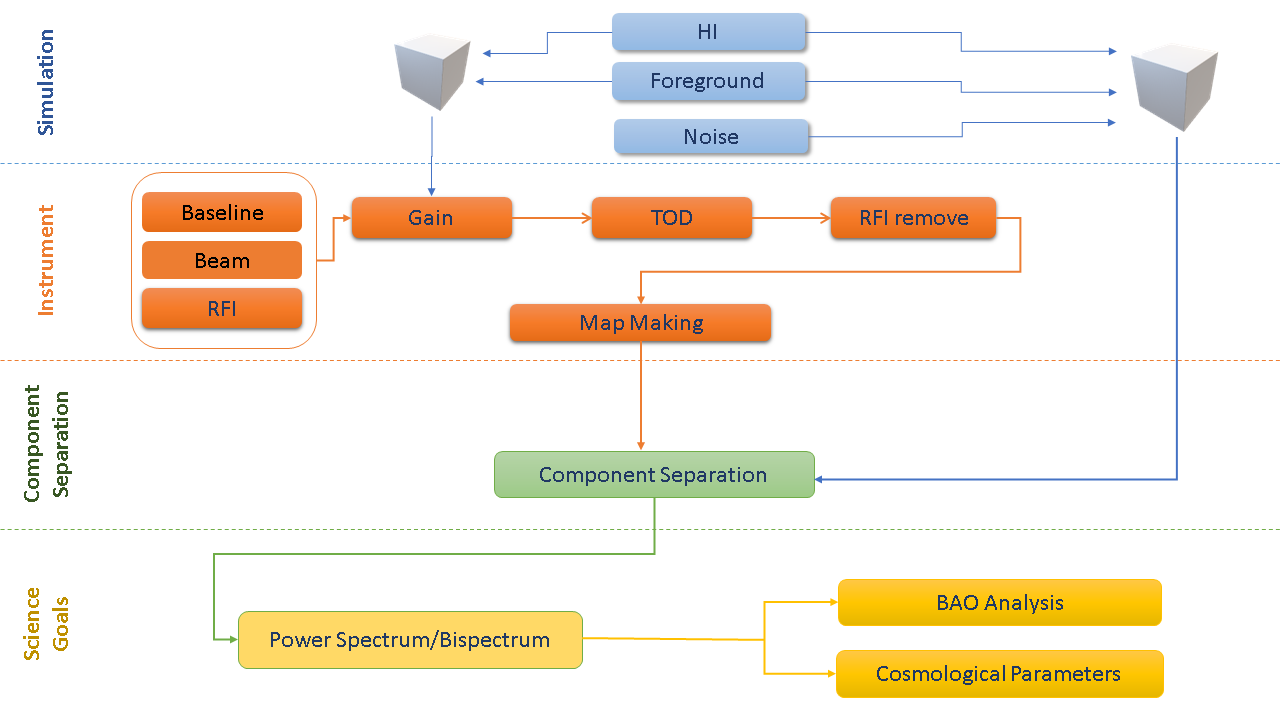}
\caption{Flowchart of the BINGO simulation pipeline.}
\label{fig:flowchart}
\end{figure*}

The BINGO simulation pipeline is an ongoing effort, being developed alongside the instrument construction, covering input modules for sky emission models and experiment characteristics, a mission simulation module, production of time ordered datasets (TOD), calibration and cleaning in the time domain, map-making, generation of multifrequency maps, component separation for foreground removal, reconstructed \hi\, sky maps and power spectrum estimation. The information flow in the BINGO mission simulation and data analysis pipeline is shown in Fig.~\ref{fig:flowchart}, and this section describes the pipeline modules according to it.

\subsection{Sky module}
\label{sec:sky_model}

The observed brightness temperature of the sky, $T_{\rm{sky}}(\nu, \mathbf{x})$, at frequency $\nu$ and direction $\mathbf{x}$, is composed by several different sources \citep{Wilson:2013trabook,Condom:2016era}
\begin{align}
\label{pipe22}
T_{\mathrm{\rm{sky}}}(\nu, \mathbf{x}) &= T_{\mathrm{\rm{gal}}}(\nu, \mathbf{x}) + T_{\mathrm{eg}}(\nu, \mathbf{x}) \nonumber \\
&+ T_{\mathrm{CMB}}(\nu, \mathbf{x}) + T_{\mathrm{atm}}(\nu, \mathbf{x}) + T_{\mathrm{cosmo}}(\nu, \mathbf{x}),
\end{align}
where $T_{\mathrm{gal}}$ is the diffuse Galactic radiation, $T_{\mathrm{eg}}$ is the emission from extragalactic sources, $T_{\mathrm{CMB}}(\nu, \mathbf{x})$ is the CMB temperature, $T_{\mathrm{atm}}(\nu, \mathbf{x})$ is the atmospheric emission, and $T_{\mathrm{cosmo}}$ is the cosmological \hi\, emission. We briefly describe their contributions to the sky module below.

\subsubsection{\hi\,emission}

The synthetic \hi\, maps used in the BINGO simulation pipeline are produced in two ways. One of them is to use the {\tt FLASK} (Full-sky Log-normal Astro-fields Simulation Kit) package\footnote{ Available at \url{ http://www.astro.iag.usp.br/~flask/}.} \citep{Xavier:2016}. {\tt FLASK} uses as input the angular auto- and cross-power spectra, $C^{ij}(\ell)$, calculated for each of the $i$ and $j$ fields and redshifts, to produce  two-dimensional tomographic realizations (spherical shells around the observer) of an arbitrary number of random astrophysical fields, reproducing the desired cross-correlations between them, derived either from a multivariate Gaussian distribution or a multivariate log-normal distribution.



An option to using {\tt FLASK}  as the \hi\,input data is to simulate 21cm IM mock maps, based upon N-body simulations and assuming all the \hi\,gas is  contained within galaxies or dark matter halos. We used the N-body simulation Horizon Run 4 \citep[HR4;][]{kim2015horizon}, which consists of $6300^3$ particles in a cubic box of volume $3150^3 \, \mathrm{Mpc}^3/h^{3}$ to produce a full sky mock map using the halo and galaxy light cone catalog where the halo mass resolution is $2.7\times10^{11}M_{\odot}/h$. We use  the \hi\,HOD model and abundance matching to generate the full sky 21cm IM mock map. The details of the mock catalog production are discussed in companion paper VI \citep{2020_mock_simulations}. 

The project plans to combine the fast log-normal realization with {\tt FLASK} and the more optimal N-body simulation mock for pipeline testing.

\subsubsection{Galactic foregrounds}
\label{Sec.:foregrounds}

It is well known that the faint \hi\, signal expected to be detected in IM experiments will be dominated by different astrophysical foregrounds, broadly classified in Galactic diffuse emission and the extragalactic, point source emission. Foregrounds removal and  component separation methods  have been explored by the Planck Collaboration \citep{Adam:2015wua} and here we use some of these techniques.

The Galactic foreground is mainly constituted by four components. The main contribution comes from synchrotron emission from relativistic cosmic ray electrons interacting with the Galactic magnetic field, producing a power law that dominates the Galactic foreground at the BINGO frequencies. Results from the WMAP team present strong evidence of a flatter spectral index in the plane, steepening with Galactic latitude \citep{WMAP:Bennett2013}. Typically, it is expected that the synchrotron spectral index varies from $\beta_s \approx 2.6$ in most of the galactic plane to $\beta \approx 3.1$ in the halo around 1 GHz \citep{WMAP:Bennett2003}.

The second strongest contribution comes from free-free emission caused by electrons scattering off ions in the interstellar medium. The $H\alpha$ emission line is usually used to trace the free-free emission as they both depend on the emission measure $\text{EM} = \int n_{e}^{2}dl$ \citep{Dickinson:2003}. The free-free spectral index is a slowly varying function of frequency with a slight dependence on the local value of $T_{e}$ \citep{Bennett:1992}. At $1\, \mathrm{GHz}$ and considering $T_{e} = 7000$ K as a typical value for the electron temperature, the spectrum has a temperature spectral index of $\beta_{\rm{ff}} = 2.1$. Both synchrotron \textbf{($C_{\ell}^{Synch}$)} and free-free emission \textbf{($C_{\ell}^{Free - Free}$)}  can be seen to dominate the signal power spectrum, together with the contribution from strong radio sources, in Fig. \ref{fig.foregrounds}.

Anomalous microwave emission (AME) produced by the the electric dipoles in the small spinning dust grains ~\citep{Leitch, Draine99} is one of the dominant foregrounds in the frequency range of $10-60\, \mathrm{GHz}$~\citep{Dickinson:2018, PlanckX}. AME is not an important contaminant to the 21-cm in BINGO frequency range, being subdominant with power spectrum well-bellow the one from signal. The same applies for the thermal dust emission from dust grains heated by interstellar radiation field \citep{Planck2011:dust}. Between 980 and 1260 MHz, thermal dust and AME ($C_{\ell}^{AME}$) contributions are well below the \hi\, signal ($C_{\ell}^{21_{40}}$), where the \hi\, power spectrum was obtained from maps convolved with a 40 arcmin beam. Their contribution in the BINGO central frequency is also shown in Fig. \ref{fig.foregrounds}.

\subsubsection{Extragalactic point sources and CMB}
 Extragalactic emission in the BINGO band mainly comes from discrete radio point sources. They comprise all radio emissions from astrophysical objects, such as the inhomogeneous mix of radio galaxies, quasars, star-forming galaxies, etc. Strong sources whose emission can be well characterized by a power-law in the radio band can be identified and masked out during the analysis, essentially eliminating their contribution to the incoming signal.

\begin{figure}[h]
\centering
\includegraphics[width=3.5in]{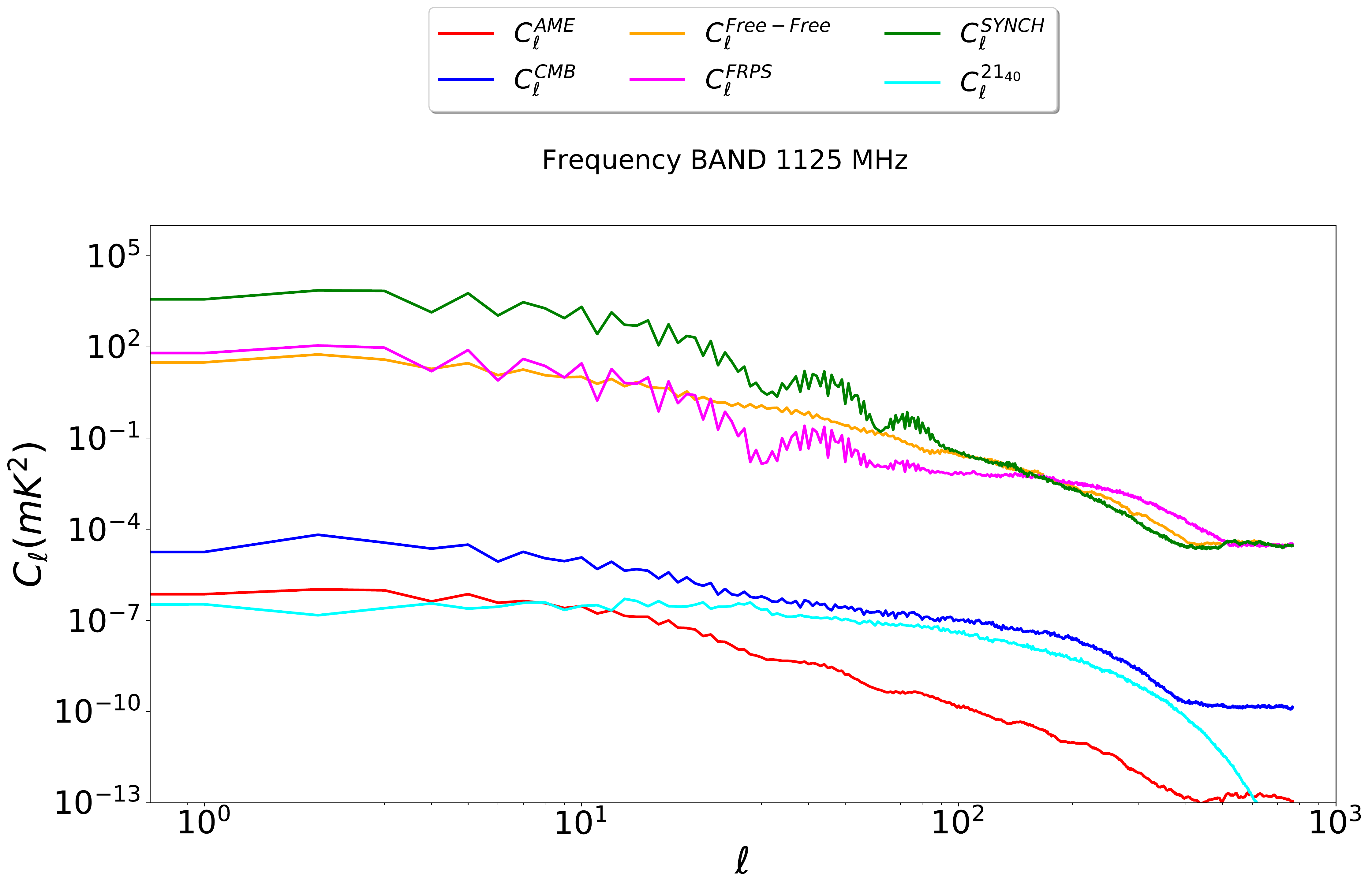}
\caption{Angular power spectra (up to $\ell=10^3)$ of the foregrounds described above, compared with the signal for the central BINGO frequency band ($1125 \, \mathrm{MHz}$) \textbf{($C_{\ell}^{21_{40}}$)}, considering a sky coverage of about 3500 square degrees ($f_{\rm{sky}} \approx 8.7 \%$).}
\label{fig.foregrounds}
\end{figure}

There is also a diffuse contribution from weaker, unresolved, radio sources that is indistinguishable from the Galactic, spatially continuous, foreground. It contaminates the incoming signal to an intensity level slightly lower than the Galactic free-free emission for lower ($\ell \lesssim 100$) multipoles and overcoming the Galactic synchrotron emission for higher ($\ell \gtrsim 300 - 400$) multipoles. It is expected to be significant at high Galactic latitudes, but can be simulated as another diffuse background and combined with the other continuous foreground emissions. The contribution of the faint radio point sources in the power spectrum ($C_{\ell}^{FRPS}$) can be seen in Fig.~\ref{fig.foregrounds}.

The CMB, with a mean temperature today of $T_{\rm{CMB}} \simeq 2.72548 $ K \citep{fixsen2009temperature}, contributes to the \hi\, signal with its temperature fluctuations $\Delta T/T \sim 10^{-5}$, being a relevant foreground at $\approx 1 \, \mathrm{GHz}$. 
Even though it is not the main contaminant for the 21-cm signal, its temperature power spectrum \textbf{($C_{\ell}^{\mathrm{CMB}}$)} is still higher than the 21-cm power spectrum \textbf{($C_{\ell}^{21_{40}}$)} for the multipole range up to $\ell \simeq 300$ for the BINGO central frequency band as shown in Fig. \ref{fig.foregrounds}.

A detailed description of the sky simulations, including the foreground models, can be found in the BINGO companion papers IV and V \citep{2020_sky_simulation,2020_component_separation}.

\subsection{Mission simulation}




This module takes as input the simulated sky maps ({\tt HEALPix} maps)   described in section \ref{sec:sky_model} and reproduces the observation processes to create TODs. The mission simulation includes various instrument information, such as receiver temperature, focal length, horn coordinates in the focal plane, number of receivers, beam profile, number of frequency channels, thermal and $1/f$ noise, noise power amplitude, correlations among different channels etc. More details on the parameters used in the simulations can be found in the companion paper II \citep{BINGO.Instrument:2020}. Different horn configurations, mission duration and different foreground combinations have been tested and are discussed in the companion paper IV \citep{2020_sky_simulation}.

\subsubsection{Beam profiles}

A useful simulated sky map is produced by convolving the frequency-dependent beam response pattern with the sky signal creating a TOD as similar to the real data as possible. The BINGO observing strategy of letting the sky drift across the beam makes the beam summations much simpler, effectively it occurs in only one direction. Frequency response is currently modeled with a Gaussian beam, with plans to use the beam profile measured for the prototype horn \citep{Wuensche:2019} in future works. At this point we have not implemented polarization effects yet (leakage etc.).

\subsubsection{Instrumental noise} 
\label{sec:noise}

Thermal noise in the TOD can be accurately modeled in the simulations from a Gaussian white noise distribution. The stochastic variations in the amplifier voltage are characterized by the receiver system temperature $T_{\rm{sys}}$ using the radiometer equation \citep[e.g.,][]{Wilson:2013trabook}. BINGO receivers are also subject to gain variations, which, incorporated to the radiometer equation, is presented in the form \citep{burke2019introduction}
\begin{equation}
	{\sigma _N} = K{T_{\rm{sys}}}\sqrt {\frac{1}{{\tau \Delta \nu }} + {{\left( {\frac{{\Delta G}}{G}} \right)}^2}} \,,
	\label{eq:1/f_2}
\end{equation}
where $\Delta G$ describes the stochastic variations in the mean gain. $\sigma_N$ is the minimum noise variation that can be detected in the TOD, $\tau$ is the receiver sampling rate and $\Delta \nu$ is the receiver bandwidth. For the BINGO correlation receiver, $K=\sqrt{2}$. 

The $1/f$ noise is a different phenomenon from thermal
noise, being a form of correlated noise present in radio receiver systems, caused by small gain fluctuations \citep[see, e.g., ][]{maino99,Seiffert:2002,Meinhold09}. It appears in the data as large scale spatial fluctuations that are not trivial to separate from the true
underlying sky signal. The removal of 1/f noise has also become an area of research that has resulted in several advanced map-making methods, discussed in section \ref{sec:mapmaking}. We model the $1/f$ contribution as  \citep{Harper:2018}
\begin{equation}
P\left( \nu  \right) = {\sigma ^2}\left[ {1 + {{\left( {\frac{\nu _{\rm{k}}}{{{\nu}}}} \right)}^\alpha }} \right]\,,
\label{eq:1/f_3}
\end{equation}
where $P$($\nu$) is the spectral power of each temporal mode, $\sigma$ is the radiometer sensitivity as described by Eq. (\ref{eq:1/f_2}), $\nu$ is the sampling frequency, $\nu_{\rm{k}}$ is the temporal knee frequency, in which the correlated noise and white noise contribute to the signal with the same level of power, and $\alpha$ is the spectral index of the correlated noise. Typically, the spectral index will vary between $1$ (usually referred to as ``pink noise'' and $2$ (``brown noise''). Simulations allow the analysis of the effects of white noise, knee frequency and $1/f$ spectral index variations during a given mission length. 

\paragraph{Atmosphere}

Radio waves at $\sim 1\,$GHz  cross the atmosphere essentially without being absorbed. However, atmospheric emission, dominated by Oxygen with a small fraction of precipitable water vapor (PWV), contributes with $1.81 \textrm{K} \le T_{atm} \lesssim 1.82 \textrm{K}$ to the final system temperature \citep{Bigot-Sazy:2015jaa}. Atmospheric fluctuations around $ \bar T_{atm}$ can mimic a correlated noise, due to varying opacity from water, and can be treated as a residual $1/f $ in the later phases of the data  analysis. These fluctuations depend on on-site atmospheric conditions, observing frequency band and also on the sky scan strategy \citep[][e.g.]{Church:1995, Lay:2000}, and are caused by the  anisotropic distribution of water vapor molecules in the upper atmosphere. \cite{Bigot-Sazy:2015jaa} estimate the atmospheric fluctuations as seen by a telescope such as BINGO should be $\Delta T_{\rm atm} = 0.01 $ mK, below the instrumental noise in Table~\ref{tab:bingo_param}
of the companion paper II \cite{BINGO.Instrument:2020}. 
The atmosphere is not expected to have any intrinsic contribution to the polarization observations at BINGO frequencies \cite[e.g.,][]{Hanany03}, but it is still important as a contribution to $I$ into $Q/U$ leakage. The simulations generated for the companion paper V \citep{2020_sky_simulation} do not take the atmospheric contribution into account.

\subsubsection{Radio frequency interference}

RFI is the usual term to nominate the emission of radio waves by man-made sources, such as personal computers, cell phones, airplanes, power lines, and orbiting satellites, such as those used for radio-navigation transmissions~\citep{Harper:2018b}. 


Removal of the transient RFI spikes requires the definition of a threshold for identifying spikes. The pixel that contains the brightest observed Galactic emission within the frequency band and the survey area is a convenient choice for this threshold. Higher intensity spikes in the TOD are then flagged by iteration, to check if they are astrophysical transient events. In negative cases, the data chunks  are excised out from the next data analysis steps. Our simulations currently include the presence of ground transients appearing as an intense spike in the TOD data and contributions from geostationary satellites, for which orbit and transmitting power are well known. 

\subsection{Map-making} 
\label{sec:mapmaking}
The map making step of the BINGO pipeline deals with the conversion from TOD to a coordinate-based, pixel-stacked data. Mathematically, the map-making equation, in its simplest linear form, is given by
\begin{equation}
	m \approx \hat s = Wd\,,
	\label{eq:map-making}
\end{equation}
where \textit{m} is the map, \textit{$\hat s$}  is our estimate of the signal, \textit{d} is the data (the TOD) and \textit{W} is a matrix that when applied to the signal performs the required steps. The BINGO pipeline currently has three different map-making routines available: naive filtering and binning, destripping, and maximum likelihood.


One of the implementations of the BINGO pipeline currently uses the simple averaging and binning routines, in a process that is usually referred to as ``naive map-making''. This map-making technique performs no additional data-processing in the timelines. The data are simply re-mapped onto the image plane, by projecting the full power seen by a detector onto the nearest sky map pixel. The output is an image map, an error map (as the standard deviation of data samples divided by the square root of the number of data samples) and a coverage map, effectively a hit map for samples/pixel.

The naive technique is the fastest of the three map-making schemes. It is relatively easy to implement but might yield crude maps due to the simplified noise estimation. When applying the naive technique we assume that the filters in use in the pipeline manage to remove enough of the noise correlations, such that we can estimate the remaining data to be signal and white noise only. For the moment, we  assume that all detectors read the same noise level. In this case the data can simply be written as
\begin{equation}
	d = Ps + {n_w}\,,
	\label{eq:map-making_2}
\end{equation}
where \textit{P} is a $(n_{t},\, n_{\mathrm{pix}})$-matrix taking the signal from the time-domain into the pixel-domain, $s$ is the signal of interest, $n_{w}$ is the uncorrelated white noise. Eq.  (\ref{eq:map-making_2}), when solved for the estimated signal $\hat s$, can be written as:
\begin{equation}
	\hat s = {\left( {{P^T}P} \right)^{-1}}{P^T}d\,.
	\label{eq:map-making_3}
\end{equation}
The result is a simple average of all observations per pixel. This equation can be modified to account for individual weighting when calculating the bin-average by adding a diagonal matrix.

 \subsection{Component separation} 
 \label{Sec.:Component_separation}

The component separation module of the pipeline is a very important part of the data analysis, where the desired \hi\, signal must be separated from all foregrounds and analyzed according to the BINGO noise model. The foreground power spectra presented in Section~\ref{Sec.:foregrounds} can be approximated by power-laws, enabling their separation from the \hi\, signal. There are several component separation methods to extract the target cosmological signal from the foregrounds that can be divided into parametric and ``blind'' techniques. The parametric methods assume each component is characterized by a specific dependent emission law known a priori, whereas the ``blind'' techniques, as the name suggests, do not assume any specific parametric model for the foregrounds. 

\begin{figure}
        \centering
        \includegraphics[width=3.5in]
        {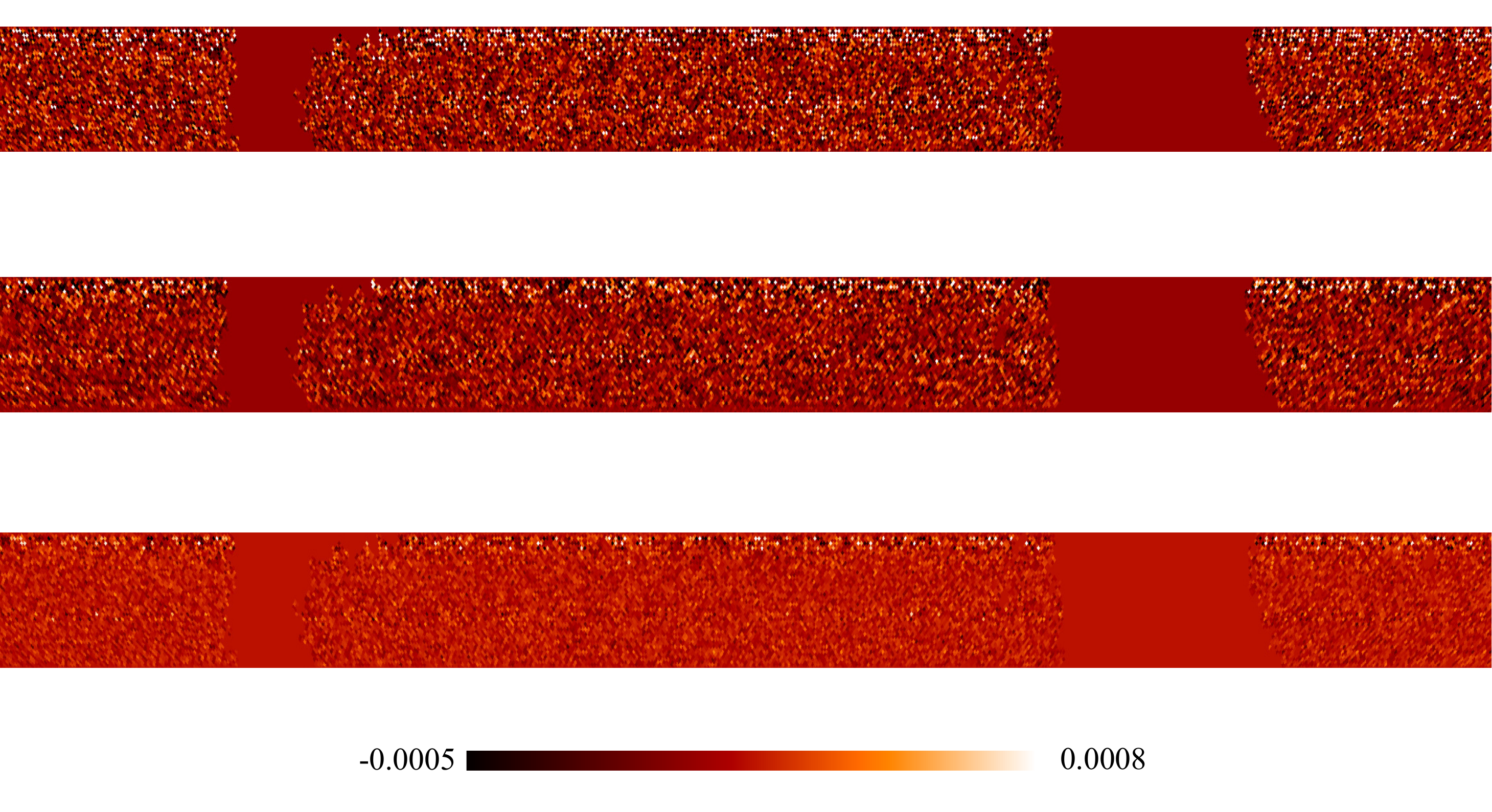}
        \caption{Comparison between BINGO \hi\, plus noise maps ($N_{\rm{side}}=128$) with a Galactic mask applied at  $\approx$ 1.1 GHz (first strip), the {\tt GNILC} reconstructed map (second strip) and the residuals map (last strip). See the companion paper IV for more details \citep{2020_sky_simulation}. Temperatures are given in K.}
        \label{fig:my_comparison}
    \end{figure}

The pipeline work currently assumes a non parametric approach, using the Generalized Needlet Internal Linear Combination ({\tt GNILC}) methodology as the standard component separation technique \citep{Remazeilles:2011,Olivari:2015} to recover the 21-cm signal. 
{\tt GNILC} takes into consideration both the frequency and spatial information of the observed (simulated) data in order to separate the components. A set of needlet windows is defined in harmonic space to isolate different ranges of angular scales in the input map. Each frequency map can be split in several needlet maps, allowing for a more localized analysis of the signal. Figure~\ref{fig:my_comparison} shows maps covering the BINGO patch of the sky with three different signals: input \hi\, plus noise, {\tt GNILC} reconstructed \hi\, plus noise, and residuals. The maps are produced at $\approx$ 1.1 GHz. It is clear how {\tt GNILC} is able to reconstruct the main features of the input signal with a r.m.s residual equal to 0.06 mK For the scenarios considered we found values off by $\leq$ 10\%. More details regarding this analysis can be be found in the companion paper V \citep{2020_component_separation}.

We intend to use other foreground cleaning methods, along with {\tt GNILC}, in our simulations in the future, as well as developing specific methods optimized for \hi\ IM experiments. The {\tt GNILC} algorithm can be improved to obtain a more efficient component separation. The BINGO collaboration is working on the optimization of the code and on a more precise estimation of the errors arising during the \hi\ recovery process. A detailed description of the current status of {\tt GNILC} performance and the results of the sky simulations, including the foreground models, can be found in the companion papers IV and V \citep{2020_sky_simulation,2020_component_separation}.

\subsection{Cosmology modules}

This final part of the pipeline is about obtaining the science products from the 21-cm \hi\ IM data. These products are science-dependent. For cosmology, for instance, we would estimate the covariance, the \hi\ power spectra, perform a Bayesian analysis to estimate the cosmological parameters of interest, and estimate the bispectrum. 

\subsubsection{Covariance estimation}
Constructing a covariance matrix is a key part for estimating observable uncertainties and further constraining the parameter space of the cosmological parameters. In our case, the observable is the angular power spectra of the \hi\ IM map. For constructing the covariance matrix of the angular power spectrum, a number of realizations are needed. However, a large scale N-body simulation is computationally very expensive. The Gaussian random realization using the calculated angular power spectra is applied in \cite{asorey2020hir4}. 

We use two methods to generate hundreds of realizations of the tomographic \hi\ IM maps: the log-normal density field realization using the {\tt FLASK} code \citep{Xavier:2016} as well as fast light cone halo catalog generation using the {\tt L-PICOLA} code \citep{howlett2015picola}. We can generate full-sky IM maps, despite BINGO actually observing only about 15\% ($6000$ deg$^{2}$) of the sky. Therefore, in every realization, we  also mask the full-sky in different random directions to achieve about eight times more maps from each realization. This reduces significantly the computational cost for constructing the covariance matrix.

\subsubsection{Bispectrum estimation}
\label{Sec.:bispectrum_module}

There is additional information in the 21-cm maps than the one that can be obtained by the angular power spectrum. Non-gaussianities (NG), which are deviations in our maps from Gaussian statistics, will be present in our data at all redsfhits, but most notably in the low redshift 21-cm signal. They might come from early-times evolution of the Universe, what we call primordial NG; and/or will get imprinted onto the 21-cm maps by gravity itself; and/or will get imprinted on the 21-cm maps when these maps are cleaned, by residuals from the galactic foreground distribution. We use higher order statistics to characterize these NG.

We focus here on the three-point correlation function, or its Fourier transform, the bispectrum. The three-point correlation function in flat sky approximation is given by \citep{Liguori, Komatsu, Zaldarriaga}
 \begin{equation}
     \langle \Delta_{T_b,\ell_1} \Delta_{T_b,\ell_2} \Delta_{T_b,\ell_3} \rangle \equiv  B_{l_{1}l_{2} l_{3}}^{m_{1} m_{2} m_{3}} = \langle B_{l_{1}l_{2} l_{3}} \rangle \Bigg (\begin{matrix} l_{1} & l_{2} & l_{3} \\ m_{1} & m_{2} & m_{3} \end{matrix}\Bigg )\,,
 \end{equation}
 where $B_{l_{1}l_{2} l_{3}}^{m_{1} m_{2} m_{3}}$ is the angular bispectrum ($B_{\ell}$ from now on), and $B_{l_{1}l_{2} l_{3}}$ is the averaged angular bispectrum. The matrix denotes the Wigner-$3j$ symbol, which is invariant under permutations. It describes three angular momenta that form a triangle $\mathbf{L}_1+\mathbf{L}_2+\mathbf{L}_3=0$, where $m_1+m_2+m_3 =0$.
The bispectrum can be written as
 \begin{equation}
     B_{\ell} = G_{l_{1}l_{2} l_{3}}^{m_{1} m_{2} m_{3}} \, b_{l_{1}l_{2} l_{3}}\,,
     \label{eq.:bispectrum}
 \end{equation}
 where $b_{l_{1}l_{2} l_{3}}$ is the reduced power spectrum and $G_{l_{1}l_{2} l_{3}}^{m_{1} m_{2} m_{3}}$ is the Gaunt integral~\citep[for more details see ][]{Komatsu,2020_component_separation}, non zero only if $l_1+l_2+l_3 = \mathrm{even}$, where all the information about the Wigner-$3j$ symbol is present. 
The bispectrum can have different shapes. Here we are going to focus on two: the equilateral shape, given by $l_1=l_2=l_3$ and the isosceles shape, where $l_2 = l_3 \neq l_1$.

While the primordial NG is a signal we would like to extract from the data, the residual NG in the maps coming from foreground subtraction and radio frequency interference mitigation is a signal that is on the way of extracting information about the cosmology from the cleaned 21-cm signal. For this reason it is important to be able to detect the presence of these NG features in the maps and characterize them to understand their origin and characteristic. 

The goal of this module is to be able to identify cosmological NG information in the reconstructed maps, after the component separation module ({\tt GNILC}) removed the foregrounds and recovered the cosmological 21-cm signal. The module looks more specifically for NG with the equilateral and isosceles shapes, and can evaluate the mean bispectrum $B_{\ell}$. 

In order to verify how well our pipeline is recovering the cosmological \hi\, data, we perform a few tests to verify if the cosmological NG 21-cm information (previously included in the simulation) is recovered in the reconstructed maps from {\tt GNILC}, and, mainly, if the bispectrum module can identify this information in those maps, by comparing those with graphical results of foregrounds. To perform such tests, we compare the maps generated by {\tt FLASK}. It contains an NG information from the input log-normal distribution, with the same maps after they are reconstructed by {\tt GNILC}, following the procedure explained in Section \ref{Sec.:Component_separation}.

\begin{figure}[h]
    \includegraphics[width=4.4cm]{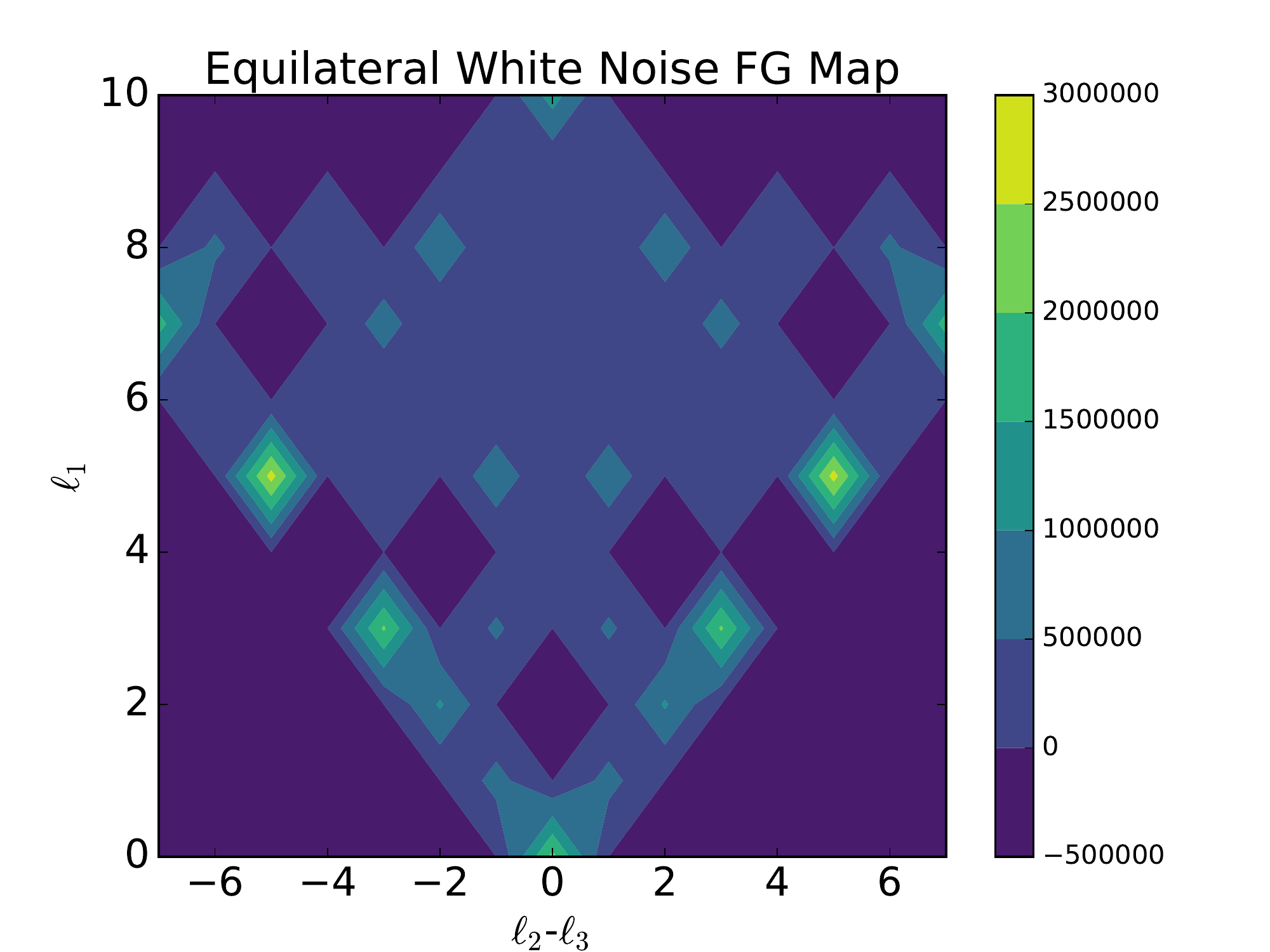}
    \includegraphics[width=4.4cm]{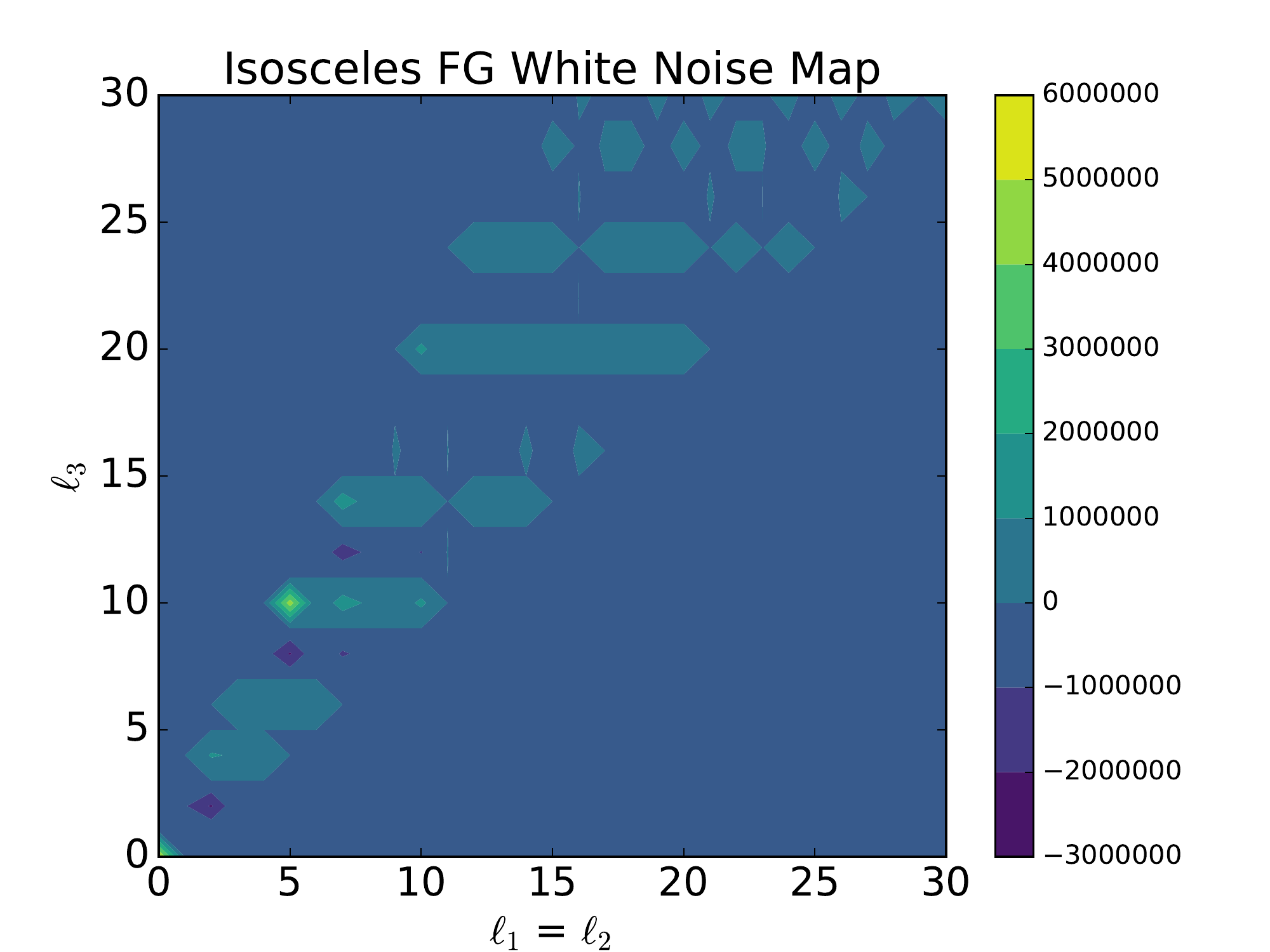}\\
    \includegraphics[width=4.4cm]{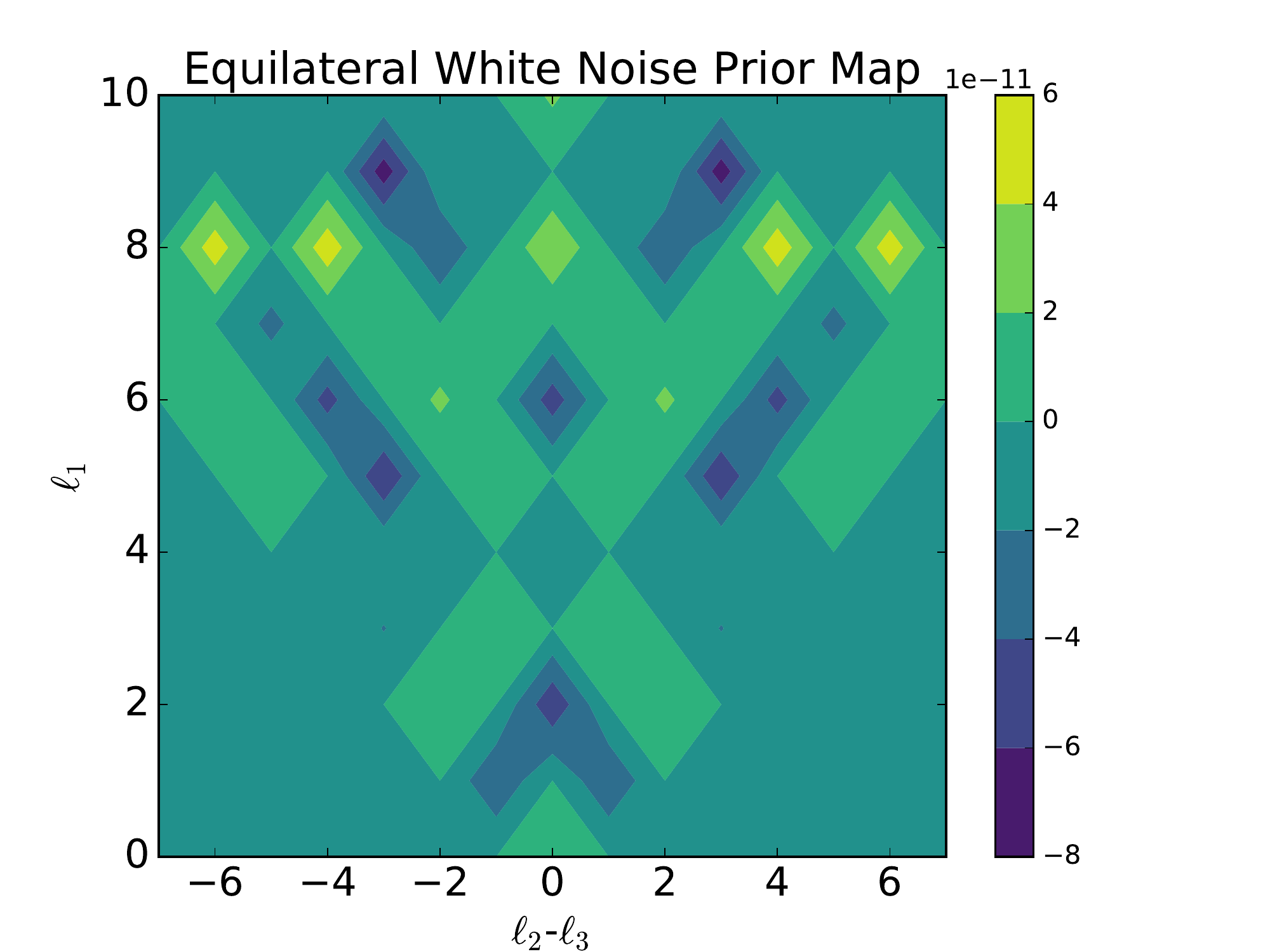}
    \includegraphics[width=4.4cm]{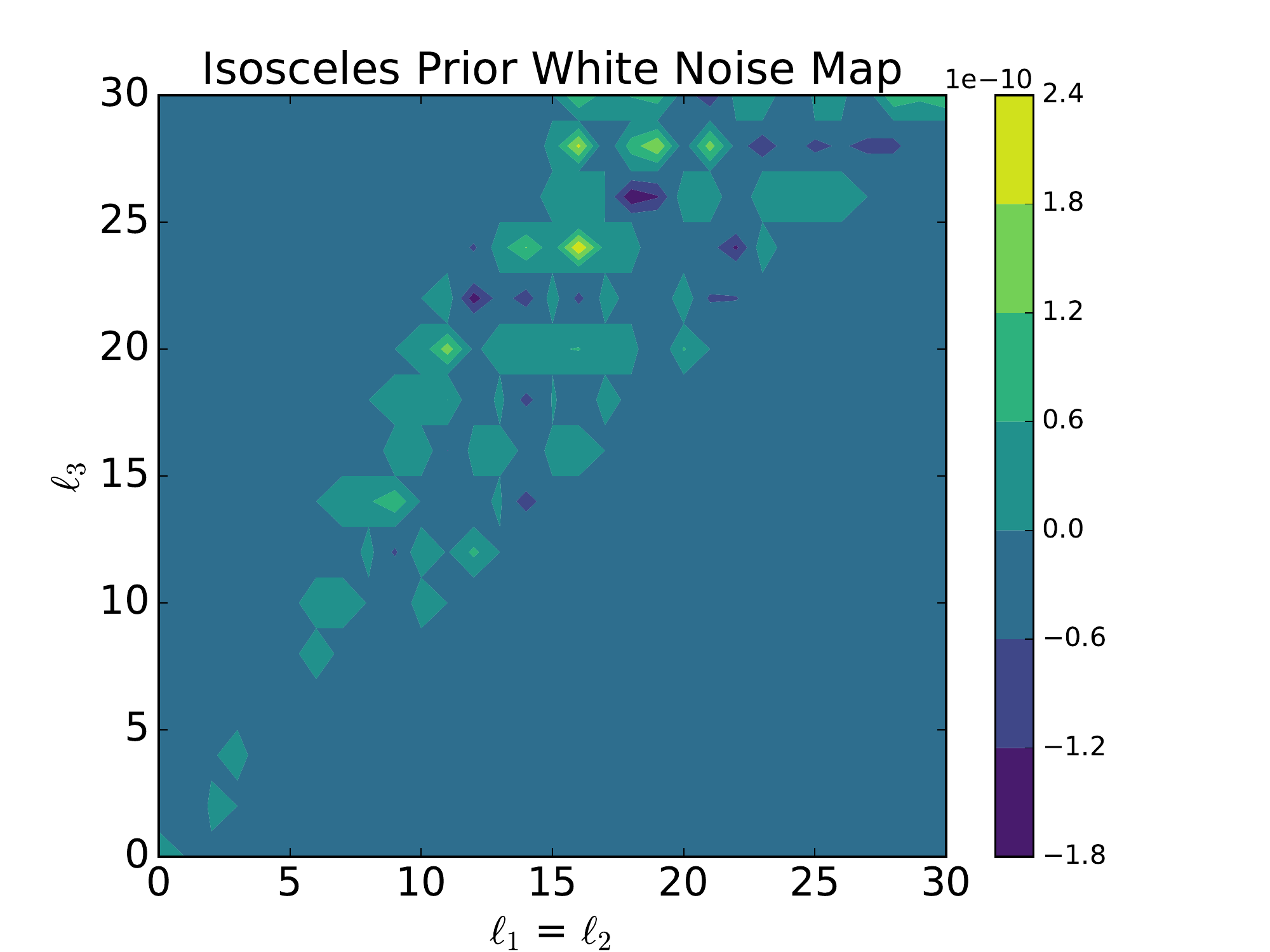}\\
    \includegraphics[width=4.4cm]{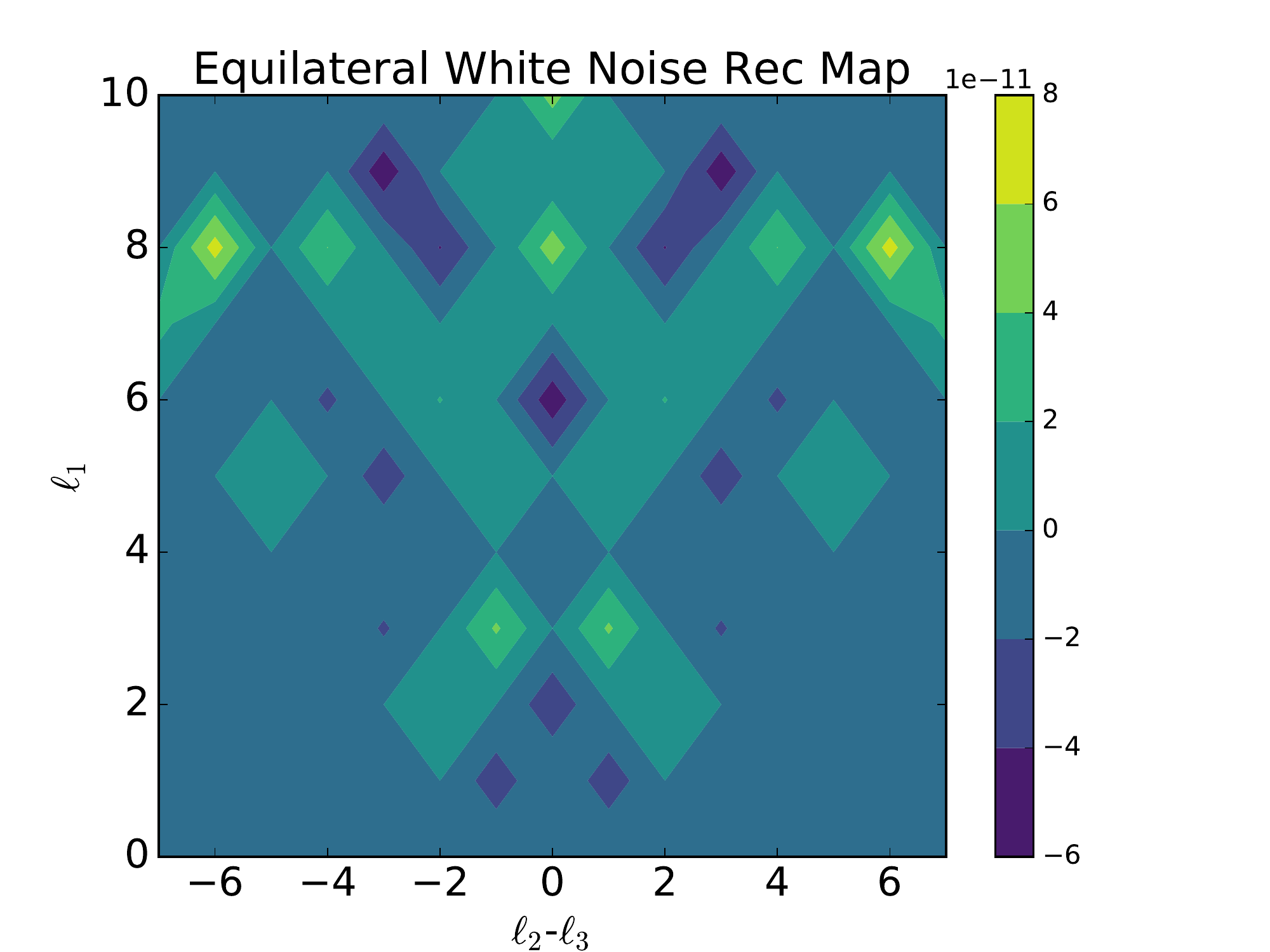}
    \includegraphics[width=4.4cm]{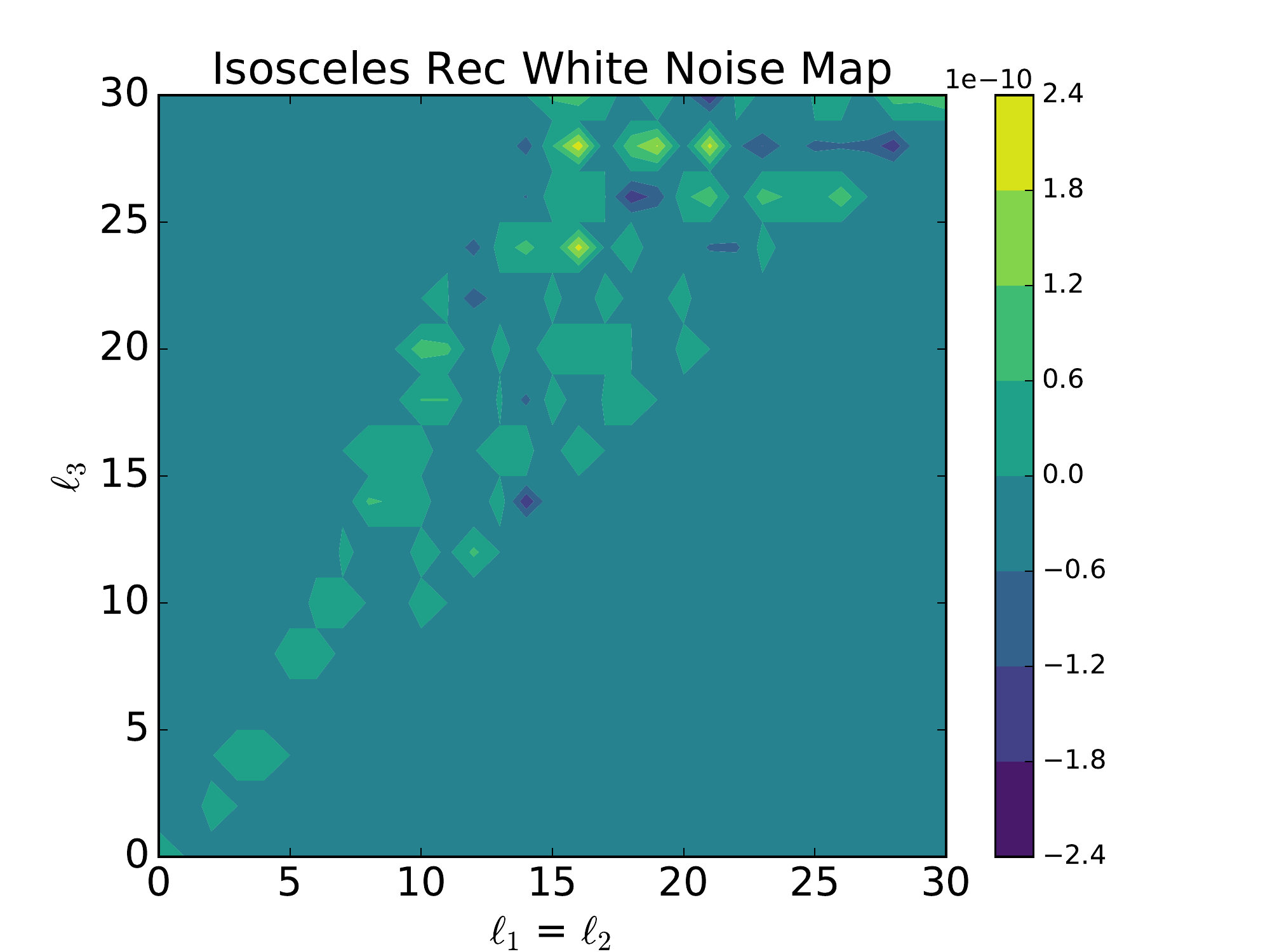}
    \caption{Contour charts for bispectrum with equilateral  and isosceles shape for central channel ($\approx$ 1125 MHz), white noise case. The first column corresponds to equilateral case and the second to isosceles one. The first row corresponds to foreground information, the second row corresponds to the 21-cm map  from the {\tt FLASK} simulation and the third row from the reconstructed map ({\tt GNILC}).}

   \label{Fig.:Bispectrum_module}
\end{figure}

To compare these different maps and test the reconstruction procedure we use the shape of the NG, and mean value of the $B_{\ell}$ obtained from each map. Using the shape, the tests showed that we cannot recover the full 21-cm information at some redshift bins, with the recovery becoming better and yielding acceptable results for higher redshift bins. The isosceles shape shows slightly more satisfactory results, as it can be seen in Fig. \ref{Fig.:Bispectrum_module}. The foreground graphs are shown to be very different from the {\tt FLASK} and recovered maps, since it contains mixed NG and Gaussian information from different types of contaminants. The reconstructed map has the 21-cm recovered information, while the so called 21-cm prior has the pure cosmological signal, with NG information.  It is expected that these two maps have the same order of magnitude and same component. However, this is not the case in our analysis, and it is still not clear if this discrepancy comes from the reconstruction procedure or from the bispectrum module. 

Comparing the average amplitude of the bispectrum confirms these results, where there is a consistent recovery, but not the entire signal. A more detailed analysis can be found in the companion paper V \citep{2020_component_separation}.



\subsubsection{Minkowski functionals estimation}
\label{Sec.:MF_module}

A different and complementary tool to the previous optimal NG estimator based on the bispectrum  is the Minkowski functionals \citep[MF;][]{Minkowski:1903, Novikov:1999,WMAP:Komatsu2003}. Sensitive to weak and arbitrary NG signals, the MF not only provide information about the spatial correlation of a random field, but can also provide morphological information and map the shape of structures.  
 
 For the 2D \hi\, brightness temperature field, defined on the sphere ${\cal S}^2$, $T = T(\theta,\phi)$, with mean $\langle T \rangle$ and variance $\sigma_0^2$, three MF are enough to provide a test of non-Gaussian features by assessing properties of connected regions in a map. The global MF are the total area $V_0(\nu_t)=(1/4\pi) \int_{\Sigma} d\Omega = \sum_{i} a_i $, perimeter $V_1(\nu_t)= (1/8\pi) \int_{\partial\Sigma} dl = \sum_{i} l_i$ and genus (number of holes) $V_2(\nu_t)= (1/8\pi^2) \int_{\partial\Sigma} \kappa~dl = \sum_{i} (1-n_i) =N_{\mathrm{over}} - N_{\mathrm{under}}$ calculated over all the connected region $R_i$ in an excursion\footnote{An excursion set is defined as the region of a sky patch where the \hi\, field exceed a given threshold $\nu_t$, $\delta T(\theta,\phi) / \sigma_0 \equiv \nu > \nu_t$, with $\delta T \equiv T(\theta, \phi) - \langle T \rangle$.} $\Sigma \subset {\cal S}^2$ ~\citep{Novikov:1999,WMAP:Komatsu2003,Novaes:2018,Ducout:2013},
 where $d \Omega$ and $dl$ are the elements of solid angle and line, respectively, $\kappa$ is the geodesic curvature, and $a_i$, $l_i$ and $n_i$ are the area, perimeter and genus of one connected region $R_i$. This method can be efficiently applied to masked skies or in small regions of the sphere, being specially useful in the case of the \hi\, measurements from BINGO, with a partial sky coverage. In fact, such versatility of the MF motivated their usage as one of the modules for \hi\ IM data analysis. 

The MF module is under development to work using two main approaches: (1) to access the cosmological information provided by the \hi\, signal in constraining cosmological parameters; and (2) to evaluate the recovered \hi\ signal by using different component separation methods, exploring the efficiency and robustness of each method and comparing their results, as a complement to the bispectrum module. 

\begin{figure}[h]
\includegraphics[width=8cm]{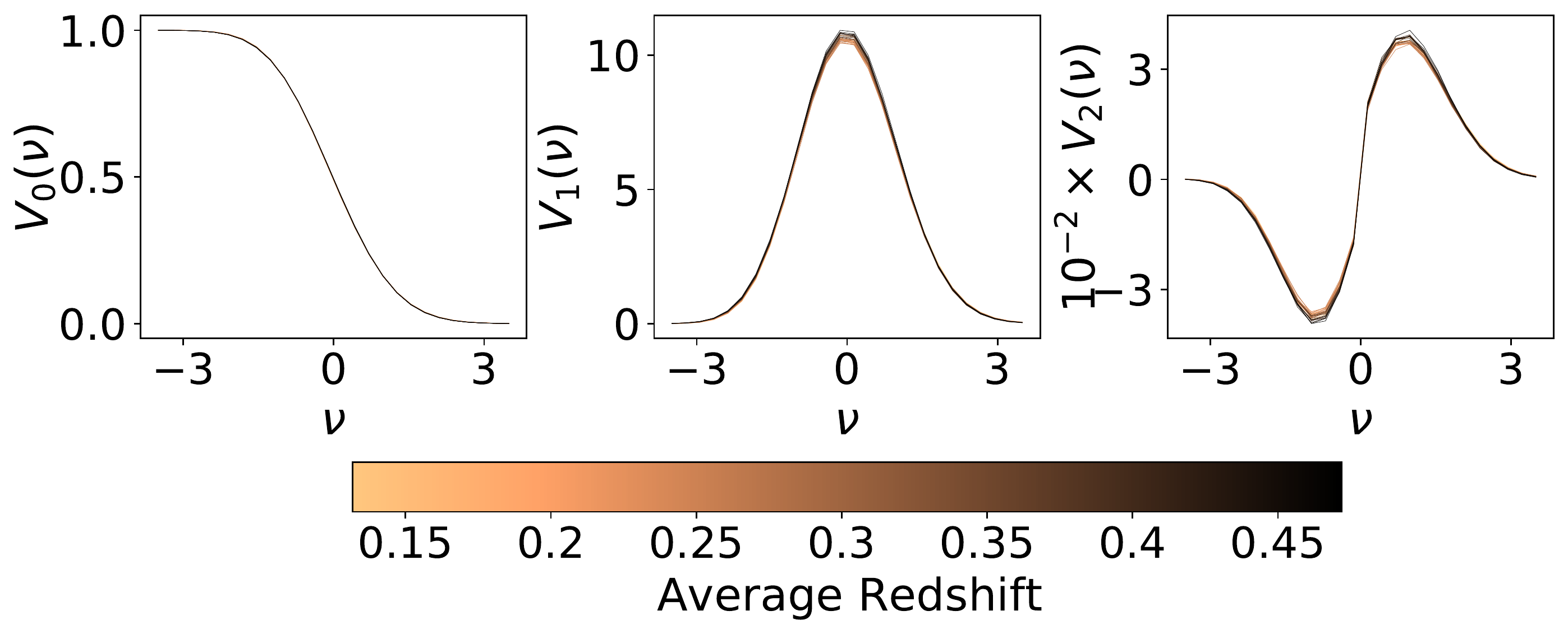} 
\caption{Three MF: area (left), perimeter (middle), and genus (right), averaged over $1000$ \hi\, IM simulations at each redshift (frequency) bin to be measured by BINGO. 
Here we consider 30 redshift bins, equally spaced in frequency, from $0.127$ ($980$ MHz) to $0.449$ ($1260$ MHz), indicated by the color scale. The simulation are generated assuming the $\Lambda$CDM cosmological parameters measured by \cite{Aghanim:2018eyx}. }
\label{fig:MFonBingoBins}
\end{figure}

Using the algorithm provided by \cite{2012/gay}, and \cite{Ducout:2013} to calculate the MF, the first approach has been tested on simulated data produced by the {\tt FLASK} code. In summary, the methodology of this approach concerns the calculation of the MF from a large set of synthetic \hi\, maps, produced assuming different cosmologies (varying a set of cosmological parameters) and redshift bins that will be observed by BINGO, and use a machine learning technique to map the MF features in terms of the cosmological parameters. As an illustrative example, Fig. \ref{fig:MFonBingoBins} shows the evolution of the \hi\ signal with redshift, as measured by BINGO, for a given fixed cosmology. After widely testing it in synthetic  data, evaluating the impacts of residual contamination and instrumental effects, we will be able to apply the methodology to the BINGO data.

\section{Scientific synergies with other instruments} \label{sec:comp}

\subsection{BINGO in the wider context}

There has been a recent burgeoning of interest in new radio astronomy instruments such as ALMA, LOFAR, and of course the SKA and its precursors MeerKAT \citep{booth2009meerkat} and ASKAP \citep{johnston2008science,deboer2009australian}. All these activities have been driven by the realization that radio astronomical observations have a unique contribution to make to cosmology, astrophysics and even fundamental physics.

\begin{figure}[h]
\centering
\includegraphics[width=9.cm]{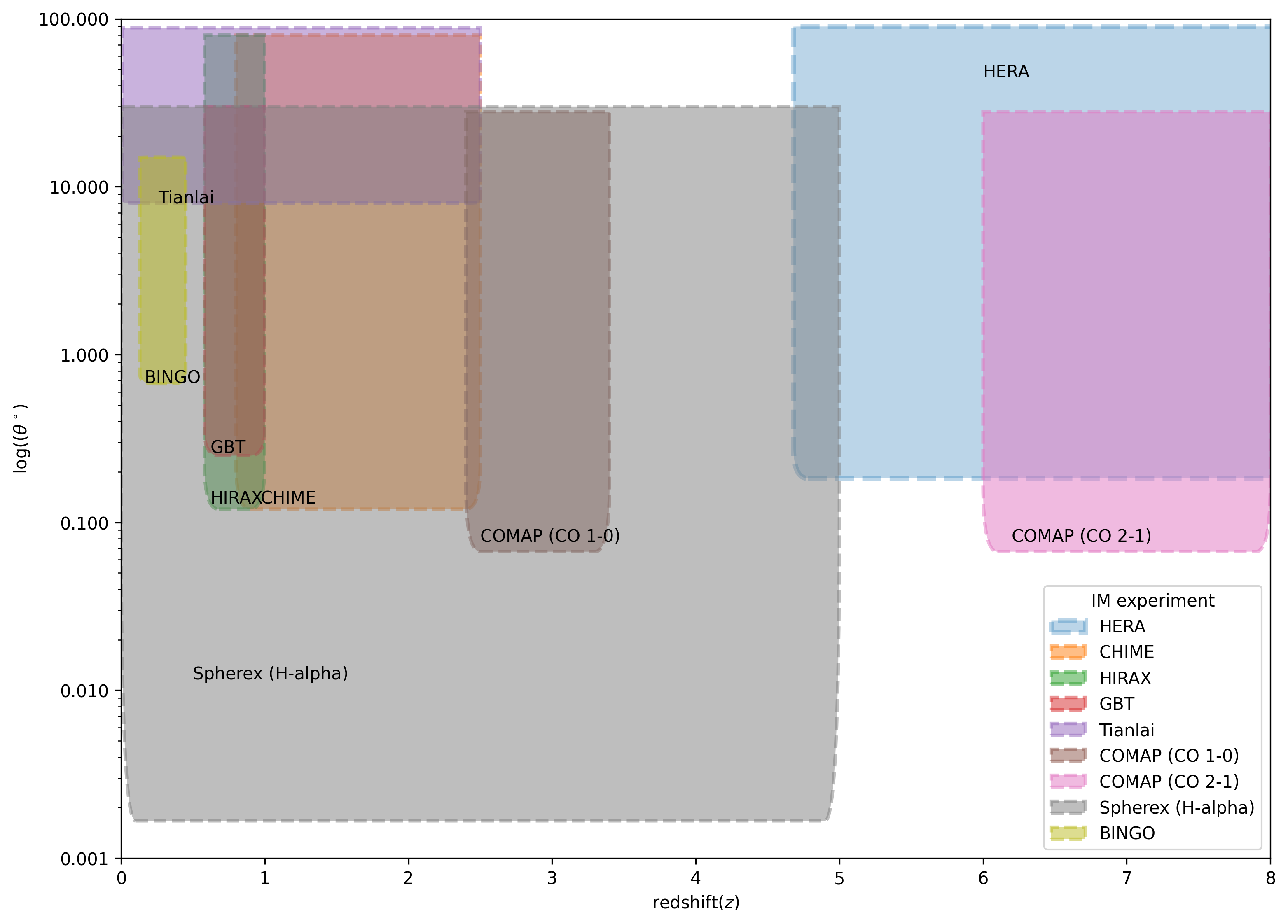}
\includegraphics[width=8.5cm]{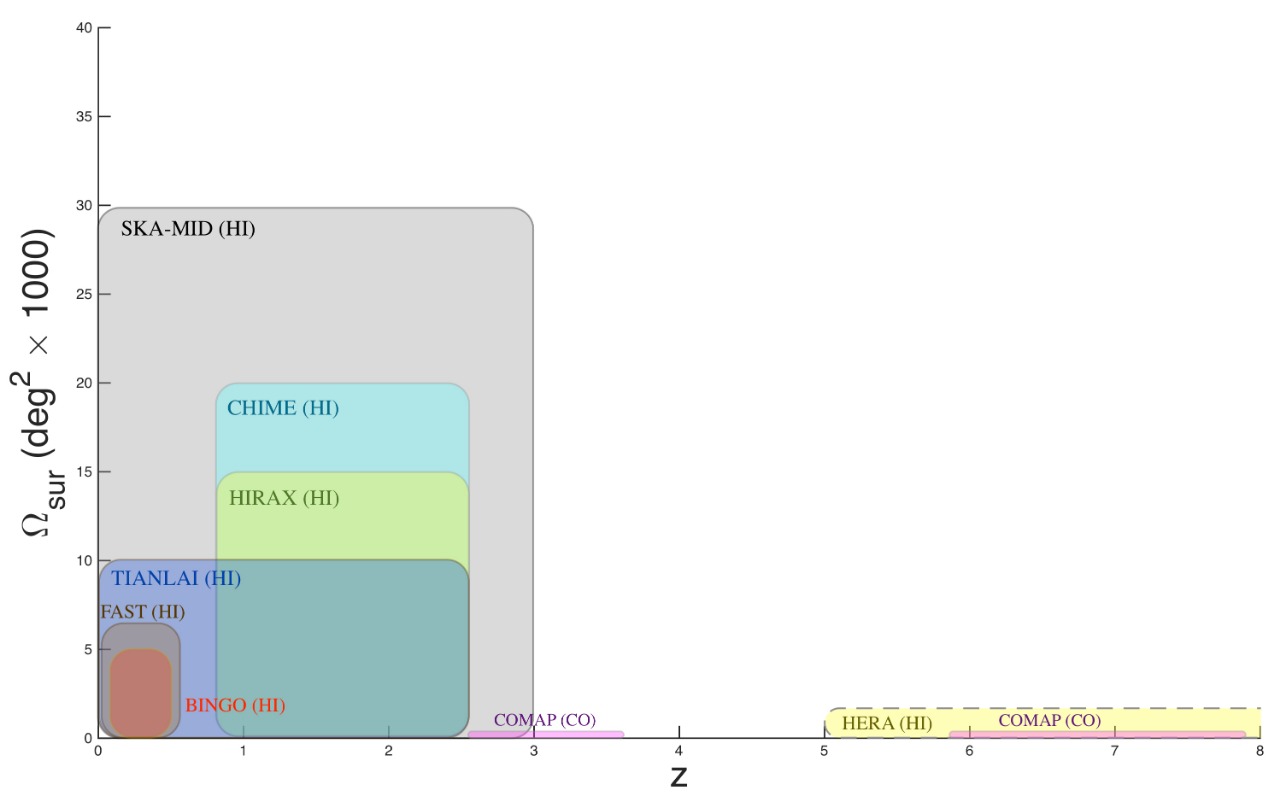}
\caption{
Top: Comparative plot of angular resolution versus redshift coverage for different IM experiments. The red rectangle with the arrow shows the approximate spot for BINGO \cite[adapted from][]{Kovetz:2017}. Bottom: 
Figure from \cite{Wuensche:2021cfs}, showing a comparative plot of the sky and redshift coverage for different IM experiments. From those, CHIME, FAST and Tianlai are already in operation.}
\label{fig:surveys}
\end{figure}

For comparison, in Fig. \ref{fig:surveys} we show some of the IM experiments already in operation or planned for the near future, with covered area in the sky as a function of redshift. Albeit operating in the same redshift range of Tianlai, FAST and SKA telescopes, BINGO can still be a strong competitor, producing good quality data very early. 

\subsection{Square kilometer array }
The SKA project is an idea under discussion since the turn of the second millennium and is well known as the largest and most important observational project in cosmology and astrophysics up to now and in the decades to come.


SKA is a project aiming to look at the cosmos from 10 million years of age, when the first generation of stars lighted, up to the DE-dominated Universe, basically today. A telescope looking at these matters should have a total size of one square kilometer, hence its name and structure. Clearly, SKA has a tremendous scope: astrobiology, tests of gravity, exotic transient phenomena (pulsars, FRBs, etc.) planet observations, magnetic fields, interplanetary medium, DE and DM, among further more specific astrophysical and cosmological aims.

A new science book has been  produced by the SKA consortium \citep{SKAWG:2020}. The previously proposed SKA surveys, which usually were thought of as the main cosmology surveys to be carried out with the SKA, were the SKA continuum survey and the SKA galaxy redshift survey \citep{2015aska.confE..17A,2015aska.confE..19S,2020PASA...37....7S}. In the recent science prioritization, three summary chapters were produced to outline the main SKA cosmology goals, two of them being those for the above mentioned surveys. An additional chapter has been produced by the SKA team to cover the science that can be done with IM \citep{Bourke2014}. This shows the growing realization that IM is one of the most powerful cosmological tools available. The main summary chapter for cosmology with the SKA argues that without using techniques to obtain intensity maps, such as the ones produced by BINGO, the project will be severely hampered in its capability of producing cosmological transformational science in its first phase. 
BINGO will be an excellent pathfinder to investigate issues that  will likely be faced with the SKA.

The main observations common to SKA and BINGO, actually the \textit{"raison d'\^etre"} of the BINGO project is the (indirect) observation of the dark sector, by means of the BAO, which can be observed by both projects, leading to a unique fingerprint on cosmological observations and in turn being used to constrain cosmological parameters. 

Good observations by BINGO can give hints to what to look for in the case of SKA, which will observe again the BINGO volume and go much beyond into the cosmic dawn. Further aims of SKA include
search of protoplanetary disks and exoplanets  as well as magnetic fields and other galaxy structures. In Galactic astrophysics, details of the synchroton emission and Faraday rotation will be clearly observed.

\subsection{MeerKAT}
The MeerKAT is a project designed as a SKA pathfinder. Located in the Karoo region, South Africa, it consists in an array of 64 antennas and is very sensitive, with a resolution of circa 1 arc second at 1420 MHz, seeking regions at $z\sim 4$ or even higher. 
A low frequency band, from 850 to 2500 MHz (which includes the BINGO band) will look at most epochs of modern Universe, and will be able to deeply map the \hi\, distribution. In particular, it covers the epoch of reionization, thus informing about DM evolution in the recent Universe. Galactic studies, magnetic fields and the tiny variations in matter distribution are also within the reach of low frequency band of MeerKAT. MeerKAT also operates in a high frequency band (8 to 14.5 GHz) searching for transients and other astrophysical objects.

\subsection{CHIME}

Another experiment aiming to measure BAO is the Canadian Hydrogen Intensity Mapping Experiment - CHIME, a radio interferometer located at the Dominion Radio Astrophysical Observatory  \citep[DRAO, CHIME;][]{Bandura:2014gwa}. CHIME operates in the frequency range of 400 MHz to 800 MHz, with a synthesized beam resolution between 20 and 40 arcminutes, in order to produce 21-cm sky maps for about 50\% of the sky, across the redshift range $z=0.8-2.5$ \citep{Shaw:2013wza}. 

CHIME is complementary to BINGO in the frequency and correspondingly redshift range ($0.8 \le z \le 2.5$) as well as in the form, consisting of 4 adjacent cylindrical reflectors of $20$ m diameter $\times 100$ meters length, oriented north-south. Similarly to BINGO, CHIME has no moving parts. 

The main cosmological goal for CHIME is also measuring BAO using \hi\, IM. However, CHIME is a very efficient telecope to detect FRBs and, at the time of this writing, has already found dozens of them, with a predicted discovery rate of $2 - 50$/day. Pulsars are also a strong target for CHIME, making it a partner for BINGO \citep{Amiri:2018qsq,Amiri:2019bjk}.

In CHIME a separate pipeline for FRBs has been constructed, with  dedicated electronics, for identifying FRBs as well as storing the data. Smaller dispersion measures are easier to handle, and CHIME has a maximum capability of handling large dispersion measures with high resolution. 

\subsection{Five-hundred meter aperture spherical radio telescope (FAST)}

The Five-hundred-meter Aperture Spherical radio Telescope (FAST) is currently the largest single dish radio telescope in the world. It is located in the Guizhou province, China \citep{Nan:2011um}. FAST is a spherical reflector with a diameter of 500 m, fit in the Dawodang depression. 
Because of its size, the FAST dish itself does not move. However, the telescope is able to observe sources within 40 degrees of the zenith because the shape of the dish’s surface made up of 4,450 aluminum panels that are tilted and turned by about 2,000 mechanical winches and the feed cabin, where the radio waves are focused, is suspended from six cables and moves around the surface of the dish. Inside the cabin, multibeam and multiband receivers are installed, covering a large frequency range of 70 MHz -- 3 GHz~\citep{Nan:2011um,2016ASPC..502...93L}.
Besides being a galaxy survey, FAST is also an IM survey~\citep{Smoot:2014oia,Bigot-Sazy:2015tot,Hu:2019okh}. As it was pointed out in these references IM cannot be done in this entire frequency range. In ~\cite{Smoot:2014oia} for a 500 m aperture, they could do IM between redshifts $0.5-2.5$, while in~\cite{Bigot-Sazy:2015tot} for the illuminated aperture os 300 m it is suggested a frequency range for IM of $350-1350$ MHz. In~\cite{Hu:2019okh} the performance of IM with FAST is studied in details for a redshift range of $0.05< z < 1.25$.  From this study, in this other mode of operation, FAST IM can provide good cosmological measurements, with forecasts showing that FAST can reach the constraining level of dark energy task force (DETF) stage IV experiments.


FAST is a Chinese contribution to the international efforts to build the SKA. It is the most sensitive single dish radio telescope ever built. 
FAST is expected to map the \hi\, gas in the Milky Way Galaxy at a very high resolution and to increase the number of known pulsars from almost 2,000 to about 6,000. 

The FAST project took 5.5 years from the commencement of work in March of 2011 and was completed in September 2016. FAST has been in final tests for three years, and it has already spotted 102 brand new pulsars (half of the pulsars in the Galaxy are in the reach of FAST). It also detected a few FRBs. In September 2019, the telescope passed a series of performance assessments and was given the green light open to astronomers.  The project was funded by the Chinese government, and FAST is operated by the National Astronomical Observatories of the Chinese Academy of Sciences.

We can see that the redshift range of FAST coincides with the one from BINGO, specially the frequency where it is possible to do IM, as shown in Fig.~\ref{fig:surveys} where we used the IM redshift range from~\cite{Hu:2019okh}. This makes FAST not only a good competitor of BINGO, but also complementary, since BINGO will be built in the southern hemisphere while FAST is in the northern hemisphere. Having different experiments measuring in the same frequency is also important to confirm the science explored by both projects.

\subsection{The Tianlai project: a 21cm radio telescope array}

The Tianlai project (``Sounds from the Heaven'') is a similar project to survey the northern sky, making IM observations of the redshifted 21-cm line from the \hi\, below redshift 3, to measure the BAO signal of LSS and measure the DE equation of state so as to constrain DE models \citep{Chen:2012xu}. It is also a pathfinder for testing the basic principles and key technologies of 21-cm IM. The project is an international effort with participants from China, France, the USA and Canada and is being built in a very quiet radio site in Balikun, Xinjiang, China. 

The Tianlai pathfinder includes a dish array with 16 dishes of $6$m diameter, each equipped with a dual linear-polarization feed, compactly arranged in two concentric rings. These dishes are equipped with electronically controlled motor drives in the altitude-azimuth mount, which allow people to steer the dishes to point to almost any desirable directions above the horizon.  The dishes operate alongside a cylinder array with three $15$-m diameter $\times \, 40$-m length, north-south oriented, cylinders. The cylinders are equipped with 96 dual linear-polarization feeds that can be moved freely along the guide rail, which allows observers to change the separation between the feeds flexibly. Tianlai array receivers are designed to have broad band response (400 -- 1500 MHz). 

The construction of the two arrays was completed in 2015, and the first trial observations were done in September 2016. An initial report of observations was done by  \cite{Wu:2020}.


\section{Conclusions}\label{sec:conclusions}

This paper is the first of a series of papers describing the developments and capabilities of the BINGO telescope. In this paper, we have described the BINGO project and its science goals: cosmology, which is the main driver of the project, and astrophysics. We summarized the most recent plans of construction and key observations strategy,  developments of the pipeline, and cosmological and astrophysical forecasts for the instrument. 

BINGO will use the IM technique to map the redshifted \hi\, between  $z=0.13$ and $z=0.45$, and measure the BAO, providing independent cosmological data. It will put complementary constraints on plenty of cosmological models and alternative cosmologies, such as dynamical DE, interacting DE and modified gravity, whose combination with \textit{Planck } results will lead to a gain in precision of the cosmological parameters. Additionally, it will be possible to constrain the 21-cm parameters (\hi\, bias and density parameter)  and the sum of neutrino masses.

Besides its main goal, BINGO has the possibility to  detect FRBs at a rate of  one each $4-5$ days for a signal-to-noise ratio greater than 3 and further investigation is going to be presented in  future work \citep{2021_frb}. Additional goals can include pulsar science and new data in an unexplored frequency range, that can be used in combination with ongoing sky mapping and radio experiments in other  frequencies.

We also present a general description of the BINGO pipeline and its different modules. It covers input modules for sky emission models and experiment characteristics, a mission simulation module, production of TOD, calibration and cleaning in the time domain, map-making, generation of multifrequency maps, component separation for foreground removal, reconstructed \hi\, sky maps and cosmology models, comprising covariance matrix, bispectrum and Minkowski functional estimations. Additional information can be found in the companion papers IV, V and VI \citep{2020_sky_simulation,2020_component_separation,2020_mock_simulations}. 


The BINGO project aims to produce a map of the nearby Universe in a redshift range that, though observing the later universe, covers a large amount of the Universe volume (about 22 Mpc$^3$), thus being able to analyze most of the important structures of the Universe at a low cost. Indeed, it covers the turn from the epoch when matter dominated over DE ($z\approx 0.45,$, when the Universe was about 9 Gyr) to the accelerated era ($z\approx 0.13$, when the Universe was 12.05 Gyr),  thus including the turning point, $z\approx 0.32$, and providing an excellent study of the properties of DE.

The BINGO results for cosmology, prime attention given to the BAO structure at radio frequencies, provide constraints to the cosmological parameters, as well as presumably hinting at alternative cosmologies and further constraints thereof. 
Moreover, a number of astrophysical questions can be addressed, such as transient objects' science, for example, FRBs and pulsars, Galaxy science, and the nearby Universe.  
Despite being a so-called "off-the-shelf" telescope, there were (and are) a number of technical/technological challenges BINGO is facing to meet the conceptual design, such as the construction of an efficient and low-noise antenna and front end, the fabrication of 4.3 m long corrugated horns and 40 m-class dishes, at affordable costs and with the required performance standards.

Finally, the project involves cultural and educational aspects as well as  outreach activities that are developed simultaneously to its technical and scientific goals. 

\begin{acknowledgements}
 The BINGO project is supported by FAPESP grant 2014/07885-0; the support from CNPq is also gratefully acknowledged (E.A.). 
 C.A.W. acknowledges a CNPq grant 2014/313.597. 
 T.V. acknowledges CNPq Grant 308876/2014-8. 
 A.A.C. acknowledges financial support from the China Postdoctoral Science Foundation, grant number 2020M671611. 
 Y.Z.M. acknowledges the support from NRF-120385, NRF-120378 and NRF-109577.
 B. W. and A.A.C. were also supported by the key project of NNSFC under grant 11835009. 
 R.G.L. thanks Rui Luo and Kejia Lee for comments and clarifications on the FRB detection rate,  CAPES (process 88881.162206/2017-01) and the Alexander von Humboldt Foundation for the financial support. 
 F.B.A. acknowledges the UKRI-FAPESP grant 2019/05687-0, and FAPESP and USP for Visiting Professor Fellowships where this work has been developed.
 J.Z. was supported by IBS under the project code, IBS-R018-D1. 
 C.P.N. and K.S.F.F. would like to thank FAPESP for grants 2019/06040-0 and 2017/21570-0. 
 F.V. and E.J.M. acknowledge the support by CAPES. 
 A.R.Q., F.A.B., L.B., J.R.L.S., and M.V.S. acknowledge PRONEX/CNPq/FAPESQ-PB (Grant no. 165/2018).
 H.X. and Z.Z. are supported by the Ministry of Science and Technology of China (grant No. 2018YFA0404601 and 2020SKA0110200), and the National Science Foundation of China (grant Nos. 11621303, 11835009 and 11973033).
 S.A. acknowledges funding for a technical internship associated to FAPESP project 2014/07885-0
 M.P. acknowledges funding from a FAPESP Young Investigator fellowship, grant 2015/19936-1.
 V.L. acknowledges the postdoctoral FAPESP grant 2018/02026-0. 
 L.S. is supported by the National Key R\&D Program of China (2020YFC2201600).
 M.R. acknowledges funding from the European Research Council Grant CMBSPEC (No. 725456).
 Some of the results in this paper have been derived using the {\tt HEALPix} package \citep{Healpix:Gorski2004}. This research made use of {\tt Astropy},\footnote{\url{http://www.astropy.org}.} a community-developed core Python package for Astronomy \citep{astropy:2018}, {\tt NumPy} \citep{harris2020array} and {\tt Healpy} \citep{Zonca2019}. 
\end{acknowledgements}

\addcontentsline{toc}{section}{References}

\bibliographystyle{aa}
\bibliography{BINGO-01_the_project}

\begin{thebibliography}{259}
\expandafter\ifx\csname natexlab\endcsname\relax\def\natexlab#1{#1}\fi

\bibitem[{Abdalla \& Marins(2020)}]{Abdalla:2020ypg}
Abdalla, E. \& Marins, A. 2020, Int. J. Mod. Phys. D, 29, 2030014

\bibitem[{Abdalla {et~al.}(2010)Abdalla, Blake, \& Rawlings}]{Abdalla:2009wr}
Abdalla, F.~B., Blake, C., \& Rawlings, S. 2010, Mon.\ Not.\ R.\ Astron.\ Soc.,
  401, 743

\bibitem[{{Abdalla} {et~al.}(2015){Abdalla}, {Bull}, {Camera},
  {Benoit-L{\'e}vy}, {Joachimi}, {Kirk}, {Kloeckner}, {Maartens}, {Raccanelli},
  {Santos}, \& {Zhao}}]{2015aska.confE..17A}
{Abdalla}, F.~B., {Bull}, P., {Camera}, S., {et~al.} 2015, in Advancing
  Astrophysics with the Square Kilometre Array (AASKA14), 17

\bibitem[{Abdalla \& Rawlings(2005)}]{Abdalla:2004ah}
Abdalla, F.~B. \& Rawlings, S. 2005, Mon.\ Not.\ R.\ Astron.\ Soc., 360, 27

\bibitem[{Abdalla {et~al.}(2015)}]{Abdalla:2015kra}
Abdalla, F.~B. {et~al.} 2015, arXiv reprint:astro-ph, 1501.04035

\bibitem[{Abdalla {et~al.}(2021)}]{2020_optical_design}
Abdalla, F.~B. {et~al.} 2021 [\eprint[arXiv]{2107.01635}]

\bibitem[{{Abergel, A.} {et~al.}(2011){Abergel, A.}, {Ade, P. A. R.}, {Aghanim,
  N.}, \& et~al}]{Planck2011:dust}
{Abergel, A.}, {Ade, P. A. R.}, {Aghanim, N.}, \& et~al, A. 2011, Astron.\
  Astrophys., 536, A25

\bibitem[{Adam {et~al.}(2016{\natexlab{a}})}]{Adam:2015wua}
Adam, R. {et~al.} 2016{\natexlab{a}}, Astron. Astrophys., 594, A10

\bibitem[{Adam {et~al.}(2016{\natexlab{b}})}]{PlanckX}
Adam, R. {et~al.} 2016{\natexlab{b}}, Astron. Astrophys., 594, A10

\bibitem[{{Aditya}(2019)}]{2019MNRAS.482.5597A}
{Aditya}, J.~N.~H.~S. 2019, \mnras, 482, 5597

\bibitem[{{Aditya} \& {Kanekar}(2018{\natexlab{a}})}]{2018MNRAS.473...59A}
{Aditya}, J.~N.~H.~S. \& {Kanekar}, N. 2018{\natexlab{a}}, \mnras, 473, 59

\bibitem[{{Aditya} \& {Kanekar}(2018{\natexlab{b}})}]{2018MNRAS.481.1578A}
{Aditya}, J.~N.~H.~S. \& {Kanekar}, N. 2018{\natexlab{b}}, \mnras, 481, 1578

\bibitem[{Aghamousa {et~al.}(2016{\natexlab{a}})}]{Aghamousa:2016zmz}
Aghamousa, A. {et~al.} 2016{\natexlab{a}}, arXiv reprint:astro-ph
  [\eprint[arXiv]{1611.00036}]

\bibitem[{Aghamousa {et~al.}(2016{\natexlab{b}})}]{Aghamousa:2016sne}
Aghamousa, A. {et~al.} 2016{\natexlab{b}}, arXiv reprint:astro-ph, 1611.00037

\bibitem[{Aghanim {et~al.}(2020)}]{Aghanim:2018eyx}
Aghanim, N. {et~al.} 2020, Astron. Astrophys., 641, A6

\bibitem[{Ahmad {et~al.}(2002)}]{Ahmad:2002jz}
Ahmad, Q.~R. {et~al.} 2002, Phys. Rev. Lett., 89, 011301

\bibitem[{Akahori {et~al.}(2016)Akahori, Ryu, \& Gaensler}]{Akahori:2016ami}
Akahori, T., Ryu, D., \& Gaensler, B.~M. 2016, Astrophys.\ J., 824, 105

\bibitem[{Alam {et~al.}(2017)}]{Alam:2016hwk}
Alam, S. {et~al.} 2017, Mon. Not. Roy. Astron. Soc., 470, 2617

\bibitem[{Albrecht {et~al.}(2006)}]{Albrecht:2006um}
Albrecht, A. {et~al.} 2006, arXiv reprint:astro-ph, 0609591

\bibitem[{{Allison} {et~al.}(2012){Allison}, {Curran}, {Emonts}, {Ger{\'e}b},
  {Mahony}, {Reeves}, {Sadler}, {Tanna}, {Whiting}, \&
  {Zwaan}}]{2012MNRAS.423.2601A}
{Allison}, J.~R., {Curran}, S.~J., {Emonts}, B.~H.~C., {et~al.} 2012, \mnras,
  423, 2601

\bibitem[{{Allison} {et~al.}(2014){Allison}, {Sadler}, \&
  {Meekin}}]{2014MNRAS.440..696A}
{Allison}, J.~R., {Sadler}, E.~M., \& {Meekin}, A.~M. 2014, \mnras, 440, 696

\bibitem[{Amendola(2000)}]{Amendola:1999er}
Amendola, L. 2000, Phys. Rev., D62, 043511

\bibitem[{Amiri {et~al.}(2018)}]{Amiri:2018qsq}
Amiri, M. {et~al.} 2018, arXiv reprint:astro-ph, 1803.11235

\bibitem[{Amiri {et~al.}(2019)}]{Amiri:2019bjk}
Amiri, M. {et~al.} 2019, Nature, 566, 235

\bibitem[{Amiri {et~al.}(2020)}]{Amiri:2020gno}
Amiri, M. {et~al.} 2020, Nature, 582, 351

\bibitem[{Andersen {et~al.}(2019)}]{Andersen:2019yex}
Andersen, B.~C. {et~al.} 2019, Astrophys.\ J.\ Lett., 885, L24

\bibitem[{Andersen {et~al.}(2020)}]{Andersen:2020hvz}
Andersen, B.~C. {et~al.} 2020, Nature, 587, 54

\bibitem[{Anderson {et~al.}(2014)}]{Anderson:2013zyy}
Anderson, L. {et~al.} 2014, Mon.\ Not.\ R.\ Astron.\ Soc., 441, 24

\bibitem[{Asorey {et~al.}(2020)Asorey, Parkinson, Shi, Song, Ahn, Kim, Yao,
  Zhang, \& Zuo}]{asorey2020hir4}
Asorey, J., Parkinson, D., Shi, F., {et~al.} 2020, Monthly Notices of the Royal
  Astronomical Society, 495, 1788

\bibitem[{{Astropy Collaboration} {et~al.}(2018){Astropy Collaboration},
  {Price-Whelan}, {Sip{\H{o}}cz}, {G{\"u}nther}, {Lim}, {Crawford}, {Conseil},
  {Shupe}, {Craig}, {Dencheva}, {Ginsburg}, {VanderPlas}, {Bradley},
  {P{\'e}rez-Su{\'a}rez}, {de Val-Borro}, {Aldcroft}, {Cruz}, {Robitaille},
  {Tollerud}, {Ardelean}, {Babej}, {Bach}, {Bachetti}, {Bakanov}, {Bamford},
  {Barentsen}, {Barmby}, {Baumbach}, {Berry}, {Biscani}, {Boquien}, {Bostroem},
  {Bouma}, {Brammer}, {Bray}, {Breytenbach}, {Buddelmeijer}, {Burke},
  {Calderone}, {Cano Rodr{\'\i}guez}, {Cara}, {Cardoso}, {Cheedella}, {Copin},
  {Corrales}, {Crichton}, {D'Avella}, {Deil}, {Depagne}, {Dietrich}, {Donath},
  {Droettboom}, {Earl}, {Erben}, {Fabbro}, {Ferreira}, {Finethy}, {Fox},
  {Garrison}, {Gibbons}, {Goldstein}, {Gommers}, {Greco}, {Greenfield},
  {Groener}, {Grollier}, {Hagen}, {Hirst}, {Homeier}, {Horton}, {Hosseinzadeh},
  {Hu}, {Hunkeler}, {Ivezi{\'c}}, {Jain}, {Jenness}, {Kanarek}, {Kendrew},
  {Kern}, {Kerzendorf}, {Khvalko}, {King}, {Kirkby}, {Kulkarni}, {Kumar},
  {Lee}, {Lenz}, {Littlefair}, {Ma}, {Macleod}, {Mastropietro}, {McCully},
  {Montagnac}, {Morris}, {Mueller}, {Mumford}, {Muna}, {Murphy}, {Nelson},
  {Nguyen}, {Ninan}, {N{\"o}the}, {Ogaz}, {Oh}, {Parejko}, {Parley}, {Pascual},
  {Patil}, {Patil}, {Plunkett}, {Prochaska}, {Rastogi}, {Reddy Janga},
  {Sabater}, {Sakurikar}, {Seifert}, {Sherbert}, {Sherwood-Taylor}, {Shih},
  {Sick}, {Silbiger}, {Singanamalla}, {Singer}, {Sladen}, {Sooley},
  {Sornarajah}, {Streicher}, {Teuben}, {Thomas}, {Tremblay}, {Turner},
  {Terr{\'o}n}, {van Kerkwijk}, {de la Vega}, {Watkins}, {Weaver}, {Whitmore},
  {Woillez}, {Zabalza}, \& {Astropy Contributors}}]{astropy:2018}
{Astropy Collaboration}, {Price-Whelan}, A.~M., {Sip{\H{o}}cz}, B.~M., {et~al.}
  2018, \aj, 156, 123

\bibitem[{Bachega {et~al.}(2020)Bachega, Costa, Abdalla, \&
  Fornazier}]{Bachega:2019fki}
Bachega, R.~R., Costa, A.~A., Abdalla, E., \& Fornazier, K. 2020, JCAP, 05, 021

\bibitem[{Bacon {et~al.}(2020)}]{Bacon:2018dui}
Bacon, D.~J. {et~al.} 2020, Publ. Astron. Soc. Austral., 37, e007

\bibitem[{Bailes {et~al.}(2020)}]{bailes2020meerkat}
Bailes, M. {et~al.} 2020, Publ. Astron. Soc. Austral., 37, e028

\bibitem[{Baker {et~al.}(2011)Baker, Ferreira, Skordis, \&
  Zuntz}]{Baker:2011jy}
Baker, T., Ferreira, P.~G., Skordis, C., \& Zuntz, J. 2011, Phys. Rev., D84,
  124018

\bibitem[{Bandura {et~al.}(2014)}]{Bandura:2014gwa}
Bandura, K. {et~al.} 2014, Proc. SPIE Int. Soc. Opt. Eng., 9145, 22

\bibitem[{Bannister {et~al.}(2017)}]{Bannister:2017sie}
Bannister, K. {et~al.} 2017, Astrophys. J. Lett., 841, L12

\bibitem[{Bartelmann \& Schneider(2001)}]{Bartelmann:1999yn}
Bartelmann, M. \& Schneider, P. 2001, Phys. Rept., 340, 291

\bibitem[{Battye {et~al.}(2016)}]{Battye:2016qhf}
Battye, R. {et~al.} 2016 [\eprint[arXiv]{1610.06826}]

\bibitem[{{Battye} {et~al.}(2012){Battye}, {Brown}, {Browne}, {Davis},
  {Dewdney}, {Dickinson}, {Heron}, {Maffei}, {Pourtsidou}, \&
  {Wilkinson}}]{Battye:2012}
{Battye}, R.~A., {Brown}, M.~L., {Browne}, I.~W.~A., {et~al.} 2012, Proceedings
  of Moriond Cosmology [\eprint[arXiv]{1209.1041}]

\bibitem[{{Battye} {et~al.}(2013){Battye}, {Browne}, {Dickinson}, {Heron},
  {Maffei}, \& {Pourtsidou}}]{Battye:2013}
{Battye}, R.~A., {Browne}, I.~W.~A., {Dickinson}, C., {et~al.} 2013, MNRAS,
  434, 1239

\bibitem[{Battye {et~al.}(2004)Battye, Davies, \& Weller}]{Battye:2004re}
Battye, R.~A., Davies, R.~D., \& Weller, J. 2004, Mon. Not. Roy. Astron. Soc.,
  355, 1339

\bibitem[{Benitez {et~al.}(2009)}]{Benitez:2008ts}
Benitez, N. {et~al.} 2009, Astrophys. J. Lett., 692, L5

\bibitem[{Benitez {et~al.}(2014)}]{J-PAS:2014hgg}
Benitez, N. {et~al.} 2014 [\eprint[arXiv]{1403.5237}]

\bibitem[{{Bennett} {et~al.}(2003){Bennett}, {Hill}, \&
  {Hinshaw}}]{WMAP:Bennett2003}
{Bennett}, C.~L., {Hill}, R.~S., \& {Hinshaw}, G. e.~a. 2003, Astrophys.\ J.\
  Supp., 148, 97

\bibitem[{{Bennett} \& {Larson}(2013)}]{WMAP:Bennett2013}
{Bennett}, C.~L. \& {Larson}, D. e.~a. 2013, Astrophys.\ J.\ Supp., 208, 20

\bibitem[{{Bennett} {et~al.}(1992{\natexlab{a}}){Bennett}, {Smoot}, {Hinshaw},
  {Wright}, {Kogut}, {de Amici}, {Meyer}, {Weiss}, {Wilkinson}, {Gulkis},
  {Janssen}, {Boggess}, {Cheng}, {Hauser}, {Kelsall}, {Mather}, {Moseley},
  {Murdock}, \& {Silverberg}}]{Bennett:1992}
{Bennett}, C.~L., {Smoot}, G.~F., {Hinshaw}, G., {et~al.} 1992{\natexlab{a}},
  \apjl, 396, L7

\bibitem[{{Bennett} {et~al.}(1992{\natexlab{b}}){Bennett}, {Smoot}, {Hinshaw},
  {Wright}, {Kogut}, {de Amici}, \& {Meyer}}]{COBE:Bennett1992}
{Bennett}, C.~L., {Smoot}, G.~F., {Hinshaw}, G., {et~al.} 1992{\natexlab{b}},
  Astrophys.\ J.\ Lett., 396, L7

\bibitem[{{Berkhuijsen}(1972)}]{1972A&AS....5..263B}
{Berkhuijsen}, E.~M. 1972, \aaps, 5, 263

\bibitem[{Bertolami \& Landim(2018)}]{Bertolami:2017opd}
Bertolami, O. \& Landim, R.~G. 2018, Phys. Dark Univ., 21, 16

\bibitem[{{Beskin} {et~al.}(2015){Beskin}, {Chernov}, {Gwinn}, \&
  {Tchekhovskoy}}]{2015SSRv..191..207B}
{Beskin}, V.~S., {Chernov}, S.~V., {Gwinn}, C.~R., \& {Tchekhovskoy}, A.~A.
  2015, \ssr, 191, 207

\bibitem[{Beutler {et~al.}(2011)Beutler, Blake, Colless, Jones, Staveley-Smith,
  Campbell, Parker, Saunders, \& Watson}]{Beutler:2011hx}
Beutler, F., Blake, C., Colless, M., {et~al.} 2011, Mon.\ Not.\ R.\ Astron.\
  Soc., 416, 3017

\bibitem[{Bharadwaj \& Sethi(2001)}]{Bharadwaj:2001vs}
Bharadwaj, S. \& Sethi, S.~K. 2001, J. Astrophys. Astron., 22, 293

\bibitem[{Bigot-Sazy {et~al.}(2015)Bigot-Sazy, Dickinson, Battye, Browne, Ma,
  Maffei, Noviello, Remazeilles, \& Wilkinson}]{Bigot-Sazy:2015jaa}
Bigot-Sazy, M.~A., Dickinson, C., Battye, R.~A., {et~al.} 2015, Mon.\ Not.\ R.\
  Astron.\ Soc., 454, 3240

\bibitem[{Bigot-Sazy {et~al.}(2016)Bigot-Sazy, Ma, Battye, Browne, Chen,
  Dickinson, Harper, Maffei, Olivari, \& Wilkinson}]{Bigot-Sazy:2015tot}
Bigot-Sazy, M.-A., Ma, Y.-Z., Battye, R.~A., {et~al.} 2016, ASP Conf. Ser.,
  502, 41

\bibitem[{{Blake} {et~al.}(2004){Blake}, {Abdalla}, {Bridle}, \&
  {Rawlings}}]{SKA:Blake2004}
{Blake}, C.~A., {Abdalla}, F.~B., {Bridle}, S.~L., \& {Rawlings}, S. 2004, New
  Astron.\ Rev., 48, 1063

\bibitem[{Blanton {et~al.}(2017)}]{Blanton:2017qot}
Blanton, M.~R. {et~al.} 2017, Astron. J., 154, 28

\bibitem[{Bonetti {et~al.}(2017)Bonetti, Ellis, Mavromatos, Sakharov,
  Sarkisyan-Grinbaum, \& Spallicci}]{Bonetti:2017pym}
Bonetti, L., Ellis, J., Mavromatos, N.~E., {et~al.} 2017, Phys. Lett., B768,
  326

\bibitem[{Bonetti {et~al.}(2016)Bonetti, Ellis, Mavromatos, Sakharov,
  Sarkisyan-Grinbaum, \& Spallicci}]{Bonetti:2016cpo}
Bonetti, L., Ellis, J., Mavromatos, N.~E., {et~al.} 2016, Phys. Lett., B757,
  548

\bibitem[{Bonoli {et~al.}(2020)}]{Bonoli:2020ciz}
Bonoli, S. {et~al.} 2020 [\eprint[arXiv]{2007.01910}]

\bibitem[{Booth {et~al.}(2009)Booth, De~Blok, Jonas, \&
  Fanaroff}]{booth2009meerkat}
Booth, R., De~Blok, W., Jonas, J., \& Fanaroff, B. 2009, arXiv preprint
  arXiv:0910.2935

\bibitem[{{Bosma}(1981)}]{Bosma:1981}
{Bosma}, A. 1981, Astron.\ J., 86, 1825

\bibitem[{{Bourke}(2014)}]{Bourke2014}
{Bourke}, T.~L. 2014, in Advancing Astrophysics with the Square Kilometre
  Array, ed. T.~L. {Bourke}

\bibitem[{{Bull} {et~al.}(2015){Bull}, {Camera}, {Raccanelli}, {Blake},
  {Ferreira}, {Santos}, \& {Schwarz}}]{Bull:2015nra}
{Bull}, P., {Camera}, S., {Raccanelli}, A., {et~al.} 2015, in {PoS AASKA14
  (2015) 024}

\bibitem[{Burke {et~al.}(2019)Burke, Graham-Smith, \&
  Wilkinson}]{burke2019introduction}
Burke, B.~F., Graham-Smith, F., \& Wilkinson, P.~N. 2019, An introduction to
  radio astronomy (Cambridge University Press)

\bibitem[{{Burke-Spolaor}(2013)}]{Burke2012}
{Burke-Spolaor}, S. 2013, in IAU Symposium, Vol. 291, Neutron Stars and
  Pulsars: Challenges and Opportunities after 80 years, ed. J.~{van Leeuwen},
  95--100

\bibitem[{Caleb {et~al.}(2016)Caleb, Flynn, Bailes, Barr, Bateman, Bhandari,
  Campbell-Wilson, Green, Hunstead, Jameson, {et~al.}}]{caleb2016fast}
Caleb, M., Flynn, C., Bailes, M., {et~al.} 2016, Monthly Notices of the Royal
  Astronomical Society, 458, 718

\bibitem[{Caleb {et~al.}(2017)}]{Caleb:2017vbk}
Caleb, M. {et~al.} 2017, Mon. Not. Roy. Astron. Soc., 468, 3746

\bibitem[{Camera {et~al.}(2018)Camera, Fonseca, Maartens, \&
  Santos}]{Camera:2018jys}
Camera, S., Fonseca, J., Maartens, R., \& Santos, M.~G. 2018, Mon.\ Not.\ R.\
  Astron.\ Soc., 481, 1251

\bibitem[{{Carilli} {et~al.}(1998){Carilli}, {Menten}, {Reid}, {Rupen}, \&
  {Yun}}]{1998ApJ...494..175C}
{Carilli}, C.~L., {Menten}, K.~M., {Reid}, M.~J., {Rupen}, M.~P., \& {Yun},
  M.~S. 1998, \apj, 494, 175

\bibitem[{{Chandola} {et~al.}(2013){Chandola}, {Gupta}, \&
  {Saikia}}]{2013MNRAS.429.2380C}
{Chandola}, Y., {Gupta}, N., \& {Saikia}, D.~J. 2013, \mnras, 429, 2380

\bibitem[{{Chandola} {et~al.}(2011){Chandola}, {Sirothia}, \&
  {Saikia}}]{2011MNRAS.418.1787C}
{Chandola}, Y., {Sirothia}, S.~K., \& {Saikia}, D.~J. 2011, \mnras, 418, 1787

\bibitem[{{Chang} {et~al.}(2010){Chang}, {Pen}, {Bandura}, \&
  {Peterson}}]{Chang:2010jp}
{Chang}, T.-C., {Pen}, U.-L., {Bandura}, K., \& {Peterson}, J.~B. 2010, Nature,
  466, 463

\bibitem[{Chang {et~al.}(2008)Chang, Pen, Peterson, \& McDonald}]{Chang:2007xk}
Chang, T.-C., Pen, U.-L., Peterson, J.~B., \& McDonald, P. 2008, Phys.\ Rev.\
  Lett., 100, 091303

\bibitem[{Chawla {et~al.}(2020)}]{Chawla:2020rds}
Chawla, P. {et~al.} 2020, Astrophys. J., 896, L41

\bibitem[{Chen(2012)}]{Chen:2012xu}
Chen, X. 2012, Int. J. Mod. Phys. Conf. Ser., 12, 256

\bibitem[{Chevallier \& Polarski(2001)}]{Chevallier:2000qy}
Chevallier, M. \& Polarski, D. 2001, Int. J. Mod. Phys., D10, 213

\bibitem[{Chimento {et~al.}(2003)Chimento, Jakubi, Pavon, \&
  Zimdahl}]{Chimento:2003iea}
Chimento, L.~P., Jakubi, A.~S., Pavon, D., \& Zimdahl, W. 2003, Phys. Rev. D,
  67, 083513

\bibitem[{{Church}(1995)}]{Church:1995}
{Church}, S.~E. 1995, \mnras, 272, 551

\bibitem[{{Condon} \& {Ransom}(2016)}]{Condom:2016era}
{Condon}, J.~J. \& {Ransom}, S.~M. 2016, {Essential Radio Astronomy} (Princeton
  University Press)

\bibitem[{Connor(2019)}]{Connor:2019wnx}
Connor, L. 2019, Mon. Not. Roy. Astron. Soc., 487, 5753

\bibitem[{Connor {et~al.}(2020)}]{Connor:2020oay}
Connor, L. {et~al.} 2020, Mon. Not. Roy. Astron. Soc., 499, 4716

\bibitem[{Copeland {et~al.}(2006)Copeland, Sami, \&
  Tsujikawa}]{Copeland:2006wr}
Copeland, E.~J., Sami, M., \& Tsujikawa, S. 2006, Int. J. Mod. Phys., D15, 1753

\bibitem[{{Corbelli} {et~al.}(1989){Corbelli}, {Schneider}, \&
  {Salpeter}}]{Corbelli:1989}
{Corbelli}, E., {Schneider}, S.~E., \& {Salpeter}, E.~E. 1989, Astron.\ J., 97,
  390

\bibitem[{Costa {et~al.}(2017)Costa, Xu, Wang, \& Abdalla}]{Costa:2016tpb}
Costa, A.~A., Xu, X.-D., Wang, B., \& Abdalla, E. 2017, JCAP, 01, 028

\bibitem[{Costa {et~al.}(2021)}]{2020_forecast}
Costa, A.~A. {et~al.} 2021 [\eprint[arXiv]{2107.01639}]

\bibitem[{Curran \& Webb(2006)}]{Curran:2006md}
Curran, S.~J. \& Webb, J. 2006, Mon. Not. Roy. Astron. Soc., 371, 356

\bibitem[{{Darling} {et~al.}(2011){Darling}, {Macdonald}, {Haynes}, \&
  {Giovanelli}}]{2011ApJ...742...60D}
{Darling}, J., {Macdonald}, E.~P., {Haynes}, M.~P., \& {Giovanelli}, R. 2011,
  \apj, 742, 60

\bibitem[{DeBoer {et~al.}(2009)DeBoer, Gough, Bunton, Cornwell, Beresford,
  Johnston, Feain, Schinckel, Jackson, Kesteven,
  {et~al.}}]{deboer2009australian}
DeBoer, D.~R., Gough, R.~G., Bunton, J.~D., {et~al.} 2009, Proceedings of the
  IEEE, 97, 1507

\bibitem[{Delubac {et~al.}(2015)}]{Delubac:2014aqe}
Delubac, T. {et~al.} 2015, Astron.\ Astrophys., 574, A59

\bibitem[{{Deng} \& {Zhang}(2014)}]{Deng:2013aga}
{Deng}, W. \& {Zhang}, B. 2014, Astrophys.\ J.\ Lett., 783, L35

\bibitem[{Di~Dio {et~al.}(2014)Di~Dio, Montanari, Durrer, \&
  Lesgourgues}]{DiDio:2013sea}
Di~Dio, E., Montanari, F., Durrer, R., \& Lesgourgues, J. 2014, JCAP, 1401, 042

\bibitem[{{Dickey}(1986)}]{1986ApJ...300..190D}
{Dickey}, J.~M. 1986, \apj, 300, 190

\bibitem[{{Dickinson}(2014)}]{Dickinson:2014}
{Dickinson}, C. 2014, Proceedings of Moriond Cosmology
  [\eprint[arXiv]{1405.7936}]

\bibitem[{{Dickinson} {et~al.}(2018){Dickinson}, {Ali-Ha{\"\i}moud}, {Barr},
  {Battistelli}, {Bell}, {Bernstein}, {Casassus}, {Cleary}, {Draine},
  {G{\'e}nova-Santos}, {Harper}, {Hensley}, {Hill-Valler}, {Hoang}, {Israel},
  {Jew}, {Lazarian}, {Leahy}, {Leech}, {L{\'o}pez-Caraballo}, {McDonald},
  {Murphy}, {Onaka}, {Paladini}, {Peel}, {Perrott}, {Poidevin}, {Readhead},
  {Rubi{\~n}o-Mart{\'\i}n}, {Taylor}, {Tibbs}, {Todorovi{\'c}}, \&
  {Vidal}}]{Dickinson:2018}
{Dickinson}, C., {Ali-Ha{\"\i}moud}, Y., {Barr}, A., {et~al.} 2018, New
  Astron.\ Rev., 80, 1

\bibitem[{{Dickinson} {et~al.}(2003){Dickinson}, {Davies}, \&
  {Davis}}]{Dickinson:2003}
{Dickinson}, C., {Davies}, R.~D., \& {Davis}, R.~J. 2003, Mon.\ Not.\ R.\
  Astron.\ Soc., 341, 369

\bibitem[{dos Santos {et~al.}(2021)dos Santos, Landim, {et~al.}}]{2021_frb}
dos Santos, M., Landim, R.~G., {et~al.} 2021, The BINGO Project VIII: Fast
  Radio Bursts detection rate and interferometry

\bibitem[{{Dragone}(1978)}]{Dragone:1978}
{Dragone}, C. 1978, AT T Technical Journal, 57, 2663

\bibitem[{Draine \& Lazarian(1999)}]{Draine99}
Draine, B. \& Lazarian, A. 1999, Astrophys. J., 512, 740

\bibitem[{Ducout {et~al.}(2013)Ducout, Bouchet, Colombi, Pogosyan, \&
  Prunet}]{Ducout:2013}
Ducout, A., Bouchet, F., Colombi, S., Pogosyan, D., \& Prunet, S. 2013, Mon.\
  Not.\ R.\ Astron.\ Soc., 429, 2104

\bibitem[{Eisenstein {et~al.}(2005)}]{Eisenstein:2005su}
Eisenstein, D.~J. {et~al.} 2005, Astrophys.\ J., 633, 560

\bibitem[{Feng {et~al.}(2008)Feng, Wang, Abdalla, \& Su}]{Feng:2008fx}
Feng, C., Wang, B., Abdalla, E., \& Su, R.-K. 2008, Phys. Lett., B665, 111

\bibitem[{Fixsen(2009)}]{fixsen2009temperature}
Fixsen, D. 2009, The Astrophysical Journal, 707, 916

\bibitem[{Fornazier {et~al.}(2021)}]{2020_component_separation}
Fornazier, K. S.~F. {et~al.} 2021 [\eprint[arXiv]{2107.01637}]

\bibitem[{Fujita {et~al.}(2017)Fujita, Akahori, Umetsu, Sarazin, \&
  Wong}]{Fujita:2016yve}
Fujita, Y., Akahori, T., Umetsu, K., Sarazin, C.~L., \& Wong, K.-W. 2017,
  Astrophys.\ J., 834, 13

\bibitem[{Fukuda {et~al.}(1998)}]{Fukuda:1998mi}
Fukuda, Y. {et~al.} 1998, Phys. Rev. Lett., 81, 1562

\bibitem[{Gajjar {et~al.}(2018)}]{Gajjar:2018bth}
Gajjar, V. {et~al.} 2018, Astrophys. J., 863, 2

\bibitem[{{Gallimore} {et~al.}(1999){Gallimore}, {Baum}, {O'Dea}, {Pedlar}, \&
  {Brinks}}]{1999ApJ...524..684G}
{Gallimore}, J.~F., {Baum}, S.~A., {O'Dea}, C.~P., {Pedlar}, A., \& {Brinks},
  E. 1999, \apj, 524, 684

\bibitem[{Gay {et~al.}(2012)Gay, Pichon, \& Pogosyan}]{2012/gay}
Gay, C., Pichon, C., \& Pogosyan, D. 2012, Physical Review D, 85, 023011

\bibitem[{{Ger{\'e}b} {et~al.}(2015){Ger{\'e}b}, {Maccagni}, {Morganti}, \&
  {Oosterloo}}]{2015A&A...575A..44G}
{Ger{\'e}b}, K., {Maccagni}, F.~M., {Morganti}, R., \& {Oosterloo}, T.~A. 2015,
  \aap, 575, A44

\bibitem[{Giannantonio {et~al.}(2010)Giannantonio, Martinelli, Silvestri, \&
  Melchiorri}]{Giannantonio:2009gi}
Giannantonio, T., Martinelli, M., Silvestri, A., \& Melchiorri, A. 2010, JCAP,
  04, 030

\bibitem[{{Gold} {et~al.}(2011){Gold}, {Odegard}, {Weiland}, {Hill}, {Kogut},
  {Bennett}, {Hinshaw}, {Chen}, {Dunkley}, {Halpern}, {Jarosik}, {Komatsu},
  {Larson}, {Limon}, {Meyer}, {Nolta}, {Page}, {Smith}, {Spergel}, {Tucker},
  {Wollack}, \& {Wright}}]{WMAP:Gold2011}
{Gold}, B., {Odegard}, N., {Weiland}, J.~L., {et~al.} 2011, Astrophys.\ J.\
  Supp., 192, 15

\bibitem[{Gorski {et~al.}(2005)Gorski, Hivon, Banday, Wandelt, Hansen,
  Reinecke, \& Bartelman}]{Healpix:Gorski2004}
Gorski, K.~M., Hivon, E., Banday, A.~J., {et~al.} 2005, Astrophys.\ J., 622,
  759

\bibitem[{Green {et~al.}(2011)}]{Green:2011zi}
Green, J. {et~al.} 2011, arXiv reprint:astro-ph, 1108.1374

\bibitem[{{Gupta} {et~al.}(2006){Gupta}, {Salter}, {Saikia}, {Ghosh}, \&
  {Jeyakumar}}]{2006MNRAS.373..972G}
{Gupta}, N., {Salter}, C.~J., {Saikia}, D.~J., {Ghosh}, T., \& {Jeyakumar}, S.
  2006, \mnras, 373, 972

\bibitem[{{Hall} {et~al.}(2013){Hall}, {Bonvin}, \& {Challinor}}]{Hall:2012wd}
{Hall}, A., {Bonvin}, C., \& {Challinor}, A. 2013, Phys.\ Rev.\ D, D87, 064026

\bibitem[{Hanany \& Rosenkranz(2003)}]{Hanany03}
Hanany, S. \& Rosenkranz, P. 2003, New Astronomy Reviews, 47, 1159

\bibitem[{{Harper} \& {Dickinson}(2018)}]{Harper:2018b}
{Harper}, S.~E. \& {Dickinson}, C. 2018, \mnras, 479, 2024

\bibitem[{{Harper} {et~al.}(2018){Harper}, {Dickinson}, {Battye},
  {Roychowdhury}, {Browne}, {Ma}, {Olivari}, \& {Chen}}]{Harper:2018}
{Harper}, S.~E., {Dickinson}, C., {Battye}, R.~A., {et~al.} 2018, Mon.\ Not.\
  R.\ Astron.\ Soc., 478, 2416

\bibitem[{Harris {et~al.}(2020)Harris, Millman, van~der Walt, Gommers,
  Virtanen, Cournapeau, Wieser, Taylor, Berg, Smith, Kern, Picus, Hoyer, van
  Kerkwijk, Brett, Haldane, del R{'{\i}}o, Wiebe, Peterson,
  G{'{e}}rard-Marchant, Sheppard, Reddy, Weckesser, Abbasi, Gohlke, \&
  Oliphant}]{harris2020array}
Harris, C.~R., Millman, K.~J., van~der Walt, S.~J., {et~al.} 2020, Nature, 585,
  357

\bibitem[{{Haskell} \& {Melatos}(2015)}]{2015IJMPD..2430008H}
{Haskell}, B. \& {Melatos}, A. 2015, International Journal of Modern Physics D,
  24, 1530008

\bibitem[{He {et~al.}(2009)He, Wang, \& Abdalla}]{He:2008si}
He, J.-H., Wang, B., \& Abdalla, E. 2009, Phys. Lett. B, 671, 139

\bibitem[{He {et~al.}(2011)He, Wang, \& Abdalla}]{He:2010im}
He, J.-H., Wang, B., \& Abdalla, E. 2011, Phys. Rev., D83, 063515

\bibitem[{He {et~al.}(2010)He, Wang, Abdalla, \& Pavon}]{He:2010ta}
He, J.-H., Wang, B., Abdalla, E., \& Pavon, D. 2010, JCAP, 1012, 022

\bibitem[{{Hobbs} \& {Dai}(2017)}]{Hobbs2017}
{Hobbs}, G. \& {Dai}, S. 2017, National Science Review, 4, 707

\bibitem[{Hojjati {et~al.}(2012)Hojjati, Pogosian, Silvestri, \&
  Talbot}]{Hojjati:2012rf}
Hojjati, A., Pogosian, L., Silvestri, A., \& Talbot, S. 2012, Phys. Rev. D, 86,
  123503

\bibitem[{{Howlett} {et~al.}(2015){Howlett}, {Manera}, \&
  {Percival}}]{howlett2015picola}
{Howlett}, C., {Manera}, M., \& {Percival}, W.~J. 2015, Astronomy and
  Computing, 12, 109

\bibitem[{Hu \& Sawicki(2007)}]{Hu:2007nk}
Hu, W. \& Sawicki, I. 2007, Phys. Rev. D, 76, 064004

\bibitem[{Hu {et~al.}(2020)Hu, Wang, Wu, Wang, Zhang, \& Chen}]{Hu:2019okh}
Hu, W., Wang, X., Wu, F., {et~al.} 2020, Mon. Not. Roy. Astron. Soc., 493, 5854

\bibitem[{{Huchtmeier}(1975)}]{Huchtmeier:1975}
{Huchtmeier}, W.~K. 1975, Astron.\ Astrophys., 45, 259

\bibitem[{Jiang {et~al.}(2020)Jiang, Tang, Hou, Liu, Kr{\v{c}}o, Qian, Sun,
  Ching, Liu, Duan, {et~al.}}]{jiang2020fundamental}
Jiang, P., Tang, N.-Y., Hou, L.-G., {et~al.} 2020, Research in Astronomy and
  Astrophysics, 20, 064

\bibitem[{Johnston {et~al.}(2008)Johnston, Taylor, Bailes, Bartel, Baugh,
  Bietenholz, Blake, Braun, Brown, Chatterjee, {et~al.}}]{johnston2008science}
Johnston, S., Taylor, R., Bailes, M., {et~al.} 2008, Experimental astronomy,
  22, 151

\bibitem[{{Keane}(2013)}]{Keane13}
{Keane}, E.~F. 2013, in IAU Symposium, Vol. 291, Neutron Stars and Pulsars:
  Challenges and Opportunities after 80 years, ed. J.~{van Leeuwen}, 295--300

\bibitem[{{Keane} \& {McLaughlin}(2011)}]{Keane:2011}
{Keane}, E.~F. \& {McLaughlin}, M.~A. 2011, Bulletin of the Astronomical
  Society of India, 39, 333

\bibitem[{Keith {et~al.}(2010)}]{Keith:2010kk}
Keith, M. {et~al.} 2010, Mon. Not. Roy. Astron. Soc., 409, 619

\bibitem[{Kim {et~al.}(2015)Kim, Park, L'Huillier, \& Hong}]{kim2015horizon}
Kim, J., Park, C., L'Huillier, B., \& Hong, S.~E. 2015, JKAS, 48, 213

\bibitem[{Komatsu(2001)}]{Komatsu}
Komatsu, E. 2001, PhD thesis, Tohoku U.

\bibitem[{Komatsu {et~al.}(2003)Komatsu, Kogut, Nolta, Bennett, Halpern,
  Hinshaw, Jarosik, Limon, Meyer, Page, {et~al.}}]{WMAP:Komatsu2003}
Komatsu, E., Kogut, A., Nolta, M., {et~al.} 2003, Astrophys.\ J.\ Supp., 148,
  119

\bibitem[{Kovetz {et~al.}(2017)}]{Kovetz:2017}
Kovetz, E.~D. {et~al.} 2017 [\eprint[arXiv]{1709.09066}]

\bibitem[{{Kumamoto} {et~al.}(2021){Kumamoto}, {Dai}, {Johnston}, {Kerr},
  {Shannon}, {Weltevrede}, {Sobey}, {Manchester}, {Hobbs}, \&
  {Takahashi}}]{2021MNRAS.501.4490K}
{Kumamoto}, H., {Dai}, S., {Johnston}, S., {et~al.} 2021, \mnras, 501, 4490

\bibitem[{Kumar {et~al.}(2019)}]{Kumar:2019htf}
Kumar, P. {et~al.} 2019, Astrophys.\ J., 887, L30

\bibitem[{Landim(2020)}]{Landim:2020ked}
Landim, R.~G. 2020, Eur. Phys. J. C, 80, 913

\bibitem[{{Lay} \& {Halverson}(2000)}]{Lay:2000}
{Lay}, O.~P. \& {Halverson}, N.~W. 2000, \apj, 543, 787

\bibitem[{Leitch {et~al.}(1997)Leitch, Readhead, Pearson, \& Myers}]{Leitch}
Leitch, E., Readhead, A., Pearson, T., \& Myers, S. 1997, Astrophys. J. Lett.,
  486, L23

\bibitem[{Levi {et~al.}(2011)Levi, Kim, Lampton, \& Sholl}]{Levi:2011dq}
Levi, M.~E., Kim, A.~G., Lampton, M.~L., \& Sholl, M.~J. 2011
  [\eprint[arXiv]{1105.0959}]

\bibitem[{Lewis {et~al.}(2002)}]{Lewis:2002yk}
Lewis, I.~J. {et~al.} 2002, Mon. Not. Roy. Astron. Soc., 333, 279

\bibitem[{{Li}(2016)}]{2016ASPC..502...93L}
{Li}, D. 2016, in Astronomical Society of the Pacific Conference Series, Vol.
  502, Frontiers in Radio Astronomy and FAST Early Sciences Symposium 2015, ed.
  L.~{Qain} \& D.~{Li}, 93--97

\bibitem[{Li {et~al.}(2020)}]{li2020tianlai}
Li, J. {et~al.} 2020, Sci. China Phys. Mech. Astron., 63, 129862

\bibitem[{Li {et~al.}(2018)Li, Gao, Ding, Wang, \& Zhang}]{Li:2017mek}
Li, Z.-X., Gao, H., Ding, X.-H., Wang, G.-J., \& Zhang, B. 2018, Nature
  Commun., 9, 3833

\bibitem[{Liccardo {et~al.}(2021)}]{2020_sky_simulation}
Liccardo, V. {et~al.} 2021 [\eprint[arXiv]{2107.01636}]

\bibitem[{{Liguori} {et~al.}(2010){Liguori}, {Sefusatti}, {Fergusson}, \&
  {Shellard}}]{Liguori}
{Liguori}, M., {Sefusatti}, E., {Fergusson}, J.~R., \& {Shellard}, E.~P.~S.
  2010, Advances in Astronomy, 2010, 980523

\bibitem[{Linder(2003)}]{Linder:2002et}
Linder, E.~V. 2003, Phys. Rev. Lett., 90, 091301

\bibitem[{Linder(2020)}]{Linder:2020aru}
Linder, E.~V. 2020, Phys. Rev. D, 101, 103019

\bibitem[{Liu \& Tegmark(2011)}]{Liu:2011hh}
Liu, A. \& Tegmark, M. 2011, Phys. Rev., D83, 103006

\bibitem[{Loeb \& Wyithe(2008)}]{Loeb:2008hg}
Loeb, A. \& Wyithe, S. 2008, Phys.\ Rev.\ Lett., 100, 161301

\bibitem[{Lorimer {et~al.}(2013)Lorimer, Karastergiou, McLaughlin, \&
  Johnston}]{Lorimer:2013roa}
Lorimer, D., Karastergiou, A., McLaughlin, M., \& Johnston, S. 2013, Mon. Not.
  Roy. Astron. Soc., 436, 5

\bibitem[{Lorimer {et~al.}(2007)Lorimer, Bailes, McLaughlin, Narkevic, \&
  Crawford}]{Lorimer:2007qn}
Lorimer, D.~R., Bailes, M., McLaughlin, M.~A., Narkevic, D.~J., \& Crawford, F.
  2007, Science, 318, 777

\bibitem[{Lorimer {et~al.}(2012)Lorimer, Lyne, McLaughlin, Pavlov, \&
  Chang}]{Lorimer:2012mn}
Lorimer, D.~R., Lyne, A.~G., McLaughlin, M.~A., Pavlov, G.~G., \& Chang, C.
  2012, Astrophys.\ J., 758, 141

\bibitem[{Luo {et~al.}(2018)Luo, Lee, Lorimer, \& Zhang}]{Luo:2018tiy}
Luo, R., Lee, K., Lorimer, D.~R., \& Zhang, B. 2018, Mon. Not. Roy. Astron.
  Soc., 481, 2320

\bibitem[{Luo {et~al.}(2020)Luo, Men, Lee, Wang, Lorimer, \&
  Zhang}]{Luo:2020wfx}
Luo, R., Men, Y., Lee, K., {et~al.} 2020, Mon. Not. Roy. Astron. Soc., 494, 665

\bibitem[{Maartens {et~al.}(2015)Maartens, Abdalla, Jarvis, \&
  Santos}]{Maartens:2015mra}
Maartens, R., Abdalla, F.~B., Jarvis, M., \& Santos, M.~G. 2015, PoS, AASKA14,
  016

\bibitem[{{Maccagni} {et~al.}(2017){Maccagni}, {Morganti}, {Oosterloo},
  {Ger{\'e}b}, \& {Maddox}}]{2017A&A...604A..43M}
{Maccagni}, F.~M., {Morganti}, R., {Oosterloo}, T.~A., {Ger{\'e}b}, K., \&
  {Maddox}, N. 2017, \aap, 604, A43

\bibitem[{Madau {et~al.}(1997)Madau, Meiksin, \& Rees}]{Madau:1996cs}
Madau, P., Meiksin, A., \& Rees, M.~J. 1997, Astrophys.\ J., 475, 429

\bibitem[{{Maino} {et~al.}(1999){Maino}, {Burigana}, {Maltoni}, {Wandelt},
  {G{\'o}rski}, {Malaspina}, {Bersanelli}, {Mand olesi}, {Banday}, \&
  {Hivon}}]{maino99}
{Maino}, D., {Burigana}, C., {Maltoni}, M., {et~al.} 1999, \aaps, 140, 383

\bibitem[{{Manchester} {et~al.}(2005){Manchester}, {Hobbs}, {Teoh}, \&
  {Hobbs}}]{2005AJ....129.1993M}
{Manchester}, R.~N., {Hobbs}, G.~B., {Teoh}, A., \& {Hobbs}, M. 2005, \aj, 129,
  1993

\bibitem[{Masui {et~al.}(2015)}]{Masui:2015kmb}
Masui, K. {et~al.} 2015, Nature, 528, 523

\bibitem[{{Masui} {et~al.}(2013){Masui}, {Switzer}, {Banavar}, {Bandura},
  {Blake}, {Calin}, {Chang}, {Chen}, {Li}, {Liao}, {Natarajan}, {Pen},
  {Peterson}, {Shaw}, \& {Voytek}}]{Masui:2013}
{Masui}, K.~W., {Switzer}, E.~R., {Banavar}, N., {et~al.} 2013, Astrophys.\ J.\
  Lett., 763, L20

\bibitem[{McLaughlin {et~al.}(2006)}]{McLaughlin:2005eq}
McLaughlin, M.~A. {et~al.} 2006, Nature, 439, 817

\bibitem[{{Meinhold} {et~al.}(2009){Meinhold}, {Leonardi}, {Aja}, {Artal},
  {Battaglia}, {Bersanelli}, {Blackhurst}, {Butler}, {Cuevas}, {Cuttaia},
  {D'Arcangelo}, {Davis}, {de la Fuente}, {Frailis}, {Franceschet},
  {Franceschi}, {Gaier}, {Galeotta}, {Gregorio}, {Hoyland}, {Hughes},
  {Jukkala}, {Kettle}, {Laaninen}, {Leutenegger}, {Lowe}, {Malaspina}, {Mand
  olesi}, {Maris}, {Mart{\'\i}nez-Gonz{\'a}lez}, {Mendes}, {Mennella},
  {Miccolis}, {Morgante}, {Roddis}, {Sandri}, {Seiffert}, {Salm{\'o}n},
  {Stringhetti}, {Poutanen}, {Terenzi}, {Tomasi}, {Tuovinen}, {Varis},
  {Valenziano}, {Villa}, {Wilkinson}, {Winder}, {Zacchei}, \&
  {Zonca}}]{Meinhold09}
{Meinhold}, P., {Leonardi}, R., {Aja}, B., {et~al.} 2009, Journal of
  Instrumentation, 12, T12009

\bibitem[{Minkowski(1903)}]{Minkowski:1903}
Minkowski, H. 1903, Mathematische Annalen, 57, 447

\bibitem[{{Moore} {et~al.}(1999){Moore}, {Carilli}, \&
  {Menten}}]{1999ApJ...510L..87M}
{Moore}, C.~B., {Carilli}, C.~L., \& {Menten}, K.~M. 1999, \apjl, 510, L87

\bibitem[{{Morganti} \& {Oosterloo}(2018)}]{2018A&ARv..26....4M}
{Morganti}, R. \& {Oosterloo}, T. 2018, \aapr, 26, 4

\bibitem[{{Morganti} {et~al.}(2005){Morganti}, {Tadhunter}, \&
  {Oosterloo}}]{2005A&A...444L...9M}
{Morganti}, R., {Tadhunter}, C.~N., \& {Oosterloo}, T.~A. 2005, \aap, 444, L9

\bibitem[{Muñoz {et~al.}(2016)Muñoz, Kovetz, Dai, \&
  Kamionkowski}]{Munoz:2016tmg}
Muñoz, J.~B., Kovetz, E.~D., Dai, L., \& Kamionkowski, M. 2016, Phys.\ Rev.\
  Lett., 117, 091301

\bibitem[{Nan {et~al.}(2011)Nan, Li, Jin, Wang, Zhu, Zhu, Zhang, Yue, \&
  Qian}]{Nan:2011um}
Nan, R., Li, D., Jin, C., {et~al.} 2011, Int. J. Mod. Phys., D20, 989

\bibitem[{Novaes {et~al.}(2018)Novaes, Bernui, Xavier, \&
  Marques}]{Novaes:2018}
Novaes, C.~P., Bernui, A., Xavier, H.~S., \& Marques, G.~A. 2018, Mon.\ Not.\
  R.\ Astron.\ Soc., 478, 3253

\bibitem[{Novikov {et~al.}(1999)Novikov, Feldman, \& Shandarin}]{Novikov:1999}
Novikov, D., Feldman, H.~A., \& Shandarin, S.~F. 1999, International Journal of
  Modern Physics D, 8, 291

\bibitem[{Olivari {et~al.}(2018)Olivari, Dickinson, Battye, Ma, Costa,
  Remazeilles, \& Harper}]{Olivari:2017}
Olivari, L.~C., Dickinson, C., Battye, R.~A., {et~al.} 2018, Mon.\ Not.\ R.\
  Astron.\ Soc., 473, 4242

\bibitem[{Olivari {et~al.}(2016)Olivari, Remazeilles, \&
  Dickinson}]{Olivari:2015}
Olivari, L.~C., Remazeilles, M., \& Dickinson, C. 2016, Mon.\ Not.\ R.\
  Astron.\ Soc., 456, 2749

\bibitem[{{Olling}(1996)}]{Olling:1996}
{Olling}, R.~P. 1996, Astron.\ J., 112, 457

\bibitem[{{Peel} {et~al.}(2019){Peel}, {Wuensche}, {Abdalla}, {Ant{\'o}n},
  {Barosi}, {Browne}, {Caldas}, {Dickinson}, {Fornazier}, {Monstein},
  {Strauss}, {Tancredi}, \& {Villela}}]{Peel:2019}
{Peel}, M.~W., {Wuensche}, C.~A., {Abdalla}, E., {et~al.} 2019, Journal of
  Astronomical Instrumentation, 8, 1940005

\bibitem[{{Pen}(2018)}]{Pen:2018ilo}
{Pen}, U.-L. 2018, Nature Astronomy, 2, 842

\bibitem[{Percival {et~al.}(2010)}]{Percival:2009xn}
Percival, W.~J. {et~al.} 2010, Mon.\ Not.\ R.\ Astron.\ Soc., 401, 2148

\bibitem[{Peterson {et~al.}(2006)Peterson, Bandura, \& Pen}]{Peterson:2006bx}
Peterson, J.~B., Bandura, K., \& Pen, U.~L. 2006, in {Proceedings, 41st
  Rencontres de Moriond, 2006 Contents and structure of the universe: La
  Thuile, Val d'Aoste, Italy, Mar 18-25, 2006}, 283--289

\bibitem[{{Petroff} {et~al.}(2019){Petroff}, {Hessels}, \&
  {Lorimer}}]{Petroff:2019tty}
{Petroff}, E., {Hessels}, J.~W.~T., \& {Lorimer}, D.~R. 2019, The Astron.
  Astrophys. Rev., 27, 4

\bibitem[{{Pihlstr{\"o}m} {et~al.}(2003){Pihlstr{\"o}m}, {Conway}, \&
  {Vermeulen}}]{2003A&A...404..871P}
{Pihlstr{\"o}m}, Y.~M., {Conway}, J.~E., \& {Vermeulen}, R.~C. 2003, \aap, 404,
  871

\bibitem[{{Planck Collaboration} {et~al.}(2020){Planck Collaboration},
  {Akrami}, {Ashdown}, {Aumont}, \& {Baccigalupi}}]{Akrami2018}
{Planck Collaboration}, {Akrami}, Y., {Ashdown}, M., {Aumont}, J., \&
  {Baccigalupi}, C. e.~a. 2020, Astron. Astrophys., 641, A4

\bibitem[{Platts {et~al.}(2019)Platts, Weltman, Walters, Tendulkar, Gordin, \&
  Kandhai}]{Platts:2018hiy}
Platts, E., Weltman, A., Walters, A., {et~al.} 2019, Phys.\ Rep., 821
  [\eprint[arXiv]{1810.05836}], [Phys. Rept.821,1(2019)]

\bibitem[{Pober {et~al.}(2013)Pober, Parsons, DeBoer, McDonald, McQuinn,
  Aguirre, Ali, Bradley, Chang, \& Morales}]{Pober:2012zz}
Pober, J.~C., Parsons, A.~R., DeBoer, D.~R., {et~al.} 2013, Astron.\ J., 145,
  65

\bibitem[{Popov {et~al.}(2018)Popov, Postnov, \& Pshirkov}]{Popov:2018hkz}
Popov, S.~B., Postnov, K.~A., \& Pshirkov, M.~S. 2018, Phys. Usp., 61, 965

\bibitem[{{Prochaska} {et~al.}(2005){Prochaska}, {Herbert-Fort}, \&
  {Wolfe}}]{SDSS_Lya}
{Prochaska}, J.~X., {Herbert-Fort}, S., \& {Wolfe}, A.~M. 2005, \apj, 635, 123

\bibitem[{{Puche} {et~al.}(1991){Puche}, {Carignan}, \& {van
  Gorkom}}]{Puche:1991}
{Puche}, D., {Carignan}, C., \& {van Gorkom}, J.~H. 1991, Astron.\ J., 101, 456

\bibitem[{Rajwade {et~al.}(2020)Rajwade, Mickaliger, Stappers, Morello,
  Agarwal, Bassa, Breton, Caleb, Karastergiou, Keane,
  {et~al.}}]{rajwade2020possible}
Rajwade, K., Mickaliger, M., Stappers, B., {et~al.} 2020, Monthly Notices of
  the Royal Astronomical Society, 495, 3551

\bibitem[{{{Ravi}, Vikram}(2019)}]{Ravi:2018ose}
{{Ravi}, Vikram}. 2019, Astrophys.\ J., 872, 88

\bibitem[{{Refregier}(2003)}]{Refregier:2003ct}
{Refregier}, A. 2003, Ann.\ Rev.\ Astron.\ Astrophys., 41, 645

\bibitem[{{Refregier} {et~al.}(2010){Refregier}, {Amara}, {Kitching}, {Rassat},
  {Scaramella}, \& {Weller}}]{EUCLID:Refregier2010}
{Refregier}, A., {Amara}, A., {Kitching}, T.~D., {et~al.} 2010, arXiv e-prints
  [\eprint[arXiv]{1001.0061}]

\bibitem[{{Reich} \& {Reich}(1986)}]{Reich:1986}
{Reich}, P. \& {Reich}, W. 1986, Astron.\ Astrophys. \ Suppl., 63, 205

\bibitem[{{Reich} {et~al.}(2001){Reich}, {Testori}, \&
  {Reich}}]{2001A&A...376..861R}
{Reich}, P., {Testori}, J.~C., \& {Reich}, W. 2001, \aap, 376, 861

\bibitem[{{Remazeilles} {et~al.}(2011){Remazeilles}, {Delabrouille}, \&
  {Cardoso}}]{Remazeilles:2011}
{Remazeilles}, M., {Delabrouille}, J., \& {Cardoso}, J.-F. 2011, Mon.\ Not.\
  R.\ Astron.\ Soc., 418, 467

\bibitem[{Riess {et~al.}(2019)Riess, Casertano, Yuan, Macri, \&
  Scolnic}]{Riess:2019cxk}
Riess, A.~G., Casertano, S., Yuan, W., Macri, L.~M., \& Scolnic, D. 2019,
  Astrophys. J., 876, 85

\bibitem[{{Roberts}(1970)}]{1970ApJ...161L...9R}
{Roberts}, M.~S. 1970, \apjl, 161, L9

\bibitem[{{Roychowdhury} {et~al.}(2019){Roychowdhury}, {Dickinson}, \&
  {Browne}}]{2019A&A...631A.115R}
{Roychowdhury}, S., {Dickinson}, C., \& {Browne}, I. W.~A. 2019, \aap, 631,
  A115

\bibitem[{{Santos} {et~al.}(2015){Santos}, {Bull}, {Alonso}, {Camera},
  {Ferreira}, {Bernardi}, {Maartens}, {Viel}, {Villaescusa-Navarro}, {Abdalla},
  {Jarvis}, {Metcalf}, {Pourtsidou}, \& {Wolz}}]{2015aska.confE..19S}
{Santos}, M., {Bull}, P., {Alonso}, D., {et~al.} 2015, in Advancing
  Astrophysics with the Square Kilometre Array (AASKA14), 19

\bibitem[{Schechter(1976)}]{Schechter:1976iz}
Schechter, P. 1976, Astrophys. J., 203, 297

\bibitem[{{Seiffert} {et~al.}(2002){Seiffert}, {Mennella}, {Burigana},
  {Mandolesi}, {Bersanelli}, {Meinhold}, \& {Lubin}}]{Seiffert:2002}
{Seiffert}, M., {Mennella}, A., {Burigana}, C., {et~al.} 2002, Astron.\
  Astrophys., 391, 1185

\bibitem[{Sethi(2005)}]{Sethi:2005gv}
Sethi, S.~K. 2005, Mon.\ Not.\ R.\ Astron.\ Soc., 363, 818

\bibitem[{Shannon {et~al.}(2018)Shannon, Macquart, Bannister, Ekers, James,
  Os{\l}owski, Qiu, Sammons, Hotan, Voronkov, {et~al.}}]{shannon2018dispersion}
Shannon, R., Macquart, J.-P., Bannister, K.~W., {et~al.} 2018, Nature, 562, 386

\bibitem[{Shao \& Zhang(2017)}]{Shao2017}
Shao, L. \& Zhang, B. 2017, Phys.\ Rev.\ D, 95, 123010

\bibitem[{Shapiro-Albert {et~al.}(2018)Shapiro-Albert, McLaughlin, \&
  Keane}]{Shapiro2018}
Shapiro-Albert, B., McLaughlin, M., \& Keane, E. 2018, Astrophys.\ J., 866, 152

\bibitem[{Shaw \& Lewis(2008)}]{Shaw:2008aa}
Shaw, J.~R. \& Lewis, A. 2008, Phys.\ Rev.\ D, D78, 103512

\bibitem[{Shaw {et~al.}(2014)Shaw, Sigurdson, Pen, Stebbins, \&
  Sitwell}]{Shaw:2013wza}
Shaw, J.~R., Sigurdson, K., Pen, U.-L., Stebbins, A., \& Sitwell, M. 2014,
  Astrophys.\ J., 781, 57

\bibitem[{Shull \& Danforth(2018)}]{Shull:2017eow}
Shull, J.~M. \& Danforth, C.~W. 2018, Astrophys.\ J.\ Lett., 852, L11

\bibitem[{{Smith} \& {Zaldarriaga}(2011)}]{Zaldarriaga}
{Smith}, K.~M. \& {Zaldarriaga}, M. 2011, Mon.\ Not.\ R.\ Astron.\ Soc., 417, 2

\bibitem[{Smoot \& Debono(2017)}]{Smoot:2014oia}
Smoot, G.~F. \& Debono, I. 2017, Astron. Astrophys., 597, A136

\bibitem[{Spitler {et~al.}(2014)}]{Spitler:2014fla}
Spitler, L. {et~al.} 2014, Astrophys. J., 790, 101

\bibitem[{Spitler {et~al.}(2016)}]{Spitler:2016dmz}
Spitler, L.~G. {et~al.} 2016, Nature, 531, 202

\bibitem[{{Square Kilometre Array Cosmology Science Working Group}
  {et~al.}(2020{\natexlab{a}}){Square Kilometre Array Cosmology Science Working
  Group}, {Bacon}, {Battye}, {Bull}, {Camera}, {Ferreira}, {Harrison},
  {Parkinson}, {Pourtsidou}, {Santos}, {Wolz}, {Abdalla}, {Akrami}, {Alonso},
  {Andrianomena}, {Ballardini}, {Bernal}, {Bertacca}, {Bengaly}, {Bonaldi},
  {Bonvin}, {Brown}, {Chapman}, {Chen}, {Chen}, {Cunnington}, {Davis},
  {Dickinson}, {Fonseca}, {Grainge}, {Harper}, {Jarvis}, {Maartens}, {Maddox},
  {Padmanabhan}, {Pritchard}, {Raccanelli}, {Rivi}, {Roychowdhury},
  {Sahl{\'e}n}, {Schwarz}, {Siewert}, {Viel}, {Villaescusa-Navarro}, {Xu},
  {Yamauchi}, \& {Zuntz}}]{SKAWG:2020}
{Square Kilometre Array Cosmology Science Working Group}, {Bacon}, D.~J.,
  {Battye}, R.~A., {et~al.} 2020{\natexlab{a}}, Publ.\ of the Astr.\ Soc.\
  Asia, 37, e007

\bibitem[{{Square Kilometre Array Cosmology Science Working Group}
  {et~al.}(2020{\natexlab{b}}){Square Kilometre Array Cosmology Science Working
  Group}, {Bacon}, {Battye}, {Bull}, {Camera}, {Ferreira}, {Harrison},
  {Parkinson}, {Pourtsidou}, {Santos}, {Wolz}, {Abdalla}, {Akrami}, {Alonso},
  {Andrianomena}, {Ballardini}, {Bernal}, {Bertacca}, {Bengaly}, {Bonaldi},
  {Bonvin}, {Brown}, {Chapman}, {Chen}, {Chen}, {Cunnington}, {Davis},
  {Dickinson}, {Fonseca}, {Grainge}, {Harper}, {Jarvis}, {Maartens}, {Maddox},
  {Padmanabhan}, {Pritchard}, {Raccanelli}, {Rivi}, {Roychowdhury},
  {Sahl{\'e}n}, {Schwarz}, {Siewert}, {Viel}, {Villaescusa-Navarro}, {Xu},
  {Yamauchi}, \& {Zuntz}}]{2020PASA...37....7S}
{Square Kilometre Array Cosmology Science Working Group}, {Bacon}, D.~J.,
  {Battye}, R.~A., {et~al.} 2020{\natexlab{b}}, \pasa, 37, e007

\bibitem[{{Strukov} {et~al.}(1993){Strukov}, {Brukhanov}, {Skulachev}, \&
  {Sazhin}}]{Strukov1993}
{Strukov}, I.~A., {Brukhanov}, A.~A., {Skulachev}, D.~P., \& {Sazhin}, M.~V.
  1993, Physics Letters B, 315, 198

\bibitem[{{Strukov} \& {Skulachev}(1984)}]{Strukov1984}
{Strukov}, I.~A. \& {Skulachev}, D.~P. 1984, Soviet Astronomy Letters, 10, 1

\bibitem[{{Switzer} {et~al.}(2013){Switzer}, {Masui}, {Bandura}, {Calin},
  {Chang}, {Chen}, {Li}, {Liao}, {Natarajan}, {Pen}, {Peterson}, {Shaw}, \&
  {Voytek}}]{Switzer:2013}
{Switzer}, E.~R., {Masui}, K.~W., {Bandura}, K., {et~al.} 2013, Mon.\ Not.\ R.\
  Astron.\ Soc., 434, L46

\bibitem[{Tansella {et~al.}(2018)Tansella, Bonvin, Durrer, Ghosh, \&
  Sellentin}]{Tansella:2017rpi}
Tansella, V., Bonvin, C., Durrer, R., Ghosh, B., \& Sellentin, E. 2018, JCAP,
  1803, 019

\bibitem[{Tello {et~al.}(2000)Tello, Villela, Smoot, Torres, \&
  Bersanelli}]{Tello:2000}
Tello, C., Villela, T., Smoot, G.~F., Torres, S., \& Bersanelli, M. 2000, in
  {IAU Symposium 201}: {New Cosmological Data and the Value of the Fundamental
  Parameters}

\bibitem[{{Tello} {et~al.}(2013){Tello}, {Villela}, {Torres}, {Bersanelli},
  {Smoot}, {Ferreira}, {Cingoz}, {Lamb}, {Barbosa}, {Perez-Becker},
  {Ricciardi}, {Currivan}, {Platania}, \& {Maino}}]{Tello:2013}
{Tello}, C., {Villela}, T., {Torres}, S., {et~al.} 2013, Astron.\ Astrophys.,
  556, A1

\bibitem[{Tingay {et~al.}(2015)}]{Tingay:2015wdv}
Tingay, S.~J. {et~al.} 2015, Astron.\ J., 150, 199

\bibitem[{{Tyson} {et~al.}(2002){Tyson}, {Wittman}, {Hennawi}, \&
  {Spergel}}]{2002TysonAPS}
{Tyson}, T., {Wittman}, D., {Hennawi}, J., \& {Spergel}, D. 2002, in APS April
  Meeting Abstracts, APS Meeting Abstracts, Y6.004

\bibitem[{van Bemmel {et~al.}(2012)van Bemmel, van Ardenne, Geralt bij~de
  Vaate, Faulkner, \& Morganti}]{vanBemmel:2012cj}
van Bemmel, I.~M., van Ardenne, A., Geralt bij~de Vaate, J., Faulkner, A.~J.,
  \& Morganti, R. 2012, PoS, RTS2012, 037

\bibitem[{{van Gorkom} {et~al.}(1989){van Gorkom}, {Knapp}, {Ekers}, {Ekers},
  {Laing}, \& {Polk}}]{1989AJ.....97..708V}
{van Gorkom}, J.~H., {Knapp}, G.~R., {Ekers}, R.~D., {et~al.} 1989, \aj, 97,
  708

\bibitem[{{Vermeulen} {et~al.}(2003){Vermeulen}, {Pihlstr{\"o}m}, {Tschager},
  {de Vries}, {Conway}, {Barthel}, {Baum}, {Braun}, {Bremer}, {Miley}, {O'Dea},
  {R{\"o}ttgering}, {Schilizzi}, {Snellen}, \& {Taylor}}]{2003A&A...404..861V}
{Vermeulen}, R.~C., {Pihlstr{\"o}m}, Y.~M., {Tschager}, W., {et~al.} 2003,
  \aap, 404, 861

\bibitem[{Visbal {et~al.}(2009)Visbal, Loeb, \& Wyithe}]{Visbal:2008rg}
Visbal, E., Loeb, A., \& Wyithe, J. S.~B. 2009, J. Cosmol. Astropart. Phys.,
  0910, 030

\bibitem[{Walters {et~al.}(2018)Walters, Weltman, Gaensler, Ma, \&
  Witzemann}]{Walters:2017afr}
Walters, A., Weltman, A., Gaensler, B., Ma, Y.-Z., \& Witzemann, A. 2018,
  Astrophys. J., 856, 65

\bibitem[{Wang {et~al.}(2016)Wang, Abdalla, Atrio-Barandela, \&
  Pavon}]{Wang:2016lxa}
Wang, B., Abdalla, E., Atrio-Barandela, F., \& Pavon, D. 2016, Rep.\ Prog.\
  Phys., 79, 096901

\bibitem[{Wang {et~al.}(2007)Wang, Zang, Lin, Abdalla, \&
  Micheletti}]{Wang:2006qw}
Wang, B., Zang, J., Lin, C.-Y., Abdalla, E., \& Micheletti, S. 2007, Nucl.
  Phys. B, 778, 69

\bibitem[{{Wang} \& {Mohanty}(2017)}]{2017PhRvL.118o1104W}
{Wang}, Y. \& {Mohanty}, S.~D. 2017, \prl, 118, 151104

\bibitem[{Wang \& Wang(2018)}]{Wang:2018ydd}
Wang, Y.~K. \& Wang, F.~Y. 2018, Astron.\ Astrophys., 614, A50

\bibitem[{Wei {et~al.}(2015)Wei, Gao, Wu, \& Mészáros}]{Wei:2015hwd}
Wei, J.-J., Gao, H., Wu, X.-F., \& Mészáros, P. 2015, Phys. Rev. Lett., 115,
  261101

\bibitem[{Wei \& Wu(2018)}]{Wei:2018pyh}
Wei, J.-J. \& Wu, X.-F. 2018, JCAP, 1807, 045

\bibitem[{Wei {et~al.}(2018)Wei, Wu, \& Gao}]{Wei:2018cgd}
Wei, J.-J., Wu, X.-F., \& Gao, H. 2018, Astrophys. J. Lett., 860, L7

\bibitem[{Wei {et~al.}(2017)Wei, Zhang, Zhang, \& Wu}]{Wei:2016jgc}
Wei, J.-J., Zhang, E.-K., Zhang, S.-B., \& Wu, X.-F. 2017, Res. Astron.
  Astrophys., 17, 13

\bibitem[{{Wilkinson} {et~al.}(2004){Wilkinson}, {Kellermann}, {Ekers},
  {Cordes}, \& {W.~Lazio}}]{SKA:Wilkinson2004}
{Wilkinson}, P.~N., {Kellermann}, K.~I., {Ekers}, R.~D., {Cordes}, J.~M., \&
  {W.~Lazio}, T.~J. 2004, New Astron.\ Rev., 48, 1551

\bibitem[{{Wilson} {et~al.}(2013){Wilson}, {Rohlfs}, \&
  {H{\"u}ttemeister}}]{Wilson:2013trabook}
{Wilson}, T.~L., {Rohlfs}, K., \& {H{\"u}ttemeister}, S. 2013, {Tools of Radio
  Astronomy} (Springer)

\bibitem[{Wu {et~al.}(2021)}]{Wu:2020}
Wu, F. {et~al.} 2021, \mnras, 506, 3455

\bibitem[{Wu {et~al.}(2016)Wu, Zhang, Gao, Wei, Zou, Lei, Zhang, Dai, \&
  M{\'e}sz{\'a}ros}]{Wu2016}
Wu, X.-F., Zhang, S.-B., Gao, H., {et~al.} 2016, Astrophys.\ J.\ Lett., 822,
  L15

\bibitem[{Wuensche(2019)}]{Wuensche:2018}
Wuensche, C.~A. 2019, J. Phys. Conf. Ser., 1269, 012002

\bibitem[{Wuensche {et~al.}(2021{\natexlab{a}})Wuensche, Abdalla, Abdalla,
  Barosi, Wang, An, Barretos, Battye, Brito, Browne,
  {et~al.}}]{Wuensche:2021cfs}
Wuensche, C.~A., Abdalla, E., Abdalla, F., {et~al.} 2021{\natexlab{a}}, Anais
  da Academia Brasileira de Ci{\^e}ncias, 93

\bibitem[{{Wuensche} {et~al.}(2020){Wuensche}, {Reitano}, {Peel}, {Browne},
  {Maffei}, {Abdalla}, {Radcliffe}, {Abdalla}, {Barosi}, {Liccardo}, {Mericia},
  {Pisano}, {Strauss}, {Vieira}, {Villela}, \& {Wang}}]{Wuensche:2019}
{Wuensche}, C.~A., {Reitano}, L., {Peel}, M.~W., {et~al.} 2020, Experimental
  Astronomy, 50, 125

\bibitem[{Wuensche {et~al.}(2021{\natexlab{b}})}]{BINGO.Instrument:2020}
Wuensche, C.~A. {et~al.} 2021{\natexlab{b}} [\eprint[arXiv]{2107.01634}]

\bibitem[{Xavier {et~al.}(2016)Xavier, Abdalla, \& Joachimi}]{Xavier:2016}
Xavier, H.~S., Abdalla, F.~B., \& Joachimi, B. 2016, Mon.\ Not.\ R.\ Astron.\
  Soc., 459, 3693

\bibitem[{Xing {et~al.}(2019)Xing, Gao, Wei, Li, Wang, Zhang, Wu, \&
  Mészáros}]{Xing:2019geq}
Xing, N., Gao, H., Wei, J., {et~al.} 2019, Astrophys.\ J.\ Lett., 882, L13

\bibitem[{Yang \& Zhang(2016)}]{Yang:2016zbm}
Yang, Y.-P. \& Zhang, B. 2016, Astrophys. J. Lett., 830, L31

\bibitem[{{Young} {et~al.}(2015){Young}, {Weltevrede}, {Stappers}, {Lyne}, \&
  {Kramer}}]{Young15}
{Young}, N.~J., {Weltevrede}, P., {Stappers}, B.~W., {Lyne}, A.~G., \&
  {Kramer}, M. 2015, \mnras, 449, 1495

\bibitem[{Yu \& Wang(2017)}]{Yu:2017beg}
Yu, H. \& Wang, F. 2017, Astron. Astrophys., 606, A3

\bibitem[{Yu \& Wang(2018)}]{Yu:2018slt}
Yu, H. \& Wang, F. 2018, Eur. Phys. J. C, 78, 692

\bibitem[{Yu {et~al.}(2018)Yu, Xi, \& Wang}]{Yu:2017xbb}
Yu, H., Xi, S., \& Wang, F. 2018, Astrophys. J., 860, 173

\bibitem[{Zhang(2014)}]{Zhang:2013lta}
Zhang, B. 2014, Astrophys. J. Lett., 780, L21

\bibitem[{Zhang {et~al.}(2021)}]{2020_mock_simulations}
Zhang, J. {et~al.} 2021 [\eprint[arXiv]{2107.01638}]

\bibitem[{{Zhou} {et~al.}(2014){Zhou}, {Li}, {Wang}, {Fan}, \&
  {Wei}}]{Zhou:2014yta}
{Zhou}, B., {Li}, X., {Wang}, T., {Fan}, Y.-Z., \& {Wei}, D.-M. 2014, Phys.\
  Rev.\ D, 89, 107303

\bibitem[{Zimdahl \& Pavon(2001)}]{Zimdahl:2001ar}
Zimdahl, W. \& Pavon, D. 2001, Phys.\ Lett.\ B, B521, 133

\bibitem[{Zitrin \& Eichler(2018)}]{Zitrin:2018let}
Zitrin, A. \& Eichler, D. 2018, Astrophys. J., 866, 101

\bibitem[{Zonca {et~al.}(2019)Zonca, Singer, Lenz, Reinecke, Rosset, Hivon, \&
  Gorski}]{Zonca2019}
Zonca, A., Singer, L., Lenz, D., {et~al.} 2019, Journal of Open Source
  Software, 4, 1298

\end{thebibliography}

\end{document}